\documentclass[useAMS,usenatbib]{mn2e}

\usepackage{graphics,float}                
\usepackage{epsf}
\usepackage[pdftex]{graphicx}  

\usepackage{amsmath}                                
\usepackage{amssymb}
\usepackage{picinpar}                               
\usepackage{relsize,setspace,subfigure}
\usepackage{tabularx}
\usepackage[innercaption]{sidecap}                  
\usepackage{url}
\usepackage{aas_macros}

\usepackage{natbib}

\usepackage[usenames]{color}

\definecolor{MyGrey10}{rgb}{0.10,0.10,0.10}
\definecolor{MyGrey30}{rgb}{0.30,0.30,0.30}
\definecolor{MyGrey80}{rgb}{0.80,0.80,0.80}
\definecolor{MyGrey90}{rgb}{0.90,0.90,0.90}
\definecolor{MyGrey95}{rgb}{0.95,0.95,0.95}
\definecolor{MyGrey98}{rgb}{0.98,0.98,0.98}
\definecolor{MyGrey100}{rgb}{1.0,1.0,1.0}
\definecolor{MyYellow}{rgb}{1.0,1.0,0.0}

\def\etal{{\it et al}.}

\def\hmpc{h^{-1} {\rm Mpc}}

\def\Mpch{~h^{-1} {\rm Mpc}}

\title[Structural Analysis of the SDSS Cosmic Web, Density Field Reconstruction]{Structural Analysis of the SDSS Cosmic Web I.\\   
Nonlinear Density Field Reconstructions}

\author[Platen et al.]{Erwin Platen$^{1}$, Rien van de Weygaert$^{1}$\thanks{E-mail: weygaert@astro.rug.nl}, Bernard J.T. Jones$^{1}$, Gert Vegter$^{2}$, 
\newauthor  and Miguel A. Arag\'on Calvo$^{3}$ \\
  $^{1}$Kapteyn Astronomical Institute, University of Groningen, P.O.  Box 800, 9700 AV, Groningen, The Netherlands.\\
  $^{2}$Bernoulli Institute for Mathematics and Computer Science, University of Groningen, P.O. Box 407, 9700 AK Groningen, the Netherlands\\
  $^{3}$Department of Physics and Astronomy, Johns Hopkins University, 3400 North Charles Street, Baltimore,MD21218 , USA}

\date{Accepted .... Received ...; in original form ...}

\pagerange{\pageref{firstpage}--\pageref{lastpage}} \pubyear{2011}

\begin{document}

\label{firstpage}

\maketitle

\begin{abstract}
This study is the first in a series in which we analyze the structure
and topology of the Cosmic Web as traced by the Sloan Digital Sky
Survey.  The main issue addressed in the present study is the
translation of the irregularly distributed discrete spatial data in
the galaxy redshift survey into a representative density field. The
density field will form the basis for a statistical, topological and
cosmographic study of the cosmic density field in our Local Universe.

We investigate the ability of three reconstruction techniques to
analyze and investigate weblike features and geometries in a discrete
distribution of objects. The three methods are the linear Delaunay
Tessellation Field Estimator (DTFE), its higher order equivalent
Natural Neighbour Field Estimator (NNFE) and a version of Kriging
interpolation adapted to the specific circumstances encountered in
galaxy redshift surveys, the {\it Natural Lognormal Kriging}
technique. DTFE and NNFE are based on the local geometry defined by
the Voronoi and Delaunay tessellations of the galaxy distribution.

The three reconstruction methods are analysed and compared using mock
magnitude-limited and volume-limited SDSS redshift surveys, obtained
on the basis of the Millennium simulation. We investigate error trends, biases
and the topological structure of the resulting fields, concentrating on the void 
population identified by the Watershed Void Finder. Environmental
effects are addressed by evaluating the density fields on a range of
Gaussian filter scales.  Comparison with the void population in the
original simulation yields the fraction of false void mergers and
false void splits.

In most tests DTFE, NNFE and Kriging have largely similar density and
topology error behaviour. Cosmetically, higher order NNFE and Kriging
methods produce more visually appealing reconstructions. Quantitatively, 
however, DTFE performs better, even while computationally 
far less demanding. A successful recovery of the void population on small 
scales appears to be difficult, while the void recovery rate improves 
significantly on scales $> 3\Mpch$. A study of small scale voids and the 
void galaxy population should therefore be restricted to the local Universe, 
out to at most 100$\Mpch$. 
\end{abstract}


\section{Introduction}
Over the past thirty years a clear paradigm has emerged as large
redshift surveys opened the window onto the distribution of matter in
our Local Universe: galaxies, intergalactic gas and dark matter exist
in a wispy weblike spatial arrangement consisting of dense compact
clusters, elongated filaments, and sheetlike walls, amidst large
near-empty void regions, with similar patterns existing at higher
redshift, albeit over smaller scales.  The Cosmic Web is the
fundamental spatial organization of matter on scales of a few up to a
hundred Megaparsec, scales at which the Universe still resides in a
state of moderate dynamical evolution
\cite{Peebles80,Zeldovich70,Bondweb96}.  Its appearance has been most
dramatically illustrated by the recently produced maps of the nearby
cosmos, the 2dF galaxy redshift survey (2dFGRS), the Sloan Digital Sky Survey (SDSS) and 
the 2MASS redshift surveys \cite{Colless03,Tegmark04,Huchra05}.

According to the standard lore of structure formation, structures
emerged from small perturbations in the primordial field of Gaussian
density and velocity perturbations. Under the force of gravity these
fluctuations grow and cluster to become the present day observed
structures. At large scales the density field has been evolving
(quasi)-linearly and still retains much information on cosmological
parameters and structure formation. The linear density field provides
an abundance of probes from which cosmological information can be
extracted. Prominent probes are the clustering of galaxies, which has
been used to infer the underlying primordial power spectrum of density
fluctuations \citep{Peebles80,Percival01,Tegmark04}, the temperature
and polarisation anisotropies in the Cosmic Microwave Background
\citep[e.g.][]{Smoot92,Spergel03} and the shearing of galaxy images by
the gravitationally lensed photon paths through the inhomogeneous
matter distribution \citep{Mellier99,Massey07,Hoekstra08}.

While these (quasi-)linear cosmological probes have yielded an
impressive amount of cosmological information, the exploitation of the
pronounced nonlinear patterns of the cosmic web towards probing
cosmological parameters and cosmic structure formation has been less
fortuitous. Even though the morphology, shape and other statistical
characteristics of the quasi-linear cosmological density field forms a
direct reflection of the structure assembly process in the Universe,
on the corresponding small scales nonlinear growth has significantly
altered and erased some of the essential cosmological information. The
absence of an objective and quantitative procedure for identifying and
isolating clusters, filaments and voids in the cosmic matter
distribution has been a major obstacle in investigating the structure
and dynamics of the Cosmic Web. The overwhelming complexity of the
individual structures and their connectivity, the huge range of
densities and the intrinsic multi-scale nature prevent the use of
simple tools that may be sufficient in less demanding
problems. However, various interesting new approaches and methods have
been forwarded in the past few years, often based on ideas stemming
from image processing, mathematical morphology, and medical
imaging \citep[e.g.]{Aragon07,Aragon10,Sousbie10,Way11,Genovese10}.

\begin{figure*}
  \centering
    \includegraphics[width=0.9\textwidth]{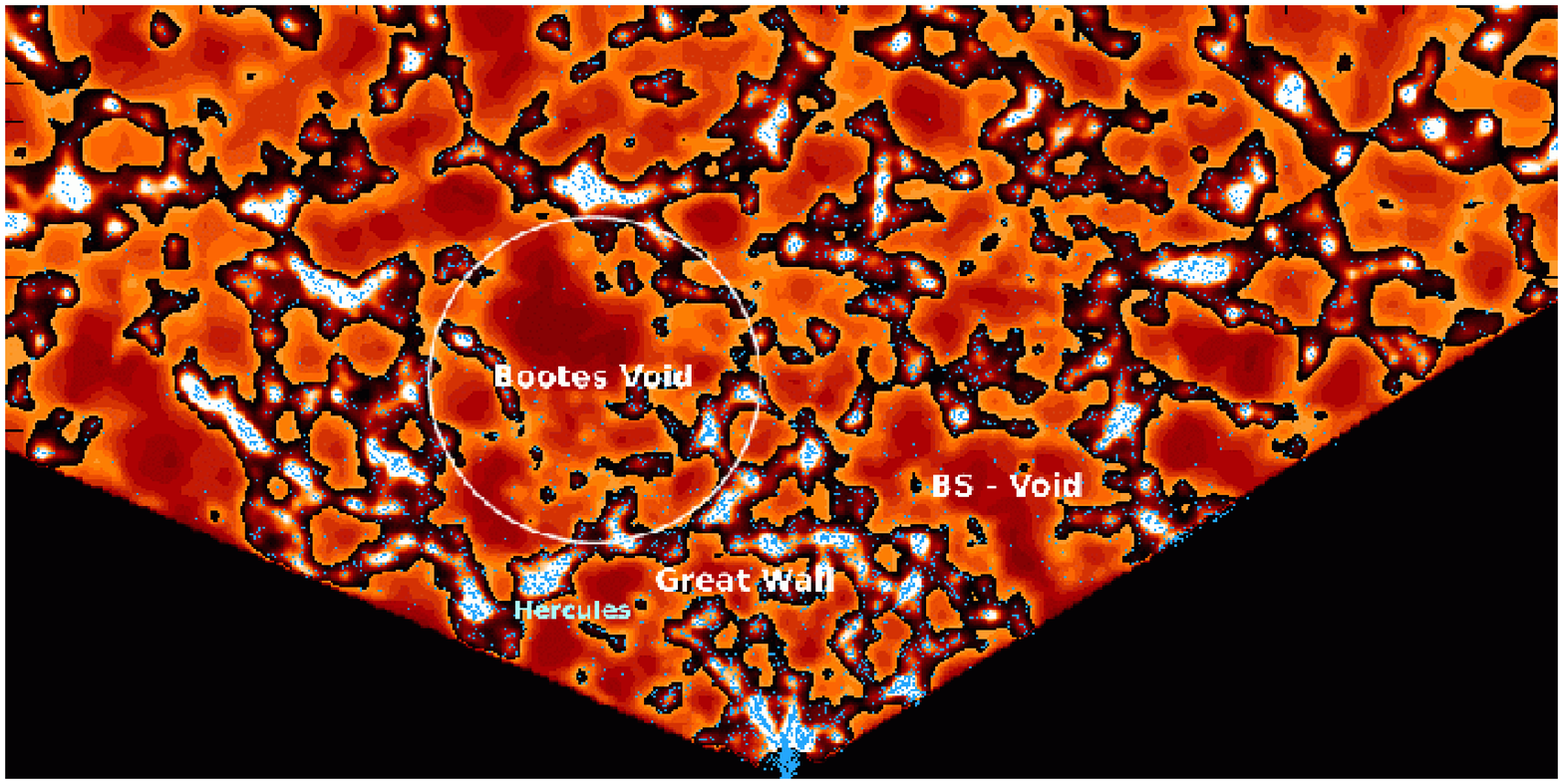}
    \caption{A visualisation of the (DTFE) SDSS density field
    (1$\Mpch$), the contour levels are divided roughly into the
    overdense and the underdense regime. Both the galaxies (blue dots)
    and the density represent a slice of 12$\Mpch$ thickness. Some of
    the most prominent features have been named, like the {\it
    Bo\"otes SuperVoid} \citep{Kirshner81} and the large supervoid
    ({\it BS SuperVoid}) identified by \citet{Bahcall82}. Also the
    largest overdense structure, the {\it(Coma) Great Wall}, and the
    location of the {\it Hercules supercluster} within the Wall are
    indicated.}
    \label{fig:nicefigure}
\end{figure*}

This study is the first in a series in which we systematically
investigate the web-related structures found in the Sloan Digital Sky
Survey (SDSS). It involves a systematic program in which we explore
the cosmography of the Local Universe, assess the statistical
characteristics of the density field, identify and categorize the
voids and filaments within the SDSS galaxy redshift sample, and study
the biasing of the galaxy population with respect to the mass
distribution and the dependence of galaxy properties on the large
scale environment.

The intention of this paper is the reconstruction of the underlying
(nonlinear) density field by translating the spatially irregularly
distributed and discrete galaxy sample in the SDSS galaxy redshift
survey into a representative density field.  The density field
reconstructions described in this study will form the basis of a
series of studies in which we address the statistical and topological
properties of the cosmic density field in our Local Universe and in
which we analyze the cosmography, void population and spinal structure
of the local Cosmic Web in the Sloan Digital Sky Survey. The
cosmographic description of the nearby Universe includes the
identification and cataloguing of the filaments, voids and clusters in
the SDSS galaxy distribution. The key motivation for the density field
reconstruction techniques should therefore be that it allows us to
probe the complex quasi-linear (and nonlinear) structures that we find
in galaxy redshift surveys. This involves the ability to reproduce the
distinct anisotropic - filamentary and wall-like - features that make
up the Cosmic Web, as well as the ability to trace the hierarchical
substructure of the cosmic matter distribution and the ability to
identify the void population that forms one of its most salient
features.

We are specifically interested in the performance of the fast and
efficient linear tessellation-based DTFE density field reconstruction
technique \citep{Schaapwey00,Schaap07,Weyschaap09}. In addition, we
investigate two additional higher-order techniques which are
potentially suited for representing the quasi-linear density field of
the cosmic web, the local natural neighbour interpolation (NNFE) and a
nonlocal Kriging interpolation technique, the Natural Lognormal
Kriging formalism. By means of an extensive comparison we evaluate
which aspects of the density field are best reproduced by either DTFE,
NNFE or Natural Lognormal Kriging, and which of these methods is best
suited to function for further structural analysis.

In order to be able to assess the reliability of the results of our
structural tools, it is crucial to understand the details and errors
in reconstructed density fields. We will therefore present a detailed
comparison study between three different reconstructions methods, and
assess their density and topological errors over a range of scales.
This will range form the small nonlinear scales, at ~1$\Mpch$, to a
scale of ~10$\Mpch$, which represents the transition from quasi-linear
to the linear regime.

We focus on the translation of the galaxy positions in the SDSS DR6
galaxy redshift survey to a representative density field within the
survey volume. Currently, SDSS encloses the largest and deepest
contiguous region of the nearby Universe mapped by a galaxy redshift
survey. This makes the SDSS an ideal data sample for a full
three-dimensional density field reconstruction. A first impression of
the resulting DTFE density field map of a region in the SDSS DR6
survey is shown in figure~\ref{fig:nicefigure}. Our techniques will be
generally applicable to any uniform galaxy redshift survey. Even
though survey limits and scales are different from that of the SDSS,
it will be straightforward to carry over the results to other redshift
surveys.

\subsection{Inferring Cosmic Density Fields}
The richest source of data for investigating the intricate web-like cosmic 
matter distribution is the distribution of galaxies in
galaxy redshift surveys.  We shall make the assumption that the galaxy
distribution is a representative tracer of the underlying mass
density field and of the underlying structure.  Different samples 
selected from such catalogues may of course trace different structures.

We set out to explore high fidelity methods for reconstruction of the
density field from the discrete galaxy distribution.  
We require accurate local density estimates that can be used to reliably 
compute structural and topological indicators. 
Computational efficiency is an important factor since  our
reconstruction technique has to be able to deal with galaxy samples of
a million of galaxies as well as with N-body simulations comprising
orders of magnitude more particles.

There are two immediate and important implications and complications
with respect to our intention of including the (quasi)-nonlinear
components in the density field. The first one is that we have to be
aware of the noise components that tend to be included in the
reconstructions as well \citep{Schaap07,Weyschaap09}. The second one is that
nonlinear data are typically not well behaved, marked by strong 
gradients in the density field. It means that we have to 
take special care of those locations marked by non-linearities in the 
data.  

\subsection{Spatial Point Processes and Continuous Density Fields}
We pursue the reconstruction of a density field from a spatial point
process consisting of irregularly distributed points.  We assume 
that the local intensity of the points is a fair tracer of the
density and that these values are samples of a continuous underlying
density field.  In doing so we appreciate that differently defined samples
may trace different structures.  The reconstruction problem we address 
here is a data processing problem.  The interpretation of the results
is an astronomical process.

When the spatial point process is defined by the galaxy distribution, 
the situation will be more complex. The cosmic web is marked by a distinct 
luminosity, colour and morphology segregation. A strong and systematic trend 
of galaxy morphology with density has been established a few decades ago 
by \cite{Dressler80}.  Early type galaxies preferentially in rich groups and 
clusters and late types galaxies residing mainly in filaments and walls 
\citep[e.g.][]{Giovanelli86}. Other strong systematic clustering trends with 
luminosity and colour have also been established, such as in the complete 
SDSS DR7 galaxy sample \citep{Zehavi10}. A fair sample of galaxies should 
in principle take this intrinsic segregation into account. We are pursuing  
this in a forthcoming publication. In this study, mainly intent on 
establishing our reconstruction technology, we will consider the total 
galaxy population. The reconstructed density maps are therefore unlikely 
to represent a fair reflection of the underlying dark matter matter network, 
but will nonetheless convey the overall pattern of the large scale structure. 

The reconstruction consists of two fundamental steps. The first is the 
estimation of the local galaxy density. The second is the interpolation 
of these density values to obtain a continuous spatial density field.

Following straightforward grid based interpolation methods and more
sophisticated adaptive filter techniques, there has been a substantial
investment into developing more advanced techniques. Examples of 
recently introduced techniques to follow the multiscale nature of the 
Cosmic Web is the Multiscale Morphology Filter \citep{Aragon07} and the 
Hierarchical SpineWeb technique \citep{Aragon10}, a Morse theory based 
formalism that is closely related to the Watershed Void 
Finder by \cite{Platen07}. An example of an alternative route involves 
the wavelet reconstruction by \cite{Martinez05}. 

Here we concentrate in particular on the density estimation and the
subsequent field interpolation, in anticipation of the post-processing
and feature detection studies of the SDSS DR6/DR7 survey, the subject
of the additional papers in this series. To deal with the shot-noise
and necessary selection effects in the resulting density fields, the
analysis will involve filtering, a necessary post-processing step.

\begin{figure}
      \includegraphics[width=0.49\textwidth]{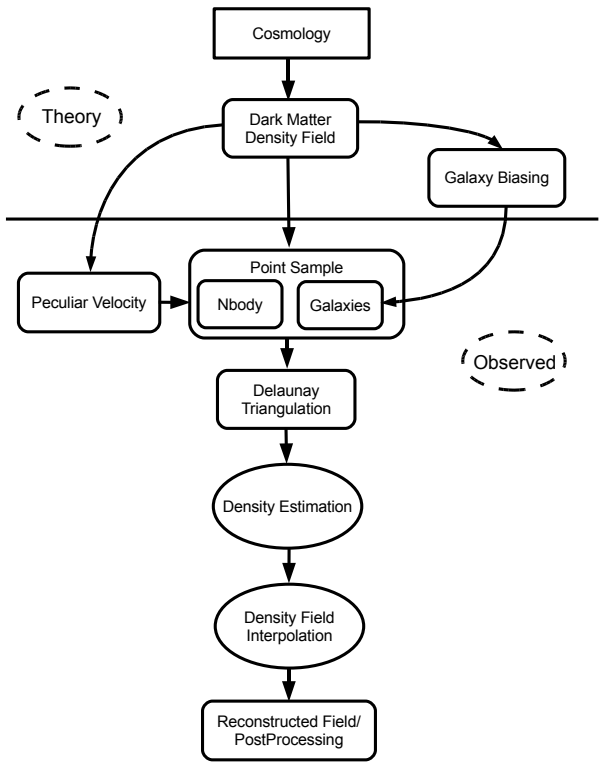}
    \caption{The schematic outline of our cosmological reconstruction
      procedure. The {\it Point Sample} is discussed in
      section~\ref{sec:data5}. The {\it Density Estimation} item will
      be treated in section~\ref{sec:density} \& \ref{sec:densdtfe}
      and the {\it Density Interpolation} methods are introduced in
      section~\ref{sec:dtfe},~\ref{sec:nnfe} and
      \ref{sec:kriging}. Section~\ref{sec:quan}, ~\ref{sec:topo} and 
      ~\ref{sec:sdssden} deal with the {\it Post-Processing}.}
    \label{fig:outline}
  \end{figure}

\subsubsection{Density Estimation}
\label{sec:density}
The first step in the reconstruction procedure is density
estimation.  We show how this fits into our general scheme in
Fig.~\ref{fig:outline}. A similar, but slightly different outline can
be found in \cite{Kitaura08}.

In the literature we may find a large variety of density
estimators \citep[e.g.][]{Silverman86,Sain02,Katkovnik02,Miller03}. Usually 
they involve filters of some kind. Dependent on
the filter kernel, density estimates may have a local character or
include the information from a wider range of points in a point
distribution. Filter kernels may have a rigid scale and shape, or they
may be adapting themselves to the local point density. An example of a
global estimator is a Gaussian kernel, which takes along the
information - be it weighted - from distant points.  A well-known
example of rigid local estimators are the CIC and TSC grid
interpolation formalisms \citep[e.g.][]{Hockney81}. A considerably
more flexible and adaptive filter kernel, frequently used in current
N-body studies, is that of the spline-based interpolation recipes
used in Smooth Particle Hydrodynamics codes \citep[e.g.][]{Monaghan92}
whose scale is determined by the distance to the $K^{th}$ nearest
neighbour.

Recently, various local and adaptive density estimators have been
shown to provide highly favourable results for complex cosmological
matter distributions, marked by a large dynamics range of scales and
density values and intricate geometric patterns. One of these exploits
the adaptivity of the kD tree
\citep{Ascasibar05,Sharma06}. We also note the use of the Epanechnikov
kernel estimator in the identification of superclusters  \cite{Einasto07}.

For the three reconstruction techniques addressed in this study, we use
the local DTFE density estimate introduced by \cite{Schaapwey00} (see
appendix~\ref{app:dtfe}). It sets the density value at a sample point
proportional to the inverse volume of the contiguous Voronoi cell
around a sample point, i.e.. the sum of the Delaunay tetrahedra to which
the sample point belongs \citep{Schaapwey00,Schaap07,Weyschaap09}.
Tessellation-based methods are based on the realization that the
optimal estimate for the spatial density at the location of a point
${\bf x}_i$ in a discrete point sample ${\cal P}$ is given by the
inverse of the volume of the corresponding Voronoi cell
\cite{Okabe2000}. They have been introduced by \cite{Brown65} and
\cite{Ord78} and were first used in astronomy, for the specific
purpose of devising source detection algorithms, by \cite{Ebeling93}.

In an extensive study, \cite{Schaap07} demonstrated that the DTFE estimates 
are substantially better than those of the rigid grid based CIC and TSC 
techniques. Also, it outperforms the adaptive SPH spline performance, in 
particular in areas with substantial density gradients. 

\subsubsection{Interpolation}
\label{sec:interpolation}
Interpolation on randomly scattered points aims to approximate a
continuous function constrained by the available data points.  A wide variety of
approaches have been put forward, each with their own advantages and
disadvantages. These methods can be roughly divided into two
categories, global and local methods. Local methods have the benefit
that they are fast and able to deal with large data-sets. Global
methods tend to produce smoother interpolated functions but are
computationally more expensive.

An overview and discussion of spatial interpolation techniques can be
found in \cite{Press07} (chapter 3.4) and in \cite{Watson92}. We refer
to \cite{Lombardi02} for a detailed review of the statistical
properties of a large number of techniques, and to \cite{Franke82} and
\cite{Amidror02} for a detailed comparison between various methods.

Amongst the more sophisticated interpolation techniques we can
distinguish various classes: inverse distance based methods (IDW),
moving least squares methods, the class of radial basis function
interpolation techniques (RBF), Kriging
interpolation methods \citep[see
  sect.~\ref{sec:kriging}]{Krige51,Matheron63}) and triangulation
based methods, both the linear DTFE technique
\citep[][sect.~\ref{sec:dtfe}]{Schaapwey00,Weyschaap09} and the higher
order NNFE natural neighbour interpolation technique
\citep[][sect.~\ref{sec:nnfe}]{Sibson81,Watson92}.

\bigskip
RBF and Kriging interpolation methods use predefined kernels to
interpolate the field. The kernel in RBF methods is a basis function
that spans the space of all the interpolating functions. Kriging
interpolation uses spatial correlations between sample points, with
its kernel being equal to the corresponding covariance function. The
method, introduced by \cite{Matheron63}, is the {\it best linear
  unbiased estimator} of a density value given a set of measured
sample points at irregularly spaced points. Given the commonly
accepted fact that the primordial density perturbation field is
Gaussian, we may therefore see the Kriging interpolator as the natural
choice for reconstructing the cosmological density fields which
emerged out of these primordial Gaussian circumstances.

There is a distant relationship of Kriging interpolation to 
Wiener filtering techniques \citep{Wiener49,Rybicki92}. However, 
Wiener filtering is based on a different philosophy than Kriging, in 
that it includes a model for the noise and is evaluated in Fourier space. 
Also, classical Wiener filtering is predicated on an underlying Gaussian 
distribution. As a result, it has the serious disadvantage of suppressing 
or substantially diluting nonlinear structures of interest. More advanced 
recent developments and applications of Wiener filters to the reconstruction 
of the density distribution have largely remedied its capacity for 
reconstructing the density distribution \citep{Kitaura08,Kitaura09,Kitaura10}.  

A rather different approach is advocated in the triangulation-based
interpolation techniques.  Both the linear Delaunay Tessellation Field
Estimator \citep[DTFE][]{Schaapwey00,Schaap07,Weyschaap09} and the
higher-order Natural Neighbour Field Estimator
\citep{Sibson81,Watson92,Braun95} use the neighbourhood relationships
defined by the Voronoi and Delaunay tessellation of the point sample
to establish a fully adaptive, irregular and local interpolation
grid. DTFE uses the Delaunay triangulation to reconstruct in a
self-adaptive, mass conservative and parameter free way the underlying
spatial (density) distribution. In combination with the DTFE
tessellation-based sample points density estimates (see previous
sect.~\ref{sec:density}), the DTFE interpolation leads to a
volume-covering density field which has been shown to recover the
hierarchical as well as the anisotropic morphology of the Cosmic Web
\citep{Schaap07,Weyschaap09}.


\subsection{DTFE, NNFE and Kriging}
The DTFE {\it Delaunay Tessellation Field Estimator} method
 will be compared to two other techniques having a similar potential
for a proper reconstruction of the cosmic web. The NNFE {\it Natural
  Neighbour Interpolation} technique shares the local nature of DTFE,
but involves higher order interpolations. The third formalism is {\it
  Natural Lognormal} Kriging is a version of Kriging interpolation,
and thus involves a nonlocal higher-order interpolation methodology.

The crucial step of importance in our investigation is the
interpolation step (see diagram Fig.~\ref{fig:outline}). Using the
same initial sample point density estimates, the first step of the
reconstruction procedure, we can assess the relative merits of
DTFE, NNFE and Kriging interpolators.

\subsection{Outline of this study}
We start by describing the data samples in section~\ref{sec:data5},
the SDSS survey sample and a mock galaxy sample which mimics the
SDSS, obtained from the Millennium simulation. The local DTFE density
estimate at the location of each of the sample galaxies is the subject
of section~\ref{sec:density}. In the subsequent section we
present and describe each of the three interpolation techniques
investigated in this study, DTFE in subsection~\ref{sec:dtfe}, NNFE in
subsection~\ref{sec:nnfe} and Lognormal Kriging in
subsection~\ref{sec:kriging}. The comparison throughout the paper is based
on one specific sample, the density field reconstruction of the SDSS
mock galaxy sample. Following
a discussion in sect.~\ref{sec:qual} of the qualitative appearance of
the density maps by each of the three methods, we turn to an intensive
quantitative error and quality analysis of the density field in
sect.~\ref{sec:quan}.  In sect.~\ref{sec:topo} this is followed by an
investigation of the topological structure of the
weblike galaxy distribution in the survey, mainly based on the void
population as traced by the Watershed Void Finder. Section~\ref{sec:sdssden} 
presents the density field reconstruction of the SDSS data sample 
and in sect.~\ref{sec:summary}.we summarize this study.

Following this first paper in a series will be a statistical study of the 
density  field, focusing in particular on the one-point probability density function. 
Subsequently, we will present and discuss the cosmography of the 
reconstructed local Universe. Later studies will analyze the void population 
in the SDSS density field, and concentrate on the properties of galaxies as function 
of the large scale environment as characterized by our technique. 

\section{The Data}
\label{sec:data5}
\begin{figure*}
  \centering
  \includegraphics[width=0.98\textwidth]{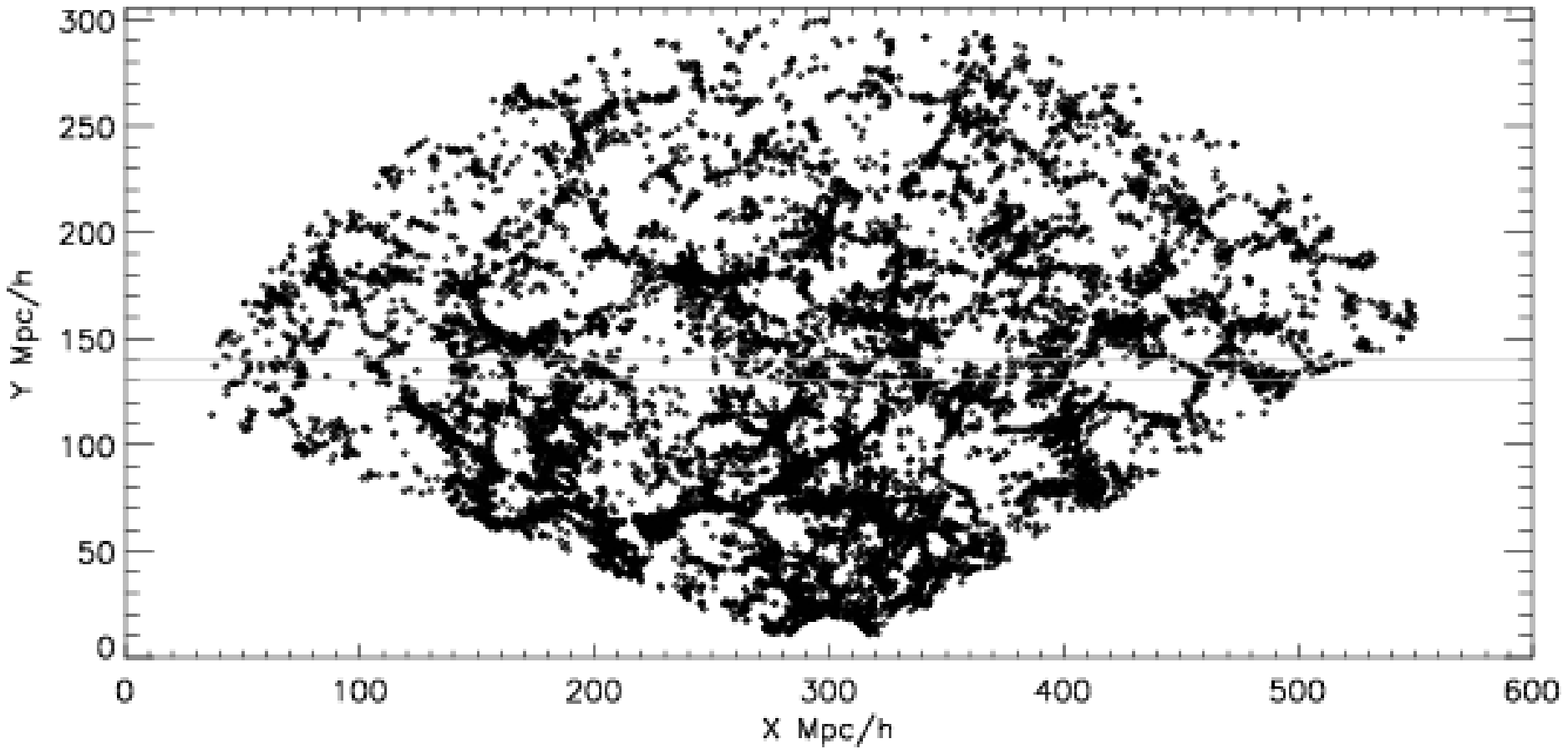}
  \includegraphics[width=0.98\textwidth]{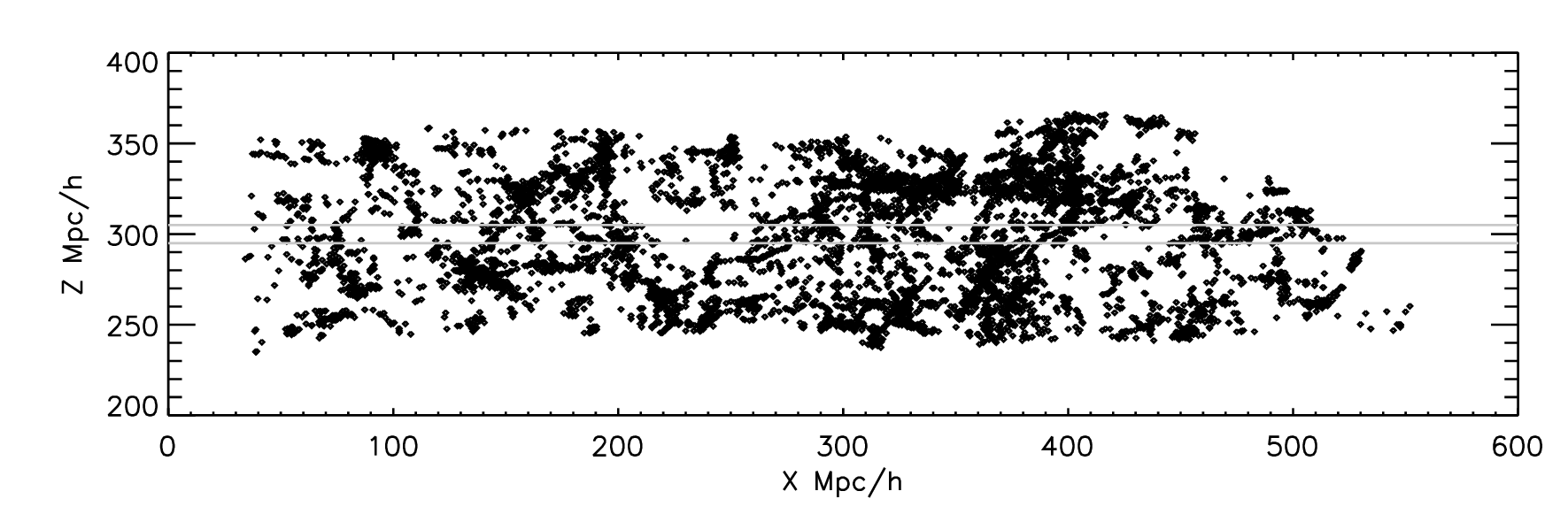}
  \caption{The top figure shows the SDSS galaxies in the X and Y
  coordinates out to a distance of 300$\Mpch$. The lower figure shows a
  XZ slice perpendicular to the XY plane at Y=135$\Mpch$. The
  corresponding boundaries of both slices are indicated in the other
  figure, e.g. grey lines in the top figure show the Y limits of the
  bottom figure and vice verse. Both slices have a thickness 10$\Mpch$.}
  \label{fig:gal_xz}
\end{figure*}
Our study is based on two major datasets. The principal one is the 
genuine SDSS DR6 galaxy redshift survey. For the purpose of understanding 
the errors and artefacts in the density field reconstruction we use a set 
of galaxy mock catalogues that model this SDSS dataset. The mock samples 
are obtained from the Millennium simulation \citep{Springel05}. 

\begin{figure*}
  \centering
  \includegraphics[width=0.8\textwidth]{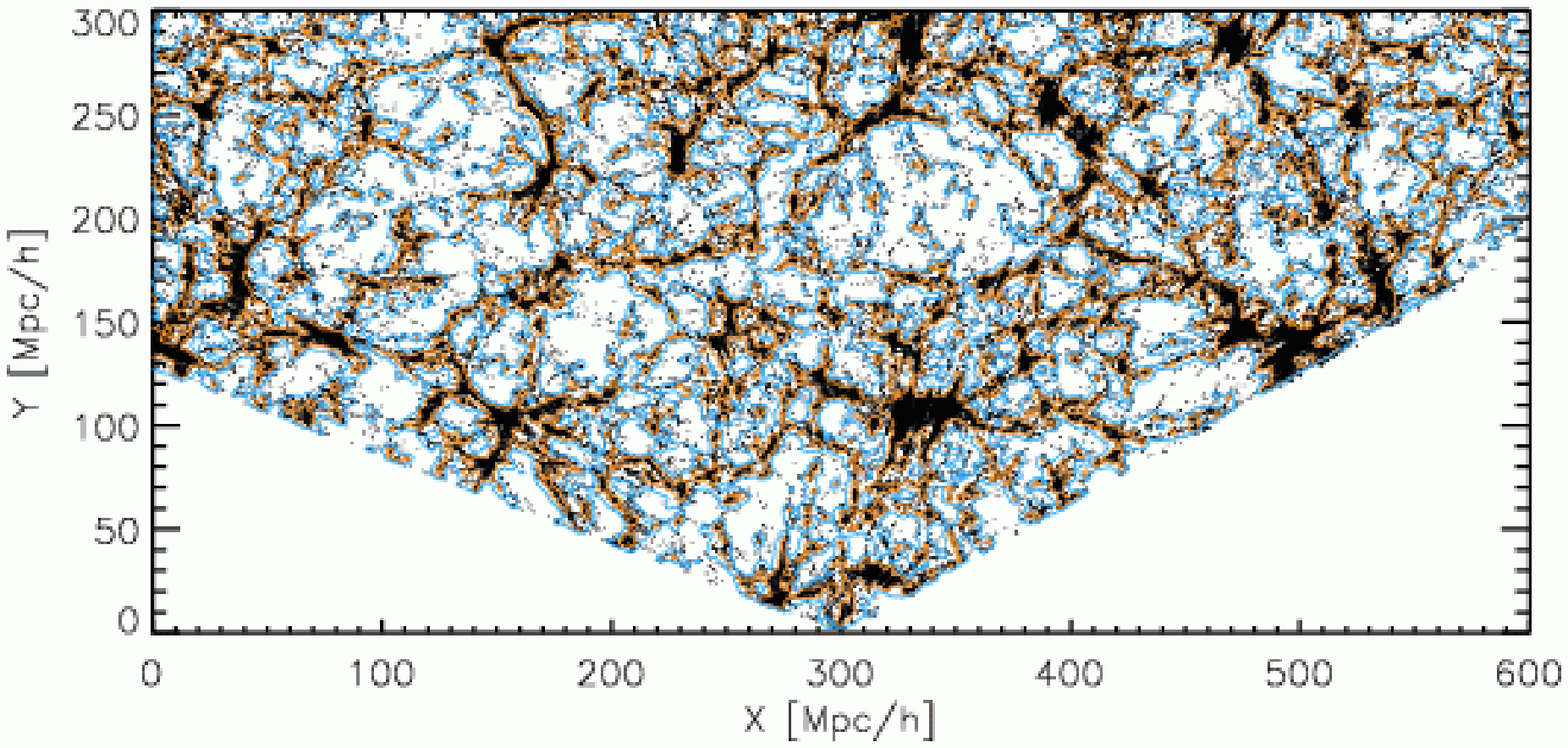}
  \includegraphics[width=0.8\textwidth]{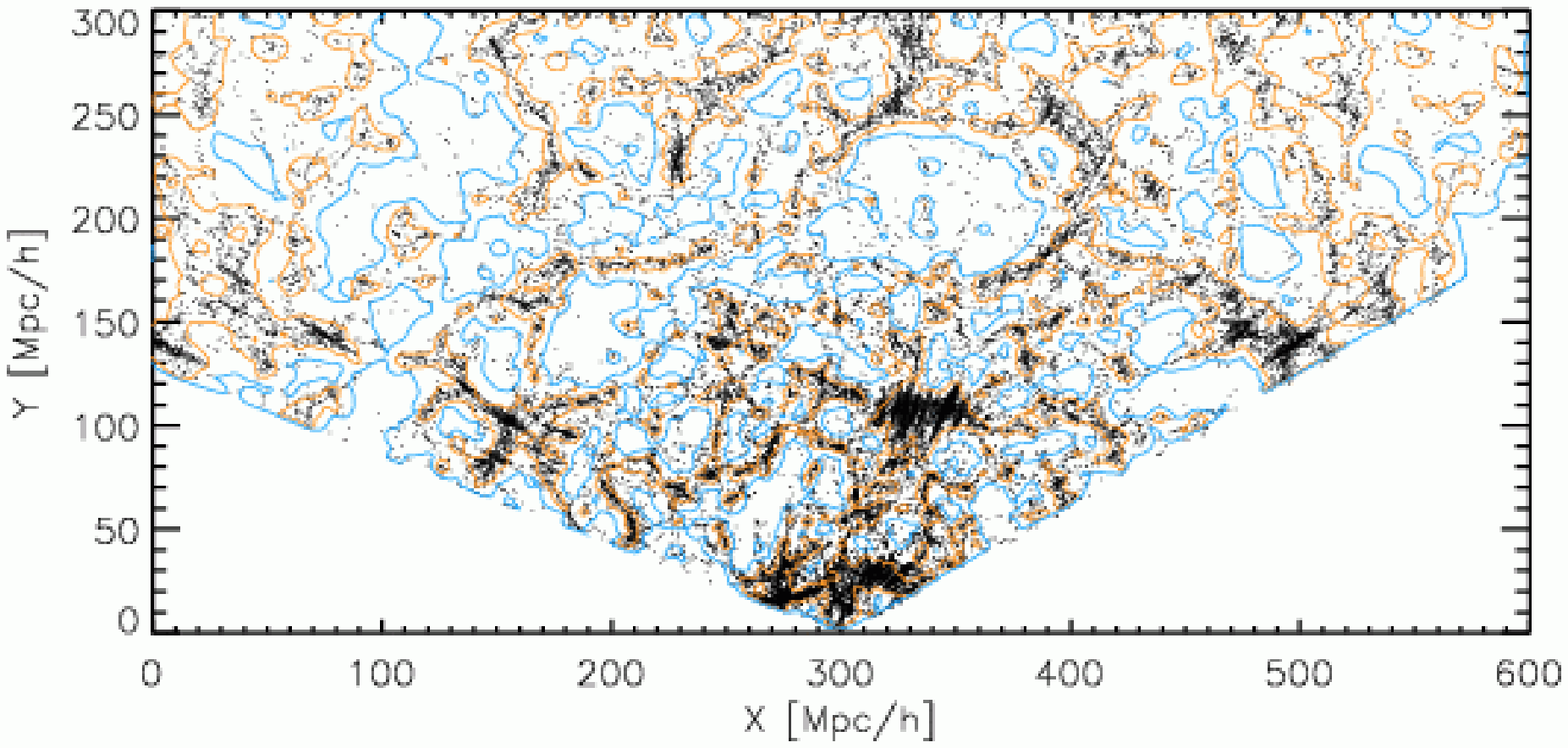}
  \includegraphics[width=0.8\textwidth]{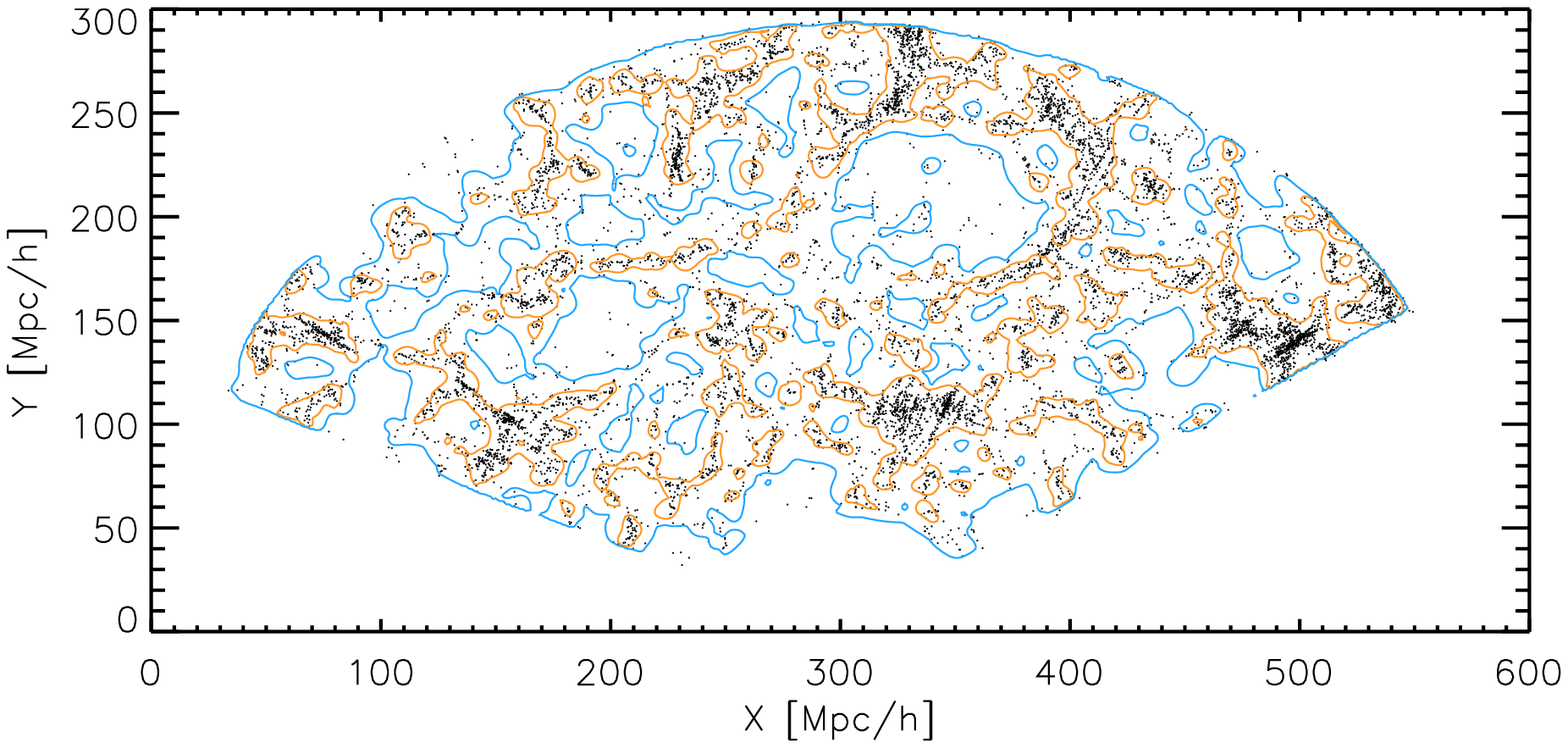}
  \caption{SDSS mock catalogues from the Millennium simulation. Top: full Millennium 
catalogue. Centre: magnitude limited mock catalogue. Bottom: volume limited mock catalogue. 
The mock galaxies are indicated by black dots. The blue and orange contour lines are 
NNFE density contour lines at density contrast $\rho/\rho_u=0.25$ (blue) and 1.0 (orange). 
For reasons of clarity, the top slice has a thickness of 6.0 $\Mpch$ while the magnitude and volume limited slices 
each 12.0 $\Mpch$: showing a thinner slice for the two last samples would show too few galaxies. }
  \label{fig:mag_vollim}
\end{figure*}
\subsection{The SDSS galaxy sample}
\label{sec:sdss}
For our analysis we use the main galaxy sample in the North
Galactic Cap from the 6th data release of the Sloan Digital Sky Survey
(SDSS) \citep{Strauss02,York00,Stoughton02,Adelmanmccarthy08}. 

The SDSS DR6 data release consists of various contiguous regions.  We  
restrict ourselves to the largest contiguous region, the northern strip 
of the North Galactic Cap (NGC) region.  The sample was retrieved from the SDSS ``casjobs server'' 
(\url{www.sdss.org}) using
the SQL query interface. Relevant properties were downloaded and most of
the post-processing was done on a workstation. We did not attempt to
assign galaxies with missing redshift due to fiber collisions. This gives
a lower density estimated at the position of the missing redshift. 
However, given the sampling scheme used to resample the density
field to a regular grid, the probability of sampling
one of the affected areas is very low.  The problem occurs only in 
the high density regions of the catalogue.

The spectroscopic SDSS galaxy sample is almost complete between a Petrosian
magnitude limit of $m_r = 14.5$ and $m_r = 17.77$. We assume that the completeness 
of the sample does not vary significantly, even though there are in fact some 
angular variations \citep{Blanton03}.

For our study, we select the SDSS galaxies that are located within 
a comoving box of 600$\Mpch$. In terms of the survey coordinates (X,Y,Z) (see 
app.~\ref{app:XYZ} for definition), the observer is located at 
(X,Y,Z)=(300.,0.300.)$\Mpch$ and the centre of the northern strip is 
rotated to lie parallel to the Y-axis starting at $(X,Z)= (300,300)\Mpch$. 
In Fig.~\ref{fig:gal_xz} the galaxies are plotted in the XY-projection 
(top) and the XZ-projection (bottom).

Within the $(600 \Mpch)^3$ volume there are a total of 311474 galaxies 
with a magnitude less than the magnitude limit $m_r=17.77$.  

\subsection{The SDSS mock samples}
\label{sec:mock}
The Millennium simulation \citep{Springel05} was used to construct
galaxy mock catalogues which emulate the SDSS galaxy redshift sample.
They are needed to get estimates of the errors induced by the
magnitude selection, redshift distortions as a result of peculiar
velocities and errors resulting from the survey mask.

We use the semi-analytical galaxy samples of \cite{DeLucia07},
\cite{Bower06} and \cite{Bertone07} to construct our own mock samples.
They are less detailed as the mock catalogue generated by
\cite{Blaizot05}, they are perfectly suited for a representation of
all necessary aspects of the SDSS sample.

\bigskip
The same (X,Y,Z) coordinate system and box size of $600 \Mpch$  have been used for the 
mock galaxy samples. The mock galaxy catalogues are constructed as follows: 
\begin{enumerate}
\item[$\bullet$] Periodically tile the Millennium cube to obtain enough cosmic volume.
\item[$\bullet$] Calculate the redshift of the model galaxies wrt. the observer
\item[$\bullet$] Compute the apparent magnitude of each model galaxy from its absolute 
magnitude and redshift. 
\item[$\bullet$] Select the model galaxies brighter than $m_r = 17.77$.
\item[$\bullet$] Add the peculiar velocity to the Hubble redshift to obtain the total redshift
\item[$\bullet$] Apply the observational mask of the DR6-NGC sample to decide whether the model 
galaxy is included in the mock catalogue. 
\end{enumerate} 

\subsection{Redshift Space Distortions}
\label{sec:zspace}
In this study we assess redshift space density maps as well as
partially corrected ``real'' space density maps.

Redshift space surveys like the SDSS are beset by distortions in the
estimated distance.  The result of large coherent cosmic flows, infall
velocities onto clusters and highly nonlinear ``thermal'' velocities
within clusters, these redshift distortions can have a dramatic effect
on the estimated distances and reshape the large scale matter
distribution. However for our purposes here, which is testing and
comparing the methods, the presence or absence of {\it fingers of God}
is immaterial.

Hence, in this paper, which deals with the Mock SDSS catalogues, 
we do not correct for the redshift distortions induced
by large scale coherent cosmic flows.  In particular the DTFE density 
reconstruction is locally adaptive and so the reconstruction 
is not corrupted by the presence of elongated radial features; 
this can be seen in figure ~\ref{fig:mag_vollim}.  

The redshift distortions will, however, be addressed 
in a following paper in which we analyse the real SDSS catalogue. 

\subsection{Magnitude- vs. Volume-limited Samples}
\label{sec:weights}
For the analysis of the SDSS sample, we extract two different samples
from the full SDSS galaxy sample. These are a volume limited (i.e. absolute magnitude limited) 
sample and an apparent magnitude limited sample. Each sample is used for
different aspects of our analysis.

\bigskip
\noindent {\it Volume Limited sample}\\ A {\it volume limited} galaxy
sample is defined in order to assure a uniform galaxy coverage over
the survey volume. A volume limited sample consists of a subset of
galaxies which are homogeneously sampled throughout the sample
volume. It has the advantage that each sample galaxy is an equal weight tracer of the
underlying density field, and the resulting field
will be statistically uniform. Our volume-limited sample has a distance limit of 
$300 \Mpch$ and includes all galaxies brighter than $M_r <
-20.45$, roughly representing the galaxies brighter than $L_{*}$. 

While the uniformity of the volume-limited samples assures a
straightforward error assessment for any analysis, it has the
disadvantage of losing the high spatial resolution represented by
fainter galaxies nearby, therefore it does not necessarily have the
{\it smallest} error.

\bigskip
\noindent {\it Magnitude Limited sample}\\ The {\it magnitude} limited
sample contains all 311474 SDSS galaxies brighter than
$m_r=17.77$. While a magnitude limited sample takes along all sampled
information, one needs to correct for the inhomogeneous selection
process.

A characteristic of magnitude-limited surveys is the change of 
intrinsic spatial resolution as we proceed out to larger distances. 
Potentially, this could be a serious issue when galaxies would be 
biased in a very complicated fashion (e.g. higher order biasing). 
This would render it very difficult to infer the density field 
at large distances, where only the most luminous objects would 
remain visible. By default, we therefore assume that all galaxies 
- independent of their luminosity- are a fair tracer of the density 
field. 

\begin{figure*}
   \includegraphics[width=0.46\textwidth,clip]{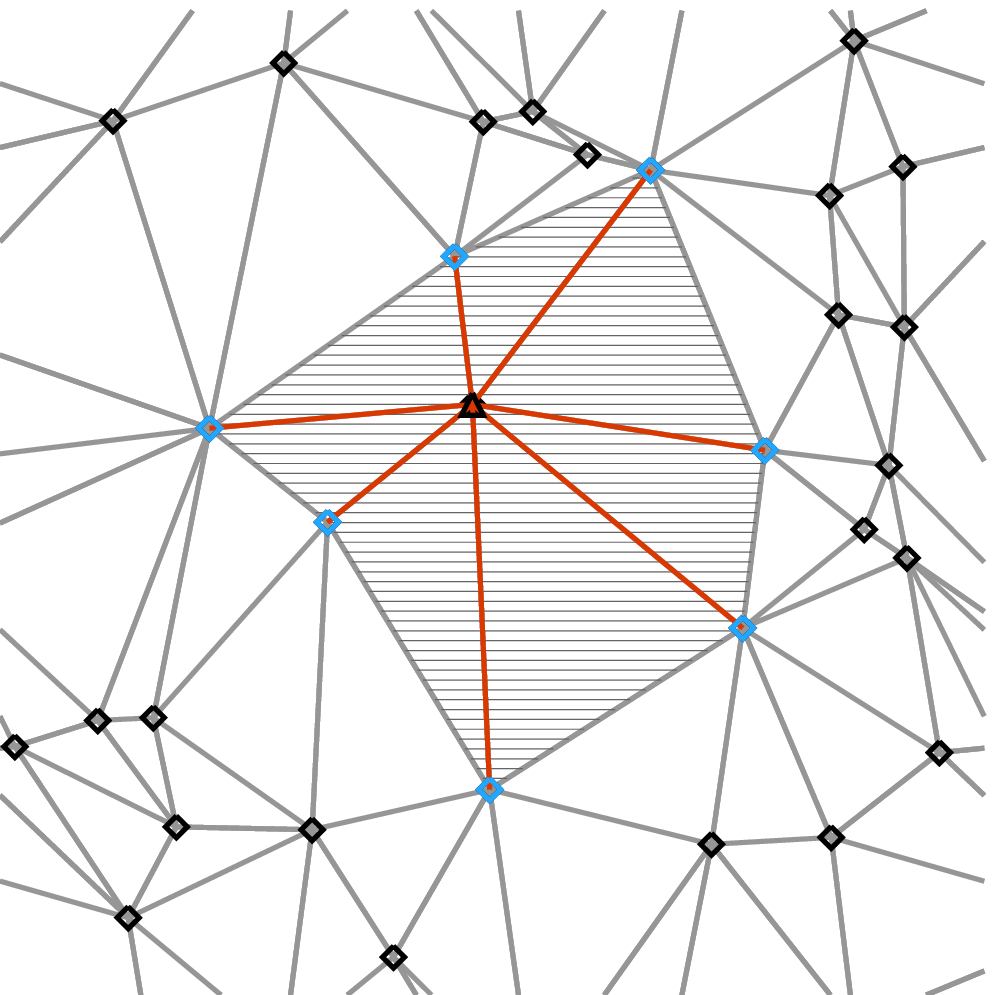}
   \hspace{1cm}
   \includegraphics[width=0.46\textwidth,clip]{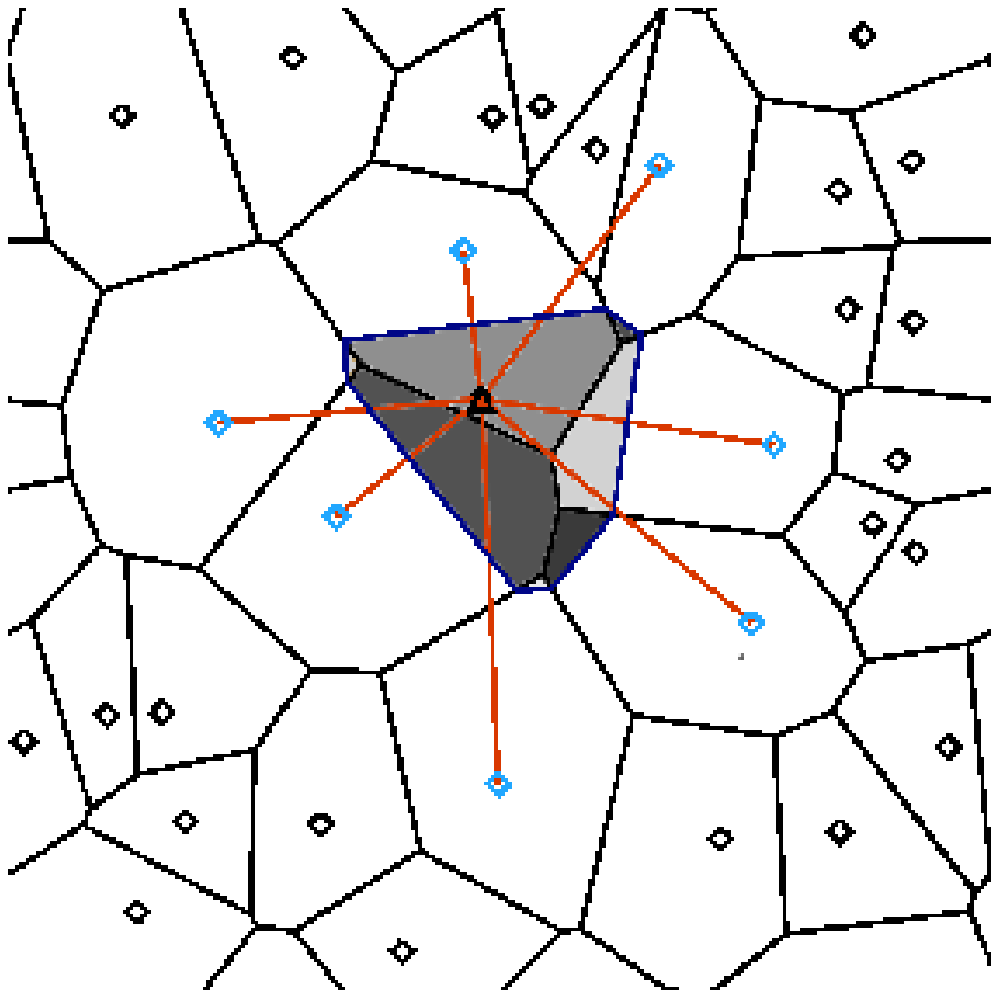}
   \caption{Voronoi tessellations, Delaunay tessellations and Natural Neighbours. 
   {\it Left:} The Delaunay triangulation (grey lines) of a point set (circles). 
   For the central point the natural neighbours are indicated by the blue points, 
   along with the corresponding Delaunay edges (red). The hatched region is the 
   contiguous Voronoi cell of the central point, the composite of all Delaunay 
   triangles shared with its natural neighbours. {\it Right:} The Voronoi 
   tessellation of a point sample (black diamonds). Relevant for Natural Neighbour 
   interpolation (NNFE): following the insertion of a central point (black triangle), 
   a new Voronoi cell (region enclosed by the dark blue polygonal boundary) is computed, 
   the {\it second order Voronoi cell}. The gray shaded areas are the overlapping regions with the
   original Voronoi tessellation.}
  \label{fig:nnbr}
\end{figure*}

Following this assumption, we correct for the dilution as a function 
of survey depth by weighing each sample galaxy by the reciprocal $w(z)$ of 
the radial selection function $\psi(z)$ at the distance 
of the galaxy.  For the SDSS, the selection function $\psi(z)$ as a function 
of redshift $z$ is well fitted by the expression forwarded by 
\cite{Efstathiou01}
\begin{equation}
\psi(z)\,=\,\exp\left\{-\left(\frac{z}{z_r}\right)^{\beta}\right\}\,,
\label{eq:psiz}
\end{equation}
where $z_r$ is the characteristic redshift of the distribution and
$\beta$ specifies the steepness of the curve. The corresponding number
density $N(z)$ of galaxies at redshift $z$ is
\begin{eqnarray}
N(z)dz&\,=\,&A z^2 \psi(z)\,dz\nonumber\\
&\,=\,& A z^2\,\exp\left\{-\left(\frac{z}{z_r}\right)^\beta\right\}\,dz\,,
\label{eq:nsurvey}
\end{eqnarray}
where $A$ is a normalization constant and the $z^2$ term represents the 
increase of volume as function of $z$. The resulting galaxy weights $w(z)$ 
are  
\begin{equation}
  w(z)\,=\,1/\psi(z)\,=\,\exp\left\{\left(\frac{z}{z_r}\right)^{\beta}\right\}.
\label{eq:surselect}
\end{equation}
When incorporating $w(z)$ the weights into the density field
reconstruction, the normalization of the resulting density field may
be achieved by modelling the details of the selection function and
calculating the appropriate normalisation constant. We chose to follow
the alternative of calculating the average of the reconstructed
density field and subtracting it from the reconstruction. This is a
simple and straightforward procedure, and perfectly valid as long as
the volume of the sample is large enough to enable the estimate of the
true average.

\bigskip
\noindent {\it Mock Samples}\\ From the mock catalogues we select 
three different samples. The {\it full mock catalogue} comprises all
Millennium (semi-analytical) model galaxies within the DR6-NGC
mask. The {\it magnitude limited} mock sample includes all model
galaxies resulting from the procedure described above. The third
sample is a {\it volume limited} set with an absolute magnitude $M_r <
-20.45$.

To appreciate the differences between the {\it magnitude-limited} and 
the {\it volume-limited} galaxy sample, and also the {\it full mock 
catalogue} within the SDSS volume, the three mock catalogues are shown 
in Fig.~\ref{fig:mag_vollim}. 

The contours superimposed on the 
corresponding galaxy distributions are the resulting NNFE density field 
contours (see sect.~\ref{sec:nnfe}). 

\section{Local DTFE Density Estimate}
\label{sec:densdtfe}
Throughout this study we use the local DTFE density estimate,
following the definition by \cite{Schaapwey00}. In
appendix~\ref{app:dtfe} one may find more details of the DTFE
procedure which we followed \citep[for an extensive review
  see][]{Weyschaap09}. Implicitly, and for simplicity, we assign to
each sample galaxy the same mass $m_i$, i.e. the density value is
predicated on the number density of galaxies.

The sample point DTFE density value is inversely proportional to the
volume of of the local neighbourhood as defined by the Voronoi
tessellation of the spatial galaxy distribution. \citep{Schaapwey00}
argued that the inverse of the volume of the {\it contiguous Voronoi
  cell} is the proper density estimate, assuring mass conservation for
the subsequent linear interpolation step. The contiguous Voronoi cell,
sometimes dubbed {\it umbrella} in the computational geometry
literature, is the region defined by all Delaunay tetrahedra of which
a given sample point is a vertex and which it shares with its {\it
  natural neighbours}.  A two-dimensional illustration of a contiguous
Voronoi cell of a point is shown as the surrounding hatched region in
the lefthand frame of Fig.~\ref{fig:nnbr}.

The density value at each sample point is determined following the 
construction of the Delaunay triangulation \citep[see e.g.][]{Delone34,Aurenhammer91}
\footnote{the Delaunay triangulation in this work has been computed
  using the CGAL library, \cite{cgal}.}. Within the triangulation we
identify for each sample point $i$ all $N_i$ neighbouring tetrahedra
$\mathcal{T}_j$ (Fig.~\ref{fig:nnbr}), which together constitute the
contiguous Voronoi cell ${\cal W}_i \cup_j \mathcal{T}_j$. Summation
of the individual tetrahedral volumes $V({\cal T}_j)$ yields the
volume of the contiguous Voronoi cell,
\begin{equation}
V({\cal W}_i)\,=\,\sum_{j=1}^{N_i}\,V({\cal T}_j)\,.
\end{equation}
For the three-dimensional SDSS sample volume, the resulting DTFE estimate 
of the density $\widehat{f_i}$ at sample point $i$ is (see equation~\ref{eq:densvornu}), 
\begin{equation}
  \widehat{f_i}\,=\,\frac{4\,w(z_i)}{V({\cal W}_i)}\,.
\label{eq:densdtfe}
\end{equation}
where the weight $w(z_i)$ is the sample selection weight at the 
galaxies' redshift $z_i$ (equation~\ref{eq:surselect}). Note that the 
factor four takes account of the fact that in three dimensions 
each sample point belongs to four tetrahedra. In practice, 
the density of all particles is calculated by looping in sequence 
over all Delaunay tetrahedra. 

\subsection{Shot noise errors}
\label{sec:shot}
A local density estimator such as the contiguous Voronoi cell has the
advantage of being very sensitive to the signal. However, this also
implies them to have a high sensitivity to shot noise present in the
data.

For appreciating the influence of shot noise in the DTFE density
estimates, we turn to the probability distribution of the estimated
intensity $\widehat{\lambda}$ for a 3D Poisson point process with
intensity $\lambda=1$. \cite{Schaap07} found that it can be very well
approximated by
\begin{equation}
  p(\widehat{\lambda})\,=\,\frac{1944}{5}\widehat{\lambda}^{-8} {\rm e}^{\left( -6/\widehat{\lambda} \right)}\,.
\label{eq:shotnoise}
\end{equation}
The first observation is that the DTFE density estimator is unbiased,
as the mean of $p(\widehat{\lambda})$ is equal to one. An estimate of
the error involved is that of the variance,
$\sigma^{2}_{\widehat{\lambda}}=1/5$, and is equal to $\sim
57\%$. Evidently, the distribution is non-Gaussian with a long tail
extending towards high density values (cf. Fig.~\ref{fig:cvt2}).

\vskip 0.5truecm
\subsection{Centroidal Voronoi tessellations}
\label{sec:cent}
The high density tail of the DTFE density estimate, and the implied shot 
noise level, can be suppressed or regularised 
by using the centroidal Voronoi tessellation (CVT) \citep[see][]{Lloyd82,
Browne07}. For a CVT the generating point distribution is such that the 
generating points are the mass centres of the resulting Voronoi cells. 

\begin{figure}
  \includegraphics[width=0.48\textwidth,clip]{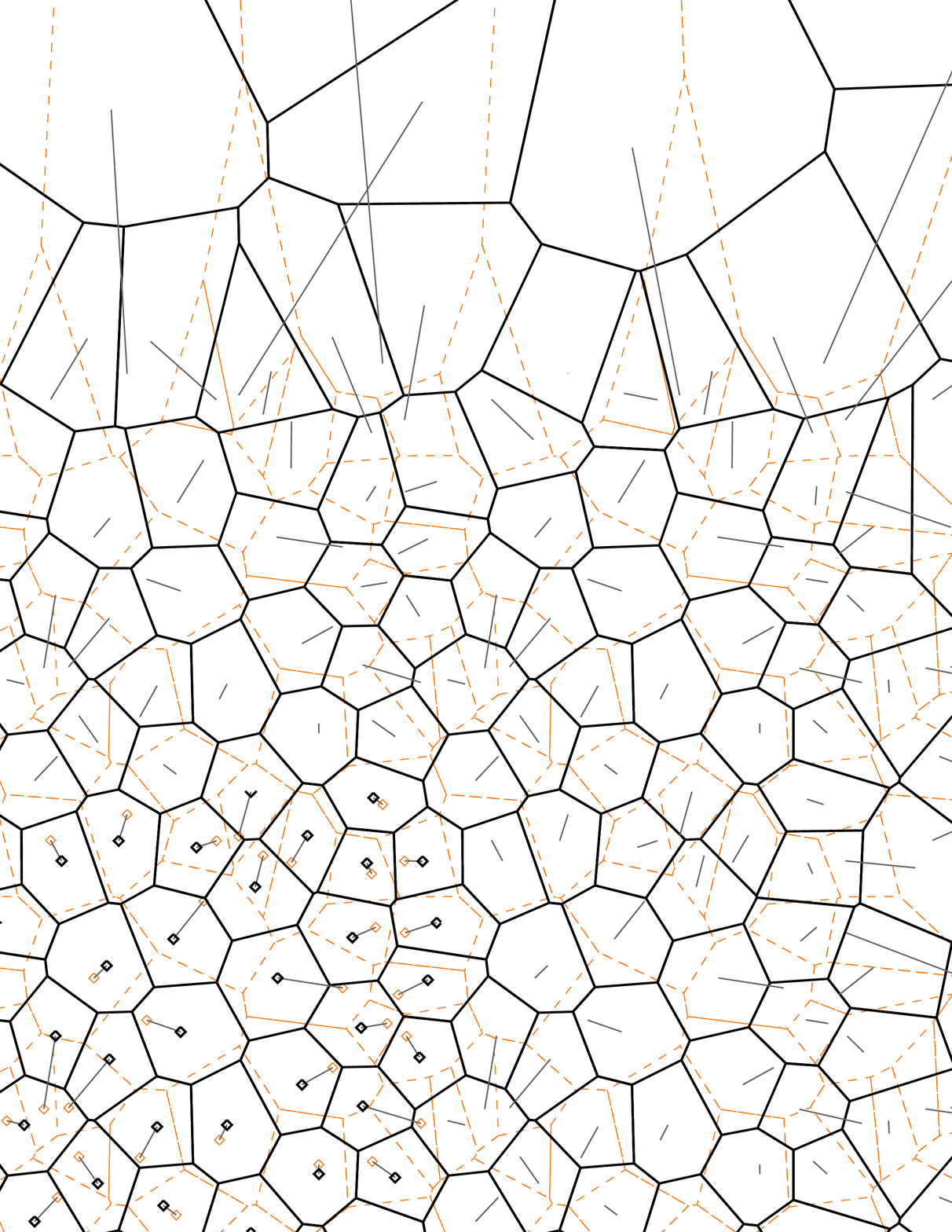}
  \hspace{1cm}
  \caption{Centroidal Voronoi Tessellation. The (original) Voronoi
    tessellation for the points of Fig.~\ref{fig:nnbr} are plotted
    in orange. In black we show the centroidal Voronoi tessellation
    with corresponding points after two Lloyd iterations. The
    displacements are indicated by the gray lines.}
  \label{fig:cvt1}
  \vspace{0.5truecm}
  \includegraphics[width=0.48\textwidth]{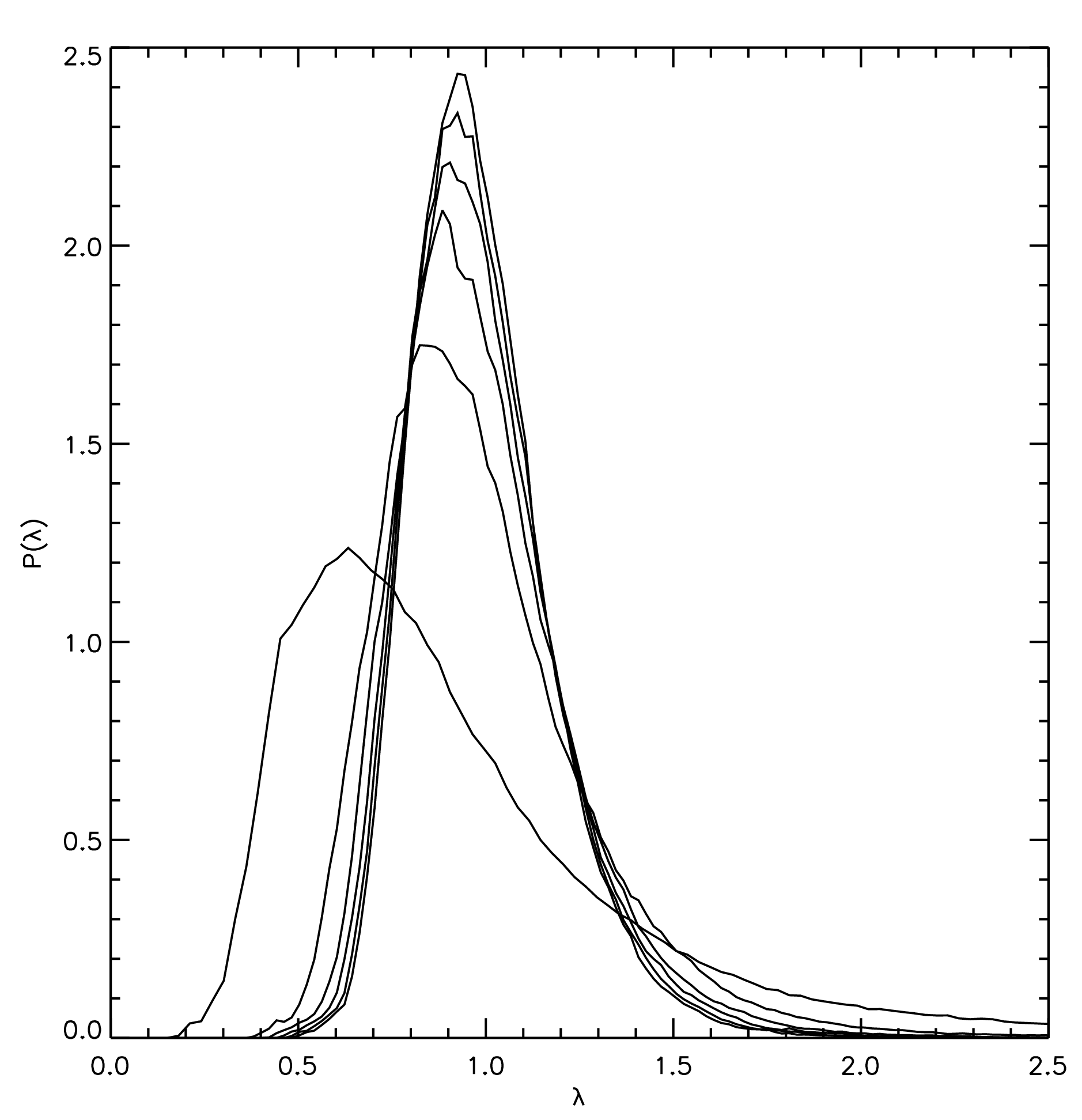}
  \caption{The error probability distribution of the centroidal Voronoi tessellation 
    based galaxy density estimator for a homogeneous Poisson sample. Arranging the curves from broadest
    to the most narrow they respectively correspond to 0, 2, 4, 6 and 10 Lloyd iterations.}
  \label{fig:cvt2}
\end{figure}

The calculation of a CVT is usually done by means of an iterative 
procedure known as Lloyd iteration. Starting with an originally random 
point distribution, the centre of mass of the corresponding Voronoi 
cells is computed. Subsequently, the points are displaced to these 
centres. After a sequence of iteration steps, the resulting point distribution 
tends to converge to a proper CVT constellation. Effectively, the points 
have been repelling each other. 

An impression of the CVT iteration procedure can be obtained from 
Fig.~\ref{fig:cvt1}. Involving an initial point 
distribution, the resulting intensity distribution $p({\widehat \lambda})$ 
for the tessellations obtained after zero, two, four, six, eight and 10 Lloyd iterations 
is shown in the righthand frame. Clearly, a CVT involves a much 
more regular distribution: after four iterations $p({\widehat \lambda})$ 
has turned into a narrow and near symmetric distribution whose high-end 
tail is almost absent (Fig.~\ref{fig:cvt2}). Potentially, a CVT might therefore help to 
suppress the density estimate error and its asymmetric distribution. 

\begin{figure*}
  \hspace{-0.5cm}
  \includegraphics[width=1.00\textwidth]{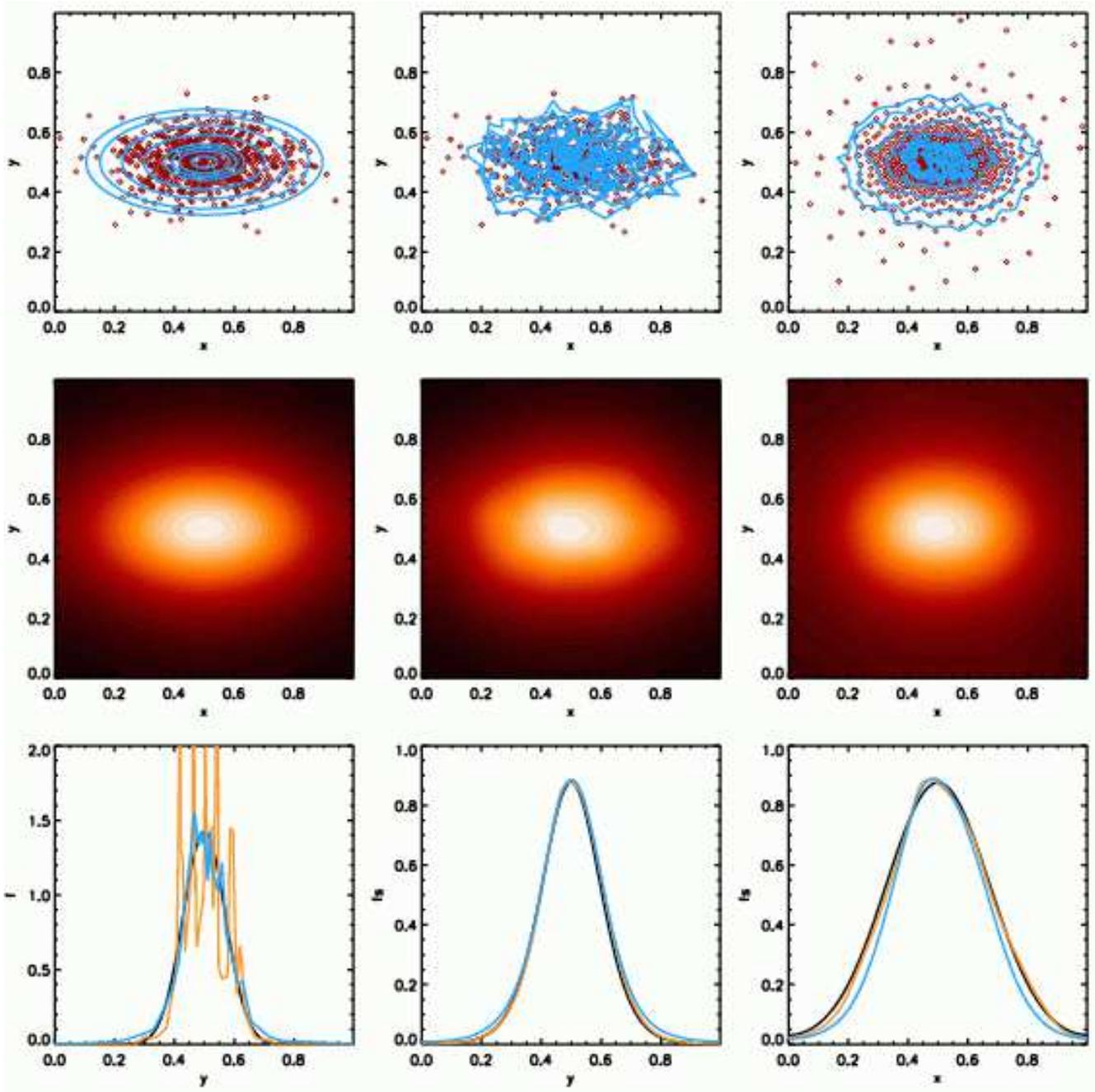}
  \caption{DTFE and DTFE/CVT reconstruction of an anisotropic Gaussian
    peak. Sampling the peak by 4500 random points, the top row shows
    the contour levels of the reconstructions: original (left), DTFE
    (centre) and DTFE/CVT with 5 Lloyd iterations. The resulting
    density maps after Gaussian smoothing, with $R_f=0.05$, is shown
    in the central row. The bottom row shows linear density profiles
    through the resulting mass distribution. Left: linear profiles
    along the $y-axis$ at $x=0.5$, for the original (black), DTFE
    (orange) and DTFE/CVT (blue). Central and Right: following the
    same colour schame, linear profiles through the filtered density
    field reconstruction, along the y-axis (at $x=0.5$) and the x-axis
    (at $y=0.5$).}
  \label{fig:gausspoiss}
\end{figure*}
\subsection{DTFE noise: a case study}
A visual impression of the shot noise involved in the density estimate 
is provided by Fig.~\ref{fig:gausspoiss}. It concerns a random sample of 4500 
points distributed according to an anisotropic Gaussian distribution. This 
configuration is more representative for what may be expected in the 
real galaxy distribution. 

Densities were estimated according to the pure DTFE procedure (see 
equation~\ref{eq:shotnoise}) and following a Lloyd CVT procedure of 
5 iterations. The top row of Fig.~\ref{fig:gausspoiss} shows the 
density contours of the original peak and the raw DTFE and DTFE/CVT 
field reconstructions. The raw DTFE reconstruction (centre) is 
highly irregular compared to the original contours (left), while the 
CVT contours are much more regular (right). The linear profile 
along the y-axis (at x=0.5, bottom left panel) emphasizes the visual 
impression: the DTFE profile (orange) is marked by salient peaks 
which reflect the high density tail of $p({\widehat \lambda})$, 
while the DTFE/CVT profile (blue) appears to adhere considerably 
better to the original profile. 

To appreciate the average trend in the density reconstruction, we
filter the original, DTFE and DTFE/CVT fields with a Gaussian filter
($R_f=0.05$). In addition to the suppressed shotnoise, the resulting
density level plots, shown in the central row of
Fig.~\ref{fig:gausspoiss}, reflect the way in which the
reconstructions affect the shape of the Gaussian peak. While the DTFE
reconstruction retains the shape of the original, entirely in
accordance with the findings by \cite{Schaap07}, the DTFE/CVT appears
not to do so. This impression is confirmed by the two linear profiles,
along the y-axis, at $x=0.5$ and along the x-axis, at $y=0.5$, shown
in the bottom central and righthand panels. While the DTFE
reconstruction is able to follow accurately the shape of the original,
the DTFE/CVT reconstruction displays considerable deviations.

We may therefore conclude that while CVT appears to suppress the
shotnoise effects on small scales, it is not able to follow the
morphology of the mass distribution on larger scales. Since we will be
filtering our fields in a similar way, this result leads us to
pursue our study on the basis of the pure DTFE density estimate
procedure {\it without} the CVT regularization.

\section{Interpolation Methods}
The key aspect in our investigation is the interpolation step of each 
of the three reconstruction methods. Here we describe the 
interpolation steps of the investigated methods, the DTFE, NNFE 
and Kriging interpolators.

\bigskip
\subsection{Interpolation Method I: \\ \ \ \ \ \ \ \ Delaunay Tessellation Field Estimator (DTFE)}
\label{sec:dtfe}
The most straightforward way to interpolate and/or reconstruct a
density field is by linear interpolation between neighbouring data
points. The linear reconstructed field is continuous throughout the sample 
volume. Within each interpolation interval the first derivative 
remains constant, although it is discontinuous at the boundaries 
between the intervals. 

The Delaunay Tessellation Field Estimator
\citep{Schaapwey00,Schaap07,Weyschaap09} is the multidimensional
equivalent of simple piecewise one-dimensional linear interpolation
from an irregularly distributed set of points.  DTFE generalizes the
concept of natural interpolation interval to any dimension $D$ by
adopting the Delaunay tetrahedra of a multidimensional point set as
such. It uses the adaptive and minimum triangulation properties of
Delaunay tessellations to use them as adaptive spatial interpolation
intervals for irregular point distributions \citep{Bernardeau96}.

Once the Delaunay tessellation has been constructed, and the densities
at each sample point determined (see sect.~\ref{sec:densdtfe}), we
determine the density gradient $\left.\widehat{\bmath{\nabla}
  f}\,\right|_j$ within each Delaunay tetrahedron $\mathcal{T}_j$ from
the density values $(f(\bmath{ r}_0),f(\bmath{ r}_1),f(\bmath{
  r}_2),f(\bmath{ r}_3))$ at its four vertices at location ${\bf
  r_0,r_1,r_2,r_3}$.

Using the density gradients in the Delaunay tetrahedra, the DTFE
density value at any point ${\bf \widehat{r}}$ can be calculated by
determining in which tetrahedron it is located and subsequently
computing its density estimate ${\widehat f}(\bmath{ \widehat{r}})$
from the linear equation,
\begin{equation}
 {\widehat f}(\bmath{ \widehat{r}})\,=\,f(\bmath{r}_0)\,+\,
\left. \widehat{ \bmath{  \nabla} f} \, \right|_j  \cdot (\bmath{ \widehat{r}}- \bmath{ r}_0)  \,.
\label{eq:dtfeintp}
\end{equation}
To obtain an image of the density field, one calculates these density
estimates at each of the voxel locations of the image grid. For a more
detailed outline of the DTFE method we refer to
section~\ref{app:dtfe}.

An impression of a DTFE interpolated field in a cosmological context
is presented in Figure~\ref{fig:recon} (top righthand panel). It
concerns a density field reconstruction from a dataset extracted from
a Millennium mock sample (top lefthand panel). The galaxy selection
follows the distant observer approximation, i.e. following parallel
lines of sight, and assumes the magnitude limit $m_r=17.77$ of the
SDSS redshift survey.

The resulting DTFE density field is the level map in the top righthand
panel. DTFE recovers the fine small-scale structures and at the same
time adapts itself to the larger scale structures at greater
distances.  It also reveals the linear interpolation artifacts, the
triangular shaped low-intensity wings. These are especially noticeable
when the data points are sparse. We must note that these wings are not
significant in mass, but arise when one takes a lower dimensional (1
or 2) section through the data.

\subsection{Interpolation Method II: \\ \ \ \ \ \ \ \ Natural Neighbour Field Estimator (NNFE)}
\label{sec:nnfe}
The DTFE is a piecewise linear interpolation ($C^0$) method. In a
sense it is a linear version of a larger class of tessellation based
interpolation methods. Of these, Natural Neighbour interpolation
\citep{Sibson81,Watson92,Braun95} is the most well known higher order
tessellation based method \citep[for more details see][]{Weyschaap09}.

The {\it Natural Neighbour Interpolation} formalism is a generic
higher-order multidimensional interpolation, smoothing and modelling
procedure utilizing the concept of natural neighbours to obtain
locally optimized measures of system characteristics. Its theoretical
basis was developed and introduced by \cite{Sibson81}, while extensive
treatments and elaborations of nn-interpolation may be found in
\cite{Watson92,Sukumarphd98}. As has been demonstrated by telling
examples in geophysics \citep{Braun95} and solid mechanics
\citep{Sukumar98,Sukumarphd98} NN methods hold tremendous potential
for grid-independent analysis and computations.

According to the Sibson natural neighbour interpolation, the
interpolated value ${\widehat f}(\widehat{\bmath{r}})$ at a position
$\widehat{\bmath{r}}$ is given by
\begin{equation}
{\widehat f}(\widehat{\bmath{r}})\,=\,\sum_i\,\lambda_{nn,i}(\widehat{\bmath{r}})\,f_i\,,
\label{eq:nnint}
\end{equation}
in which the summation is over the natural neighbours $i$ of the point
$\widehat{\bmath{r}}$ amongst the data points (see
Fig.~\ref{fig:nnbr}, righthand frame). Note that a slight movement of
the interpolation point will evoke a different set of natural
neighbours.

The Sibson natural neighbour  interpolation uses area-based (or volume
in 3D) interpolation  weights
$\lambda_{nn,i}(\widehat{\bmath{r}})$.  These are determined  from the
volumes    of    the    {\it    order-2}    Voronoi    cells    ${\cal
  V}_2(\widehat{\bmath{r}},{\bf  r}_j)$.  To  understand  the concept,
imagine we virtually insert  the location $\widehat{\bmath{r}}$ in the
spatial sample  point distribution. Around its location  a new Voronoi
cell    ${\cal     V}(\widehat{\bmath{r}})$    is    delimited    (see
Fig.~\ref{fig:nnbr}, where the cell is  traced by the blue edges). The
virtual  cell   ${\cal  V}(\widehat{\bmath{r}})$  overlaps   with  the
original Voronoi cells ${\cal V}_j$ of its natural neighbours $j$. The
{\it  order-2}   Voronoi  cells  ${\cal  V}_2(\widehat{\bmath{r}},{\bf
  r}_j)$ are  the regions of overlap,  and define the  region of space
for  whom  $\widehat{\bmath{r}}$  and  ${\bf  r}_j$  are  the  closest
``nuclei''.

According to Sibson interpolation the interpolation kernel
$\lambda_{nn,j}(\widehat{\bmath{r}})$ is equal to the normalized
area of the order-2 Voronoi cell,
\begin{equation}
\lambda_{nn,i}(\widehat{\bmath{r}})\,=\,{\displaystyle {{\cal A}_{2}(\widehat{\bmath{r}},{\bf r}_i)} \over 
{\displaystyle {\cal A}(\widehat{\bmath{r}})}}\,,
\label{eq:nnintint}
\end{equation}
in which ${\cal A}(\widehat{\bmath{r}})=\sum_j {\cal
  A}_2(\widehat{\bmath{r}},{\bf r}_j)$ is the area of the virtual
Voronoi cell of point $\widehat{\bmath{r}}$ and ${\cal
  A}_{2}(\widehat{\bmath{r}},{\bf r}_i)$ the area of the order-2
Voronoi cell ${\cal V}_2(\widehat{\bmath{r}},{\bf r}_i)$.

\begin{figure*}
  \centering
    \includegraphics[width=0.45\textwidth]{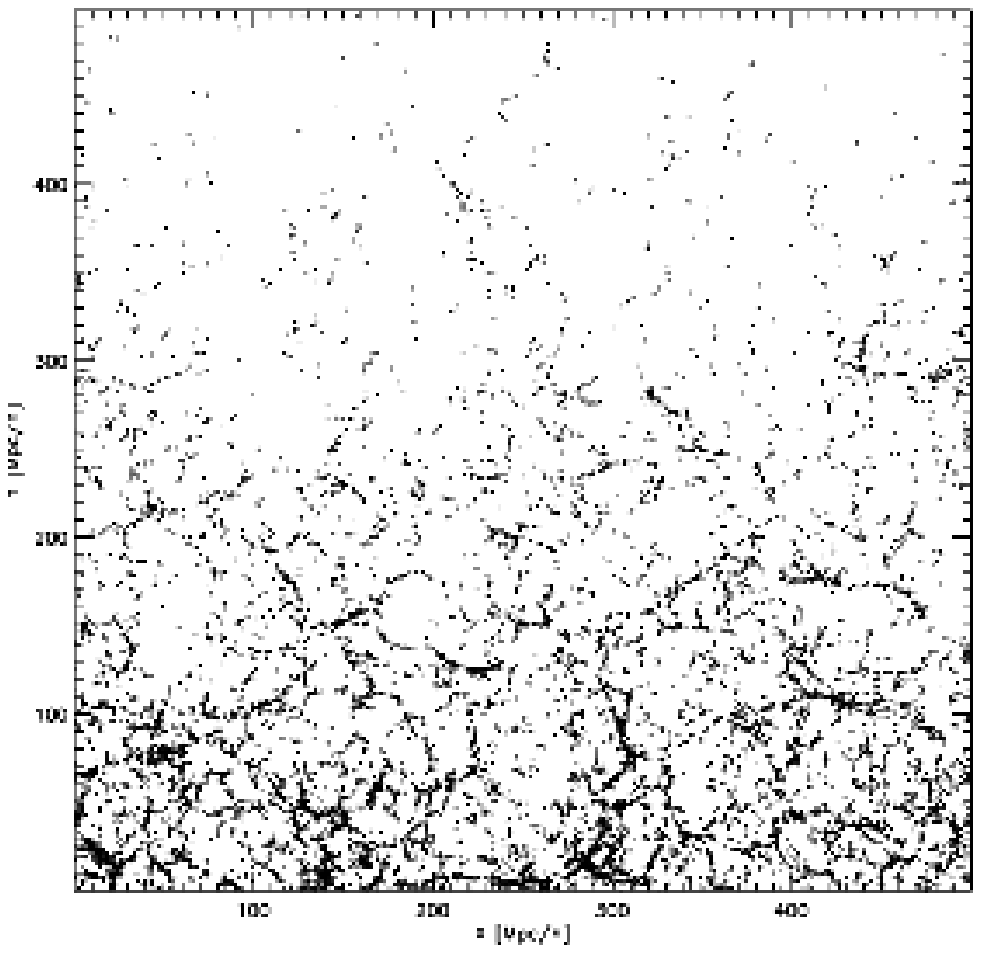}\hspace{0.15cm}
    \includegraphics[width=0.45\textwidth]{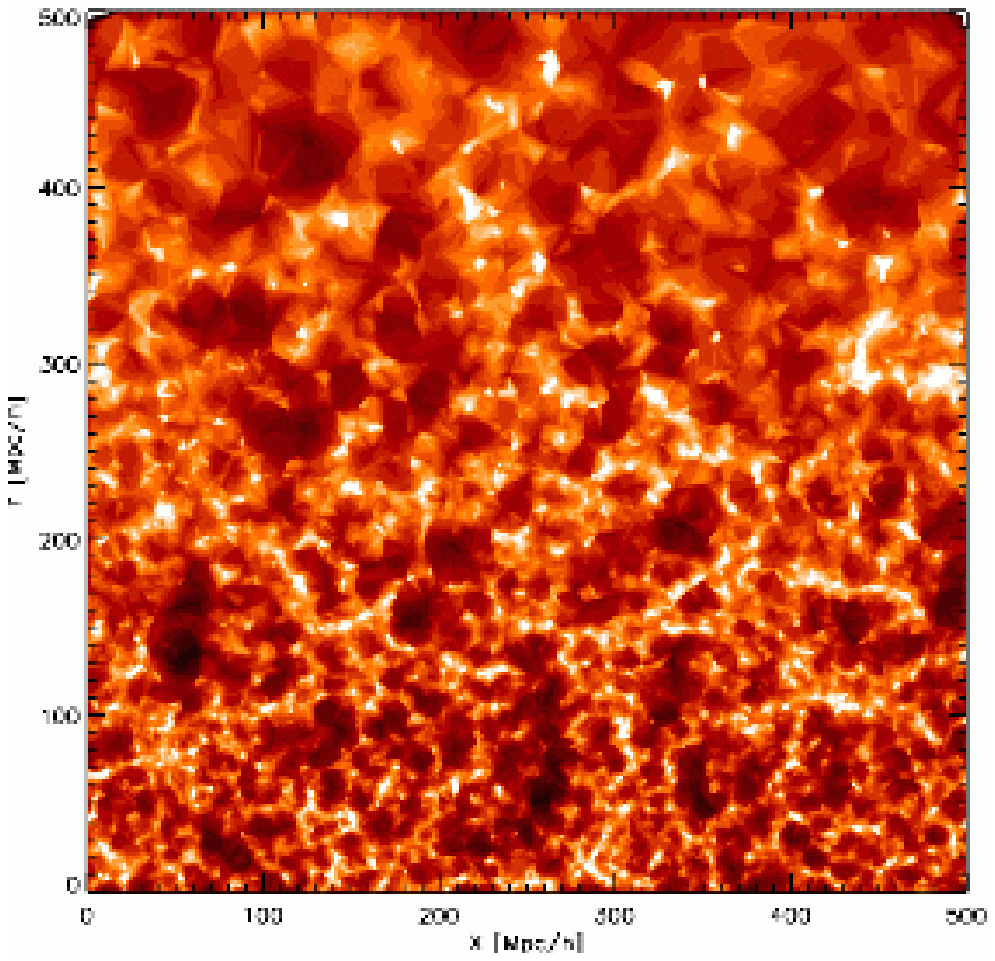}\\
    \includegraphics[width=0.45\textwidth]{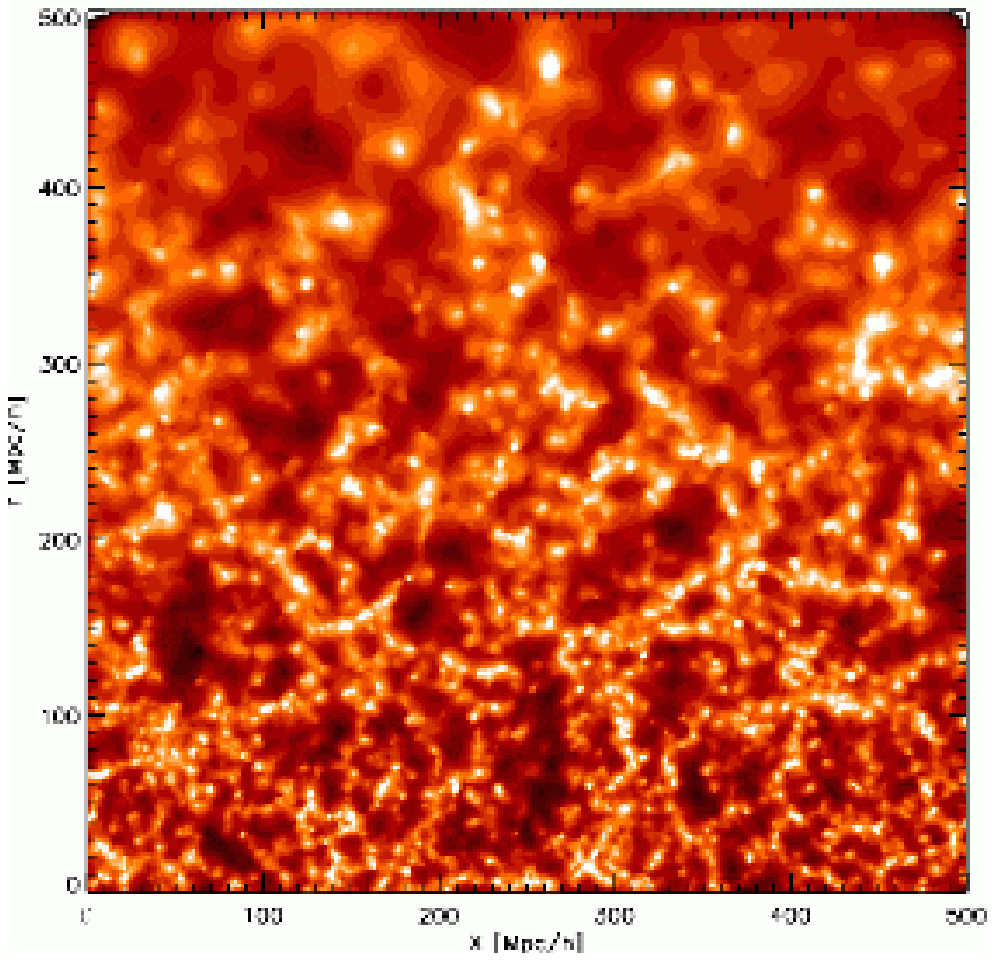}\hspace{0.15cm}
    \includegraphics[width=0.45\textwidth]{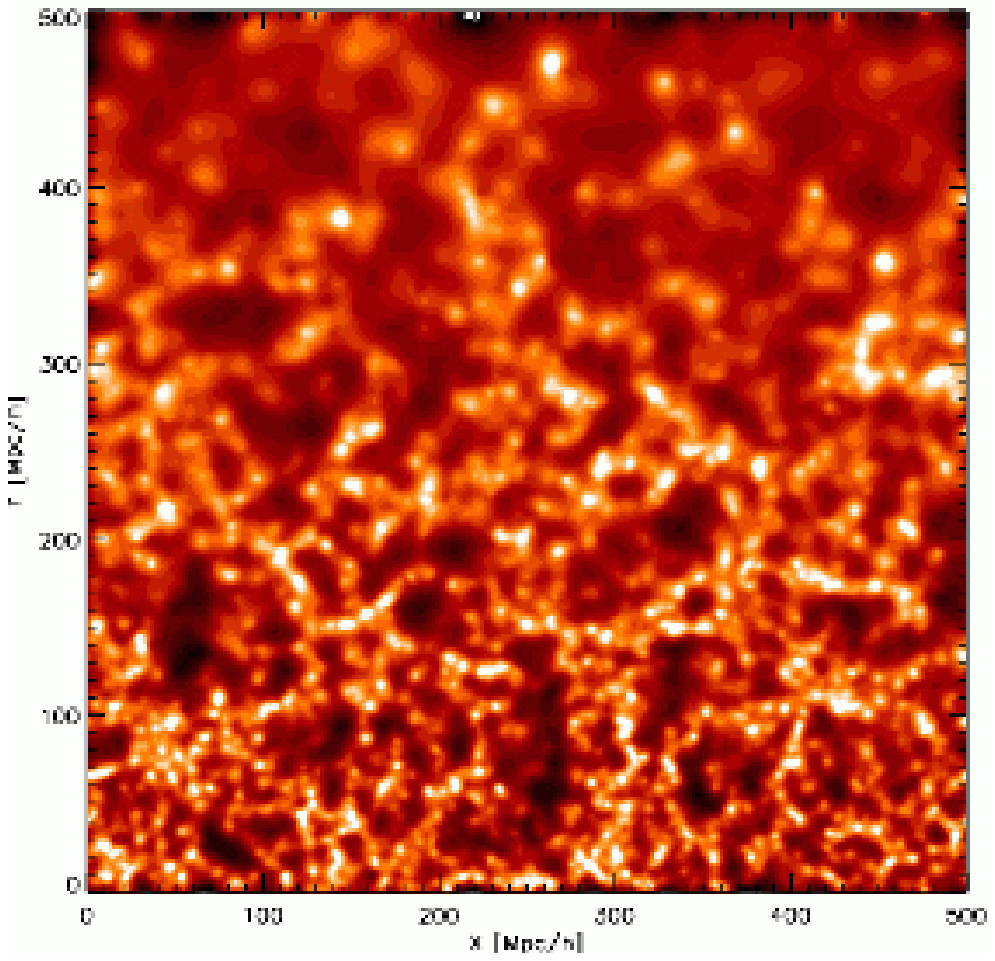}    
    \caption{Density Reconstructions of SDSS mock catalogue. Top left: the Millennium galaxy mock sample, 
    following the distant observer approximation. Top right: DTFE density field reconstruction. Bottom left: 
    NNFE density field reconstruction. Bottom right: lognormal Kriging reconstruction.}
    \label{fig:recon}
\end{figure*}

Evidently, the closer one moves a point $\widehat{\bmath{r}}$ to a
sample point ${\bf r}_j$, the more Voronoi cell ${\cal
  V}(\widehat{\bmath{r}})$ will overlap with the original Voronoi cell
${\cal V}_j$, and the larger the volume ${{\cal
    A}_{2}(\widehat{\bmath{r}},{\bf r}_i)}$ of the order-2 Voronoi
cell ${\cal V}_2(\widehat{\bmath{r}},{\bf r}_i)$ becomes, thus
increasing the weight $\lambda_{nn,i}(\widehat{\bmath{r}})$. When
$\widehat{\bmath{r}}$ finally coincides with one of the natural
neighbours, the order-2 Voronoi cell will be identical to the old
Voronoi cell at this point. The interpolated field value ${\widehat
  f}(\widehat{\bmath{r}})$ will then be equal to the field value at
that point.

Notice that the interpolation weights $\lambda_{nn,i}$ are always
positive and sum to one,
\begin{equation}
  \sum_{i=1}^{N} \lambda_{nn,i}(\widehat{\bmath{r}})\,=\,1. 
\end{equation}
This property is called {\it partition of unity}. The resulting
function ${\widehat f}({\bf r})$ is continuous everywhere within the
convex hull of the data, and has a continuous slope everywhere except
at the data themselves.  At the position of the vertices the
derivative of the interpolant is discontinuous.

In one dimension DTFE and NNFE are exactly the same. When the
data-points are given on a regular grid, the NNFE reduces to the more
familiar bi-linear (2d) or trilinear (3d) interpolation schemes. Our
NNFE implementation is that of \cite{Eldering}, a three-dimensional
adaption of the two-dimensional version available in the CGAL library.

The NNFE density field reconstruction for the same Millennium mock
sample as described in section~\ref{sec:dtfe} is shown in the bottom
lefthand panel of Fig.~\ref{fig:recon}. Some of the peaks in the
regions with sparse sampling appear somewhat anisotropic. This is a
consequence of the discontinuous derivative at the sample point. The
overall resulting NNFE density field is well-behave and smooth,
without the artifacts that beset the DTFE reconstruction.

A drawback of the DTFE and the NNFE methods is that neither take
into account the existing spatial correlations that characterize the
cosmological density field. These are explicitly taken into account by
the Kriging interpolation technique.

\vskip 2.0truecm

\subsection{Interpolation Method III: \\ \ \ \ \ \ \ \ Kriging Interpolation}
\label{sec:kriging}
By basing itself on the covariance function of the density field,
Kriging naturally includes the global spatial correlations of the
field.

The method was named by \cite{Matheron63} after D. G. Krige, who
started the development of the method \citep[see][for a historical
  overview]{Cressie90}.  The interpolator has the property that it is
a {\it best linear unbiased estimator} \citep{Cressie88, Cressie93}. Most
applications of Kriging stem from the field of geostatistics, where
Kriging found its origin.  The applications concern measurements at
irregularly scattered points which have to be translated into, for
example, gold, ore or oil field reconstructions or into altitude maps.

\medskip
There is a distant relationship of Kriging interpolation to 
Wiener filtering techniques \citep{Wiener49,Rybicki92,Zaroubi95,
Zaroubi02,Erdogdu04,Erdogdu06,Kitaura08}. However, Wiener filtering is
based on a different philosophy than Kriging, in that it includes a
model for the noise and is evaluated in Fourier space. The retrieved
field therefore corresponds to an optimally filtered field over a
range of {\it unknown} scales. The filter scale is dictated by the
locally estimated noise, with more noise corresponding to a larger
amount of smoothing. An additional disadvantage for recovering 
nonlinear weblike features of classical Wiener filtering is that it is 
predicated on an underlying Gaussian 
distribution in the construction of the least squares estimator 
for filtering the data. While advantageous for the purpose of
ascertaining the exclusive presence of significant features, it 
therefore has the serious disadvantage of suppressing or substantially 
diluting nonlinear structures of interest and may have difficulties
in reconstructing the intricacies of the nonlinear structures.

More advanced recent developments and applications of Wiener filters 
to the reconstruction of the density distribution have revived its 
potential for reconstructing the density distribution. For an extensive 
and in-depth overview of these developments we refer to \cite{Kitaura08} 
\citep[also see][]{Kitaura09,Kitaura10}.  
 
\subsubsection{The Kriging formalism}
Interpolation can be viewed as estimating the field value $\widehat{f}$ at 
location $\widehat{\bmath{r}}$ by means of a weighted linear combination of nearby
known data points $f(\bmath{r}_i)$;
\begin{equation}
  \widehat{f}(\widehat{\bmath{r}})\,=\,\sum_{i=1}^{N} \lambda_{i} f(\bmath{r}_i)\,.
  \label{eq:krig_est}
\end{equation}

\noindent The main idea of Kriging, as originally formulated, is to calculate the values of weights
$\lambda_i$ that minimise the error with respect to the data according
to the mean square variation,
\begin{equation}
\Bbb{E}(|\widehat{f}({\bmath{r}}) - f(\bmath{r})|^2),\nonumber
 \label{eq:error}
\end{equation}
\noindent where $\Bbb{E}$ is the expectation over the specified quantity.  
We show in \cite{Jones11} that this criterion can be replaced with the weaker requirement that the data and 
the errors be orthogonal in a statistical sense.  It follows that the statistical distribution of the field 
$f(\bmath{r})$ need not be Gaussian distributed in order to achieve optimal reconstruction of the density field via Kriging.  

\noindent The {\it Kriging equations} for the weights $\lambda_j$ are
\begin{equation}
\sum_{j=1}^N\,\bmath{C}(\bmath{r}_i,\bmath{r}_j)\,\lambda_{j} = \bmath{c}(\bmath{r}_i,\bmath{\widehat{r}}),
\end{equation}
where the matrix elements of $\bmath{C}$ are given by 
\begin{equation}
\bmath{C}(\bmath{r}_i,\bmath{r}_j)\,=\,\Bbb{E}(f(\bmath{r}_i)
f(\bmath{r}_j))\,,
\end{equation} 
and the vector elements $\bmath{c}(\bmath{r}_i,\bmath{\widehat{r}})$ by
\begin{equation}
\bmath{c}(\bmath{r}_i,\bmath{\widehat{r}})\,=\,\Bbb{E}(f(\bmath{r}_i)f(\bmath{\widehat{r}}))\,.
\label{eq:kriging}
\end{equation}
\noindent From this linear system of $N$ equations, it is straightforward to determine the 
$N$ unknown weights $\lambda_i$. 

While the matrix has to be inverted only once, the weights 
$\lambda_i$ have to be specifically computed for each interpolation site $\widehat{r}$. 
After the weights have been determined, one can directly obtain the interpolated field 
values $\widehat{f}$ from equation~(\ref{eq:krig_est}).

\subsubsection{The Kriging Variogram}
Usually, the covariance function $\Bbb{E}(f(\bmath{r}_i)f(\bmath{\widehat{r_j}}))$ 
depends only on distance, $d =|\bmath{r}_i-\bmath{r}_j|$. In geostatistics,  
this spatial dependence of random field is usually characterised by 
means of a variogram $\gamma(\bmath{r}_1,\bmath{r}_2)$. The variogram is the 
mean square variation of the field values as function of distance, 
\begin{equation}
  2\gamma(\bmath{r}_1,\bmath{r}_2)\,\equiv\,\Bbb{E}(|f(\bmath{r}_1)-f(\bmath{r}_2)|^2)\,,
\end{equation}
which for a stationary random field reduces to
\begin{equation}
  2\gamma(\bmath{h}) = \Bbb{E}(|f(\bmath{r})-f(\bmath{r+h})|^2)\,.
\end{equation}
The variogram is related to the covariance function $c(h)$,
\begin{eqnarray}
~~~~~~~~~~~~~~~ c(h) & = & \Bbb{E}(f(\bmath{r}_1)f(\bmath{r}_2)) \nonumber \\
~~~~~~~~~~~~~~~      & = & c(0) - \gamma(h).
\label{eq:covar}
\end{eqnarray} 

\noindent For practical purposes it is preferable to use a functional form for the 
variogram. There is variety of such variogram models 
\citep[see][for a detailed description]{Cressie93}. We use the exponential 
expression, 
\begin{equation}
 \gamma(h) = {\sigma}_{0}\,\left\{1- \exp\left(\frac{h}{h_{0}}\right)^p\right\}\,.
 \label{eq:vario}
\end{equation}
which represents a good fit for the variogram measured from a 
a Millennium SDSS mock galaxy survey (see sect.~\ref{sec:mock}).
To estimate the variogram for this galaxy distribution, we used 
the estimator \citep{Cressie93},
\begin{equation}
  \widehat{\gamma}(h)\,=\,\left(\frac{1}{N_p} \sum_{i=1}^{N_p} \sqrt{|f(r)-f(r+h)|}\right)^4
  \label{eq:varioest}
\end{equation}
We base our estimate on a large number $N_p$ of randomly chosen locations within the 
sample volume. 

\medskip
\begin{figure}
  \includegraphics[width=0.45\textwidth,height=0.40\textwidth,clip]{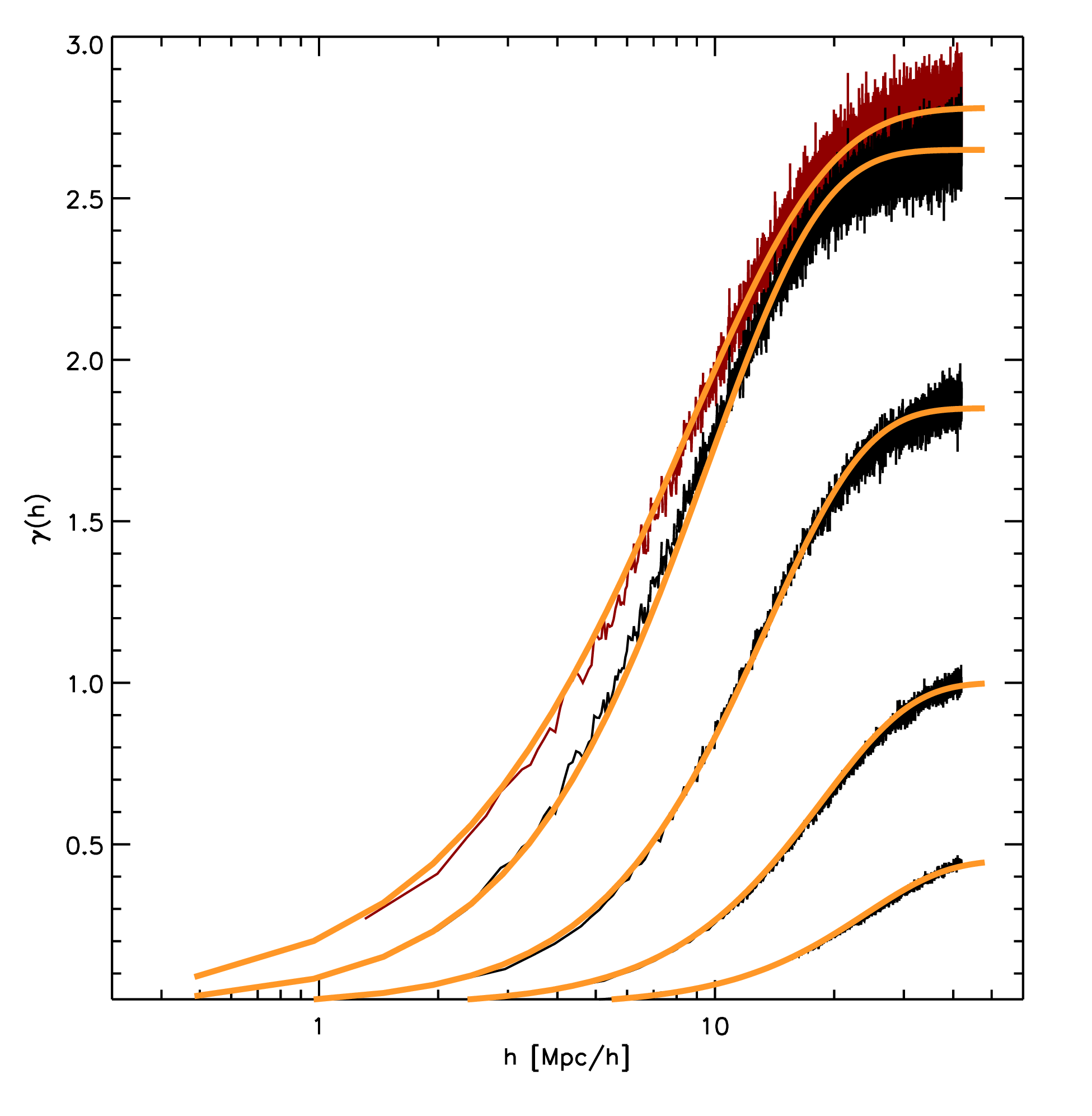}
  \caption{The measured variogram from top to bottom correspond to the
  unsmoothed field (red), and the filtered fields at scales of 1$\Mpch$
  Gaussian filtered, 3$\Mpch$, 6$\Mpch$ and 10$\Mpch$ (black). The fitted
  variogram models according to equation~(\ref{eq:vario}) are shown in
  orange.} 
  \label{fig:vario}
\end{figure}
\begin{table}
  \center
 \caption{Kriging Variogram Parameters.  
Parameters obtained from simulated Millennium SDSS mock 
catalogue (see text).}
 \label{symbols}
 \begin{tabular}{@{} lcccccc}
  \hline
  $R_f$ $(\Mpch)$   & $\sigma_0$ & $h_0$ & $p$  & $R_{max}$ $(\Mpch)$\\
  \hline
  0  & 2.78 & 8.4  & 1.2  & 200\\
  1  & 2.65 & 9.6  & 1.5  & 200\\
  3  & 1.85 & 13.5 & 1.72 & 300\\
  6  & 1.0  & 18.6 & 1.9  & 400\\
  10 & 0.45 & 24.0  & 2.1  & 500\\
  \hline
  \label{table:params}
 \end{tabular}
  \center
\end{table}
\subsubsection{Lognormal density fields}
A lognormally distributed density field $f$ has distribution
\begin{equation}
  P_{LN}(f)\,=\,\frac{1}{\sqrt{2 \pi S^2 }} \exp\left(-\frac{\left[\log (1+f) +S^2/2 \right]^2}{2S^2}\right) \frac{1}{1+f},
\end{equation}
where $S$ is the $S = \log (1+\sigma)$ and $\sigma^2=\langle f^2 \rangle$ is the variance of
the density field, \citep[see][]{Coles91}. 

What might simply be referred to as the ``Lognormal Kriging'' procedure uses the logarithm of the density field value to transform the density field data, 
\begin{equation}
\phi(f)\,=\,\log(1+f)
\end{equation}
Since the density $\rho=\bar{\rho}(1+f)$ is always positive, the application of 
the lognormal approach is valid everywhere and guarantees a positive definite reconstruction of the density. 

The final interpolation values are obtained by taking the inverse transformation, 
ie. the exponential of the interpolated data values,
\begin{equation}
  \widehat{f}(\widehat{\bmath{r}})\,=\,\exp \left\{ \sum_{i=1}^{N} \lambda_{i} \log \left( f(\bmath{r}_i)\right)\right\}.
  \label{eq:log_krig_est}
\end{equation}
We will use the logarithmic value of the DTFE-interpolated field in what follows. 

For such a field, figure ~\ref{fig:vario} shows the variogram for a DTFE interpolated field (red) and 
for fields Gaussian smoothed on scales $R_f=1,~3,~6,~10 \Mpch$. The fitted variogram model parameters 
(equation~\ref{eq:vario}) are listed in table~\ref{table:params}. 

In the second paper of this series on SDSS density field reconstructions, we will demonstrate that the galaxy distribution is indeed very well modelled by a lognormal distribution at scales in excess of $3 \Mpch$.

\subsubsection{Localized Kriging}
The value of $N$ to be used in equation~(\ref{eq:krig_est}) has so far not been defined.
We chose the local
neighbourhood to be the {\it tetrahedral natural neighbourhood}. In 3 dimensions this is
the union of all natural neighbours of the four vertices of the
Delaunay tetrahedron in which a point ${\widehat {\bf r}}$ is
located. This choice exploits the self-adaptivity to density and local
shape of the Delaunay triangulation, and does not suffer the adverse
effects mentioned above for the options of distance or number of
neighbour selection \citep[also cf. the discussions in][]{Schaap07,Weyschaap09}.
Our experiments indeed confirm that the tetrahedral natural
neighbourhood choice is superior to that of the 2 options listed
above.

We found that in 3 dimensions the {\it tetrahedral natural
  neighbourhood} on average contains approximately 57 particles. One
may extend the neighbourhood by adding a third or even more layers
around it.  An additional third layer would involve an average of 284
neighbours: however, the overall quality of the field reconstruction is not 
significantly better than with two layers despite a substantial increase in 
computational effort. See \cite{Jones11} for more details on this.

Thus in our key equation~(\ref{eq:log_krig_est}) we use the value of N that is the number of first and second layer vertices surrounding each point.


\section{Qualitative Density Comparison}
\label{sec:qual}
For an assessment of the performance of the three reconstruction 
techniques, we turn to the mock catalogues modelling the SDSS 
DR6 galaxy survey sample. On the basis of the knowledge of the 
underlying density field of the mock samples, we will be able to 
infer absolute statements about the quality of the reconstructions. 

In this section will first address the 
visual appearance of the density field reconstructions, which 
will allow a qualitative and global judgement on their ability 
to reproduce the true density field. A quantitative error and 
correlation analysis will follow in the subsequent section~\ref{sec:quan}, 
while the topological properties of the reconstructions will be discussed 
in sect.~\ref{sec:topo}.

\begin{figure*}
 \centering
  \includegraphics[width=.85\textwidth]{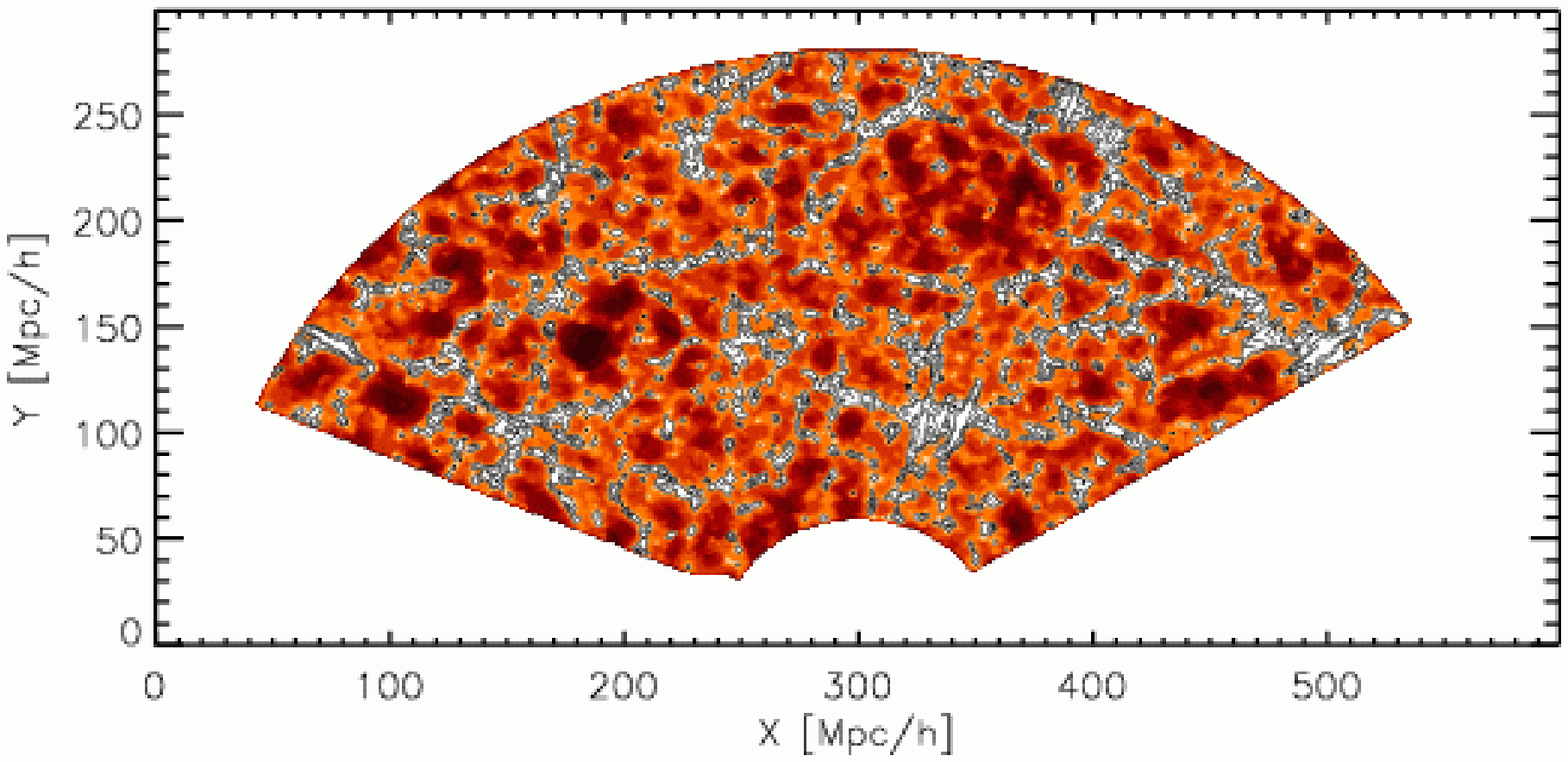}
    \includegraphics[width=0.85\textwidth]{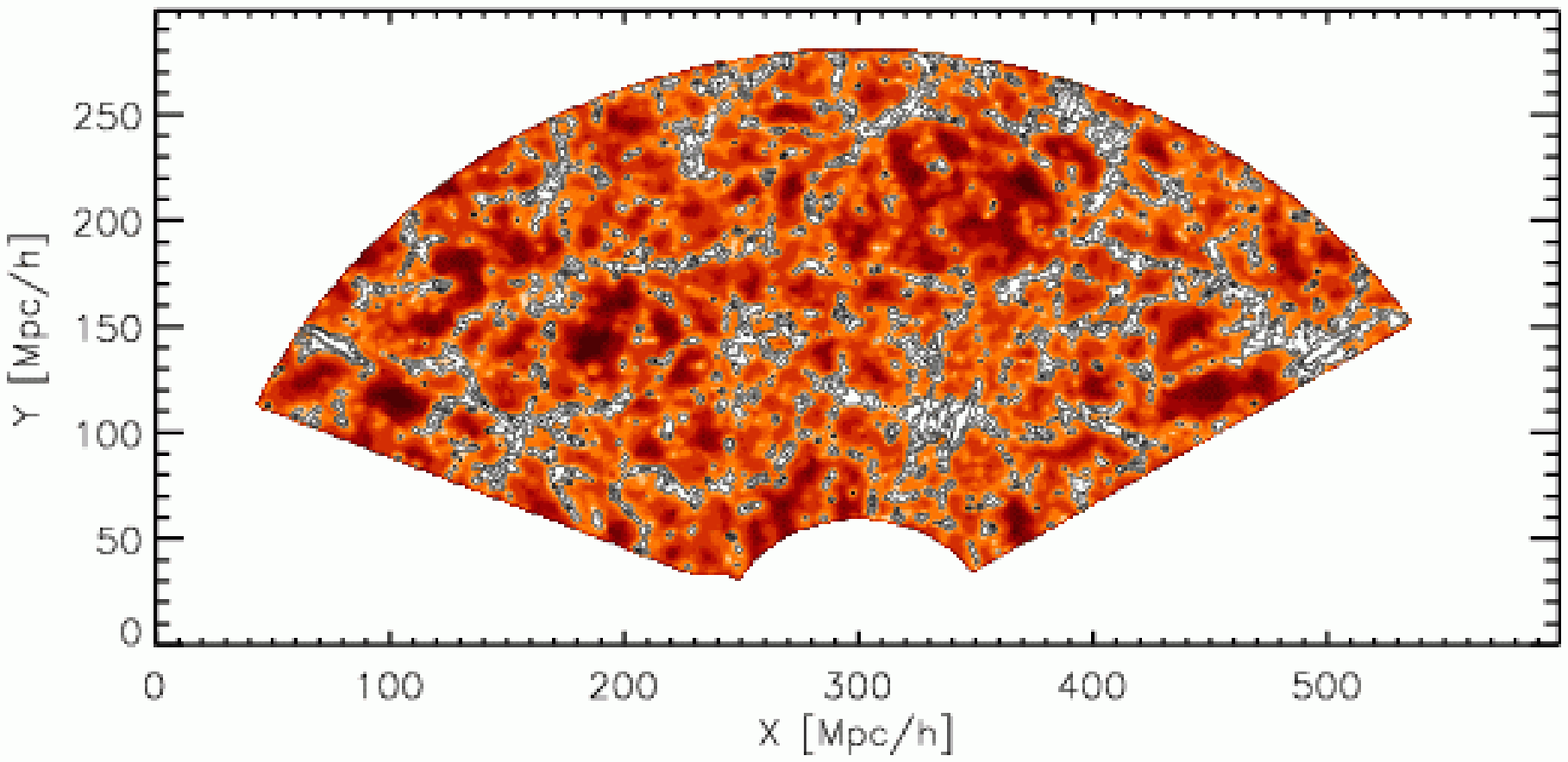} 
    \includegraphics[width=0.85\textwidth]{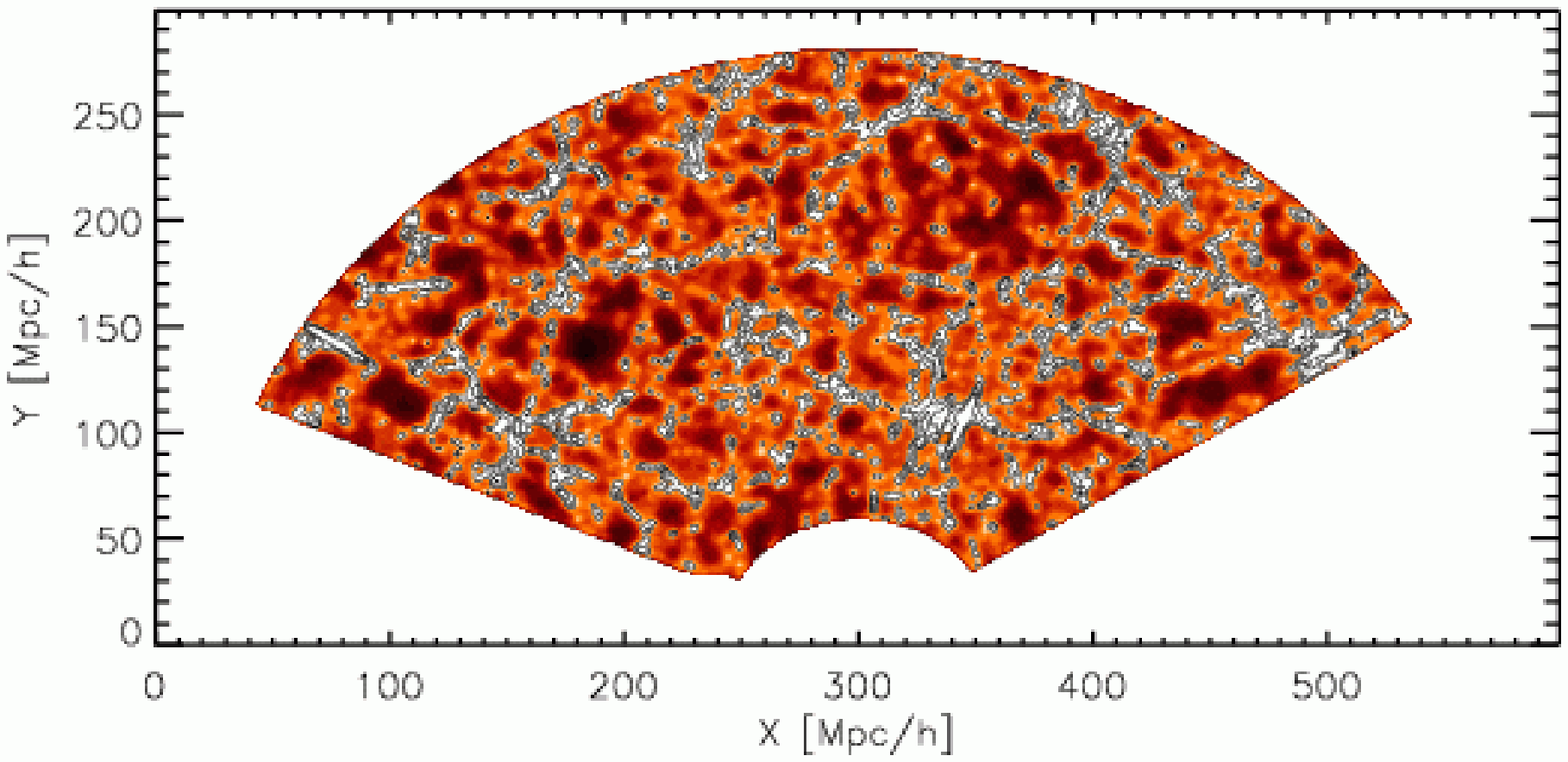}
    \caption{Comparison SDSS-DR6 reconstructions for the magnitude-limited galaxy sample. 
    Shown are - zero width - sections through the density field reconstructions. 
    {\it Top}: DTFE; {\it Centre}: NNFE; 
    {\it Bottom}: Natural Lognormal Kriging. The coloured contour levels represent the
    underdense regions, at $\rho/\rho_u=[0.001, 0.002, 0.005, 0.01,
    0.02, 0.03, 0.05, 0.07, 0.1, 0.2, 0.3, 0.6, 0.7, 0.8, 1]$. The
    white areas are the overdense regions and the black contour lines
    represent a density contrast $\rho/\rho_u=[1., 3., 10.]$. }
    \label{fig:full_comparison}
\end{figure*}
\begin{figure*}
  \centering
    \includegraphics[width=0.85\textwidth]{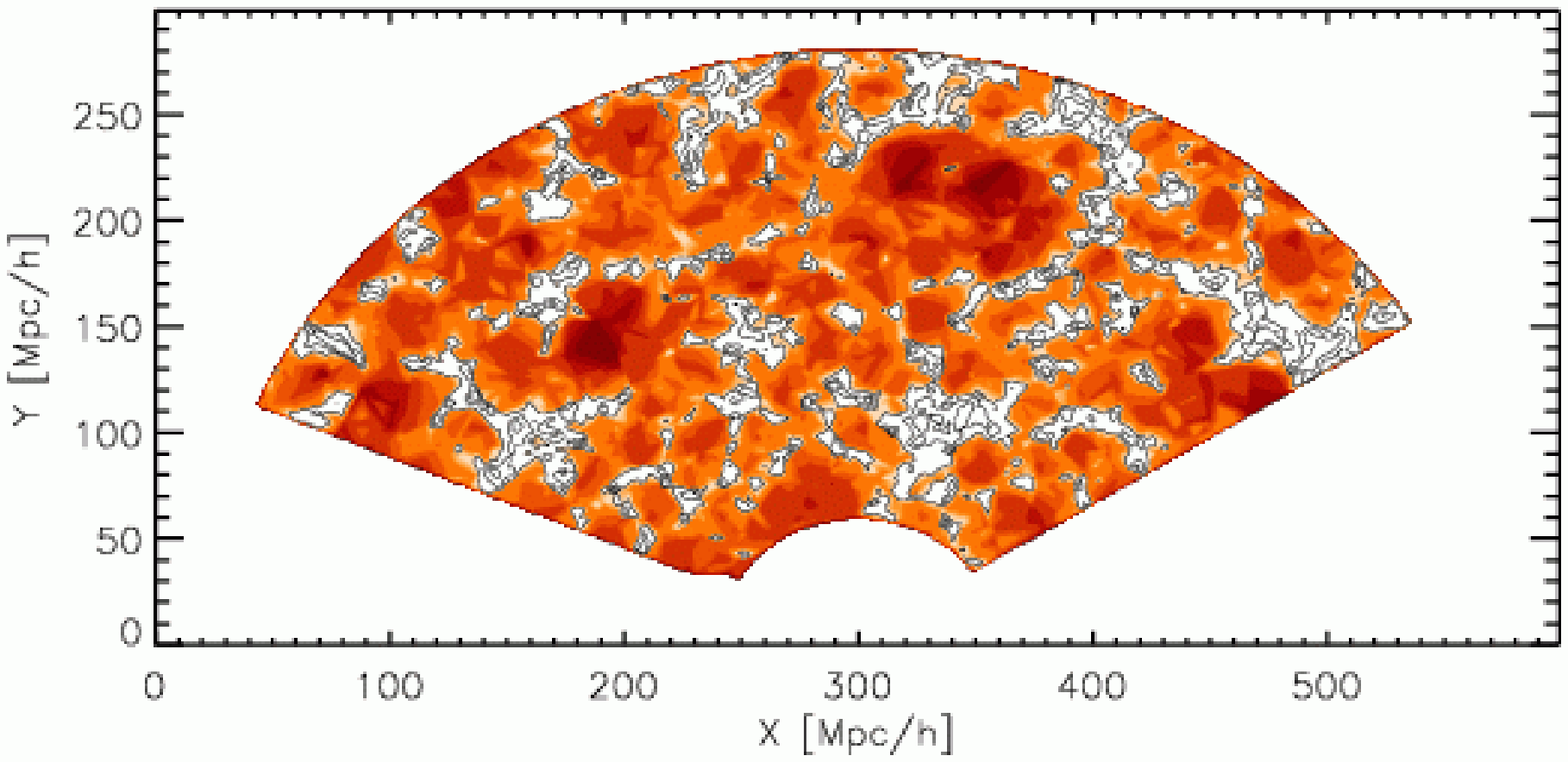}\\
    \includegraphics[width=0.85\textwidth]{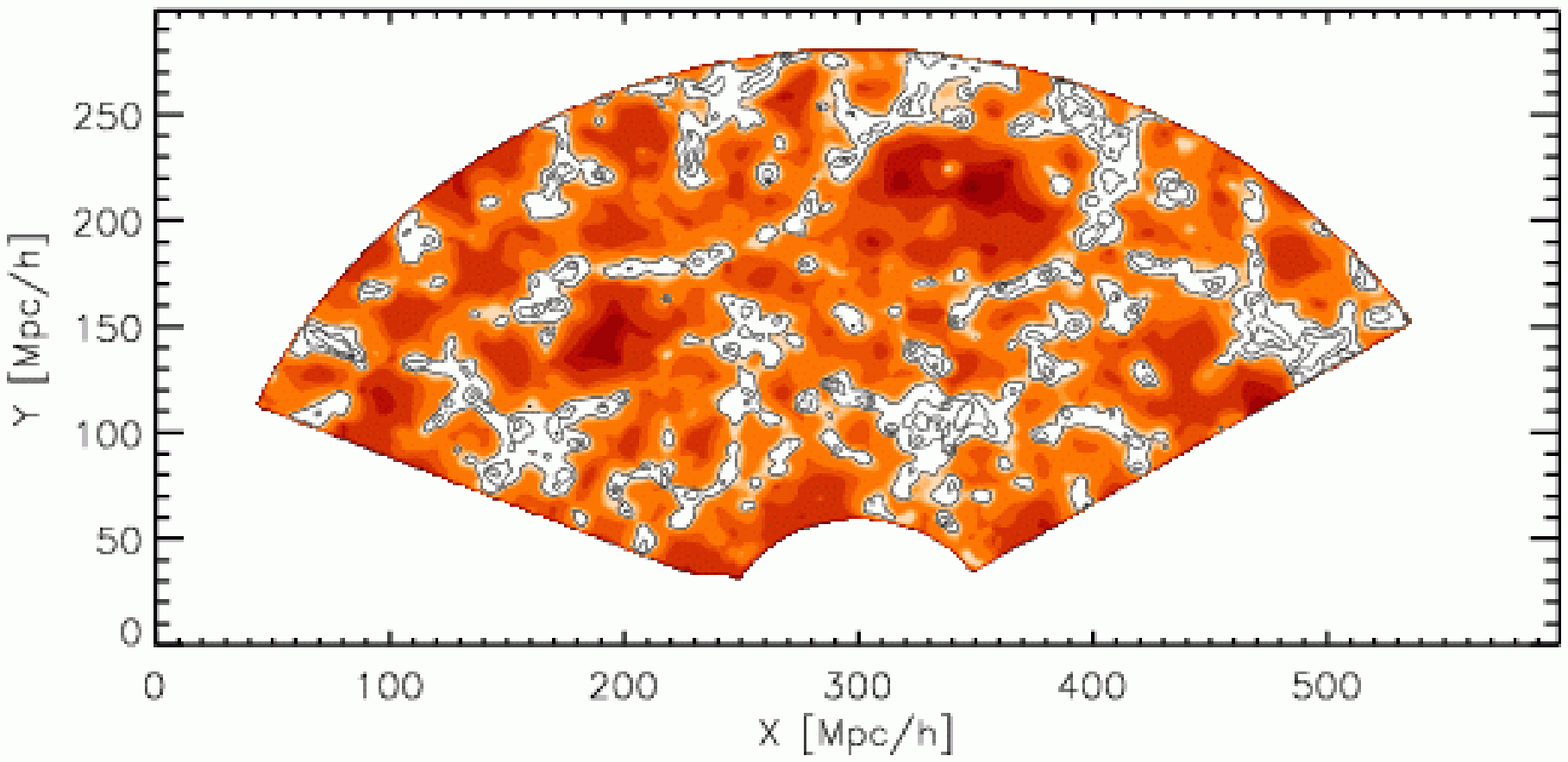}\\
    \includegraphics[width=0.85\textwidth]{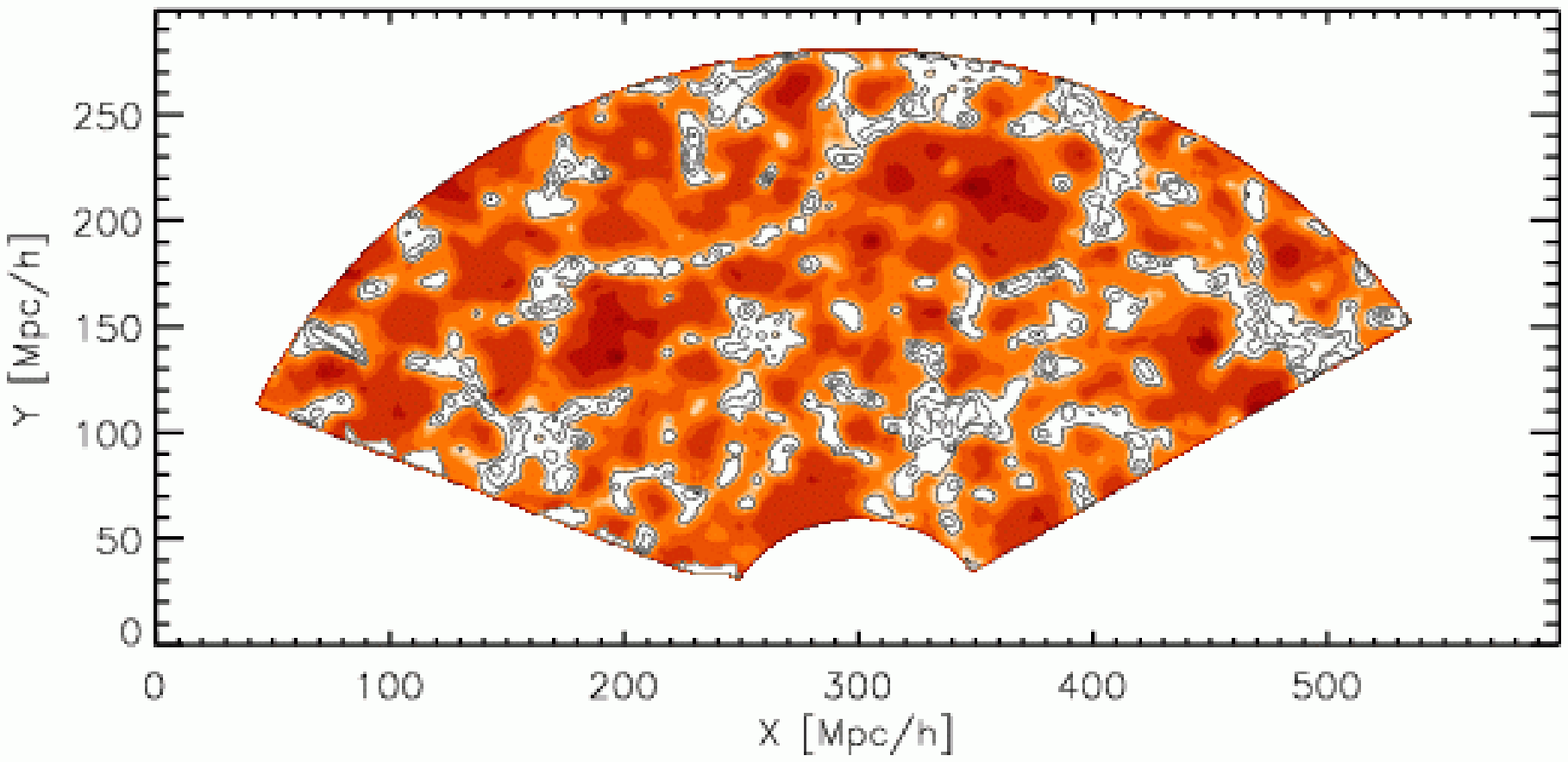}
    \caption{Comparison SDSS-DR6 reconstructions for the volume-limited galaxy sample. 
    Shown are - zero width - sections through the density field reconstructions. 
    {\it Top}: DTFE; {\it Centre}: NNFE; {\it Bottom}: Natural Lognormal Kriging. The 
    coloured contour levels represent the
    underdense regions, at $\rho/\rho_u=[0.001, 0.002, 0.005, 0.01,
    0.02, 0.03, 0.05, 0.07, 0.1, 0.2, 0.3, 0.6, 0.7, 0.8, 1]$. The
    white areas are the overdense regions and the black contour lines
    represent a density contrast $\rho/\rho_u=[1., 3., 10.]$. }
    \label{fig:vollim_comparison}
\end{figure*}
\begin{figure*}
  \centering
  \includegraphics[width=0.9\textwidth]{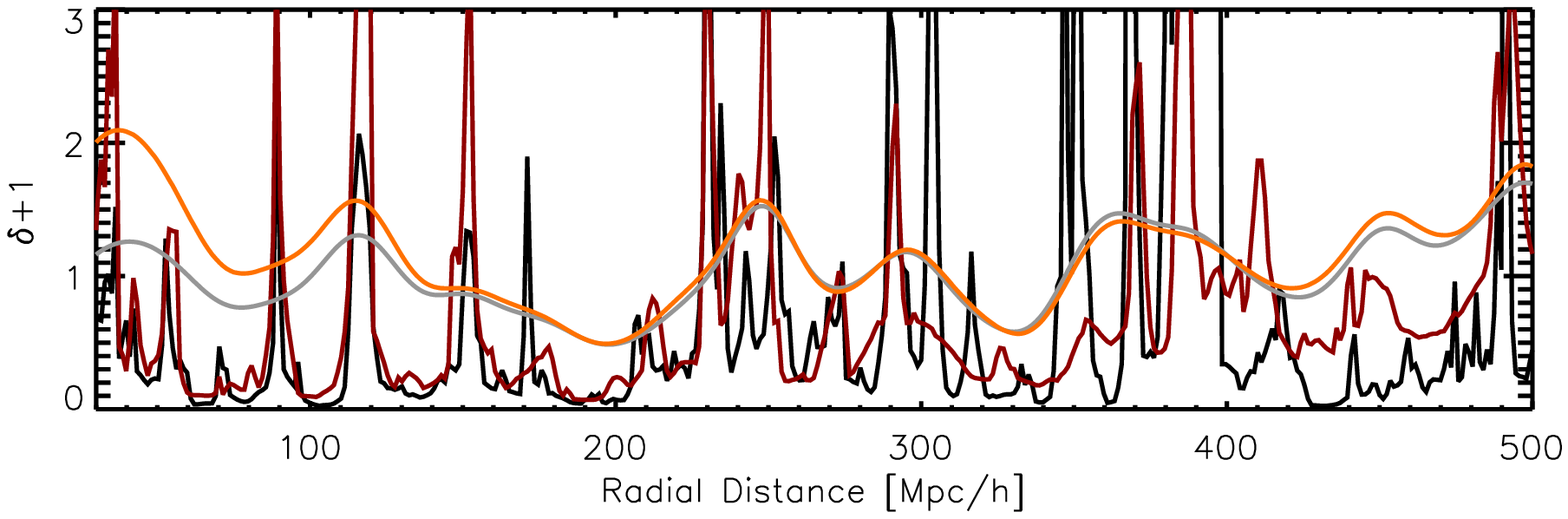}
  \includegraphics[width=0.9\textwidth]{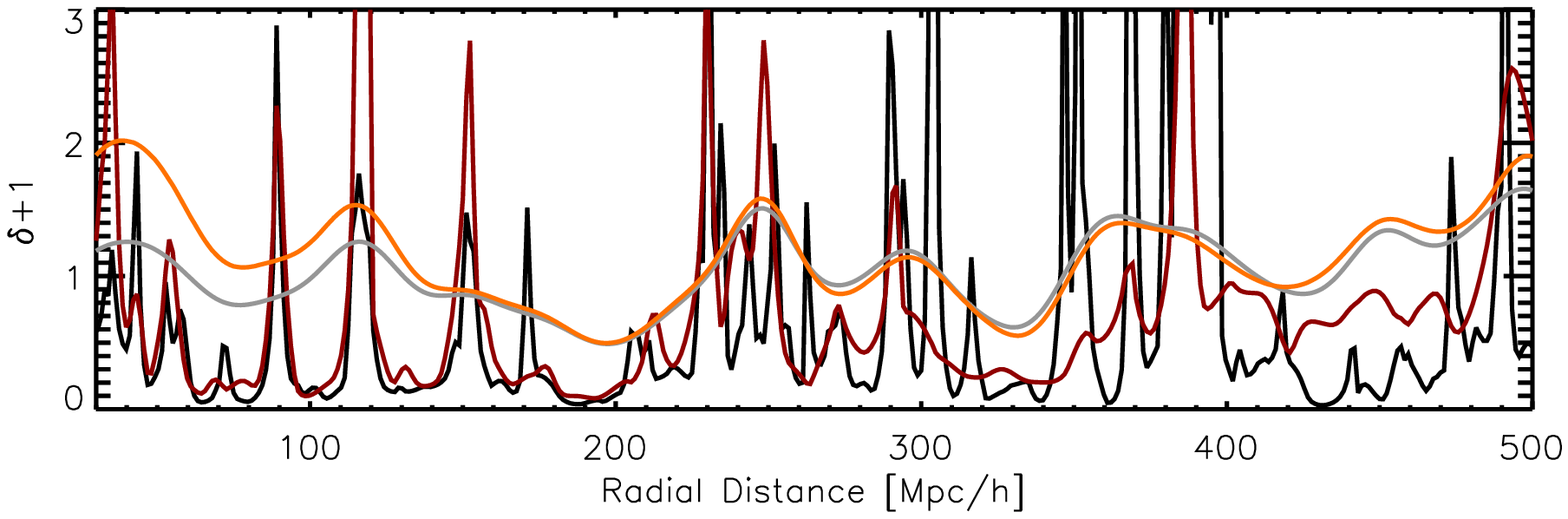}
  \includegraphics[width=0.9\textwidth]{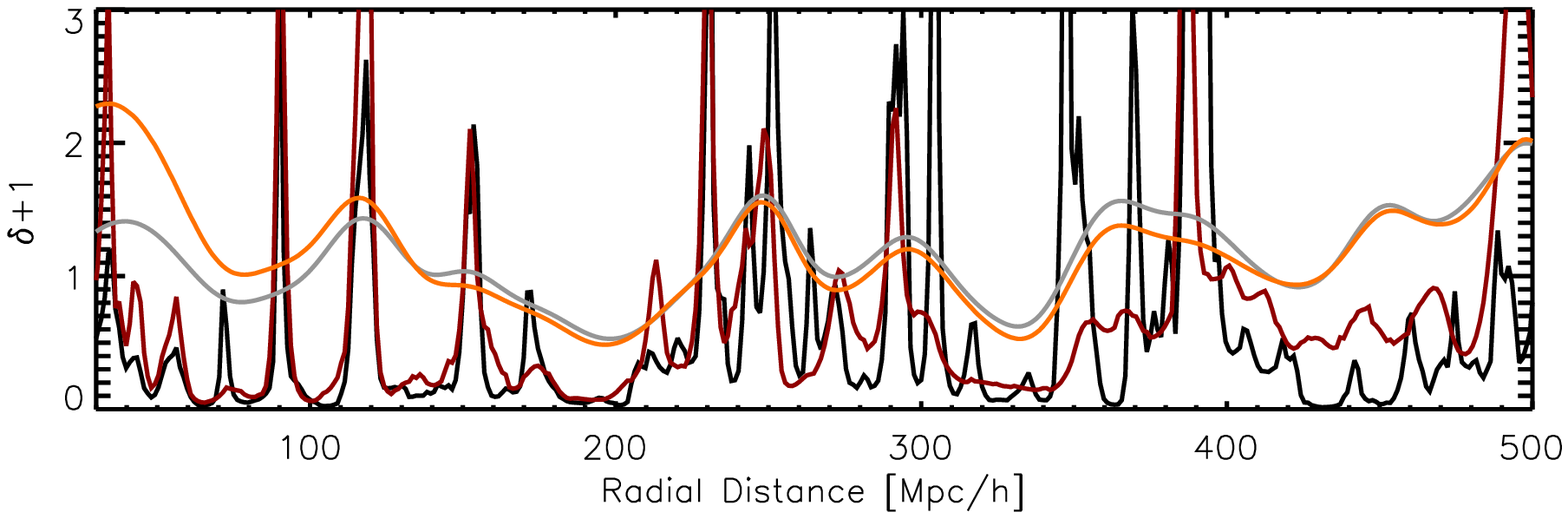}
  \caption{Linear profiles along line of sight. All profiles are taken 
  along the Y-axis, at $(X,Z)=(300,300)\Mpch$. {\it Top}: DTFE density field 
  reconstruction; {\it Centre}: NNFE density field reconstruction; 
  {\it Bottom}: Natural Lognormal Kriging reconstruction. In each panel: 
  {\it black line}: density field reconstruction full galaxy sample; 
  {\it red lines}: density field reconstruction volume-limited galaxy sample; 
  {\it gray lines}: Gaussian filtered density field reconstruction full galaxy sample, 
  $R_f=10.0\Mpch$; {\it orange lines}: Gaussian filtered density field reconstruction 
  volume-limited sample.}
  \label{fig:profile}
\end{figure*}
\subsection{Maps of the density field}
Density contour maps for the three reconstructions are shown in 
Fig.~\ref{fig:full_comparison} and Fig.~\ref{fig:vollim_comparison}. 
The maps concern thin slices through the density field reconstructions 
(and thus have zero intrinsic width). 

The first set of three maps concern the reconstruction on the 
basis of the {\it full} Millennium mock galaxy sample within the 
boundaries of the SDSS DR6 North Galactic Cap region. The second 
set concerns the same region, but for the {\it volume-limited} 
mock sample. 

To distinguish underdense and overdense regions, the underdense 
regions are indicated by means of a colour contour map while the 
overdense regions are represented by black contours. The contour 
levels in all maps are the same. 

The DTFE, NNFE and Kriging maps all display the same global 
structure. The overall appearance of both the DTFE, NNFE and 
Kriging density maps is strikingly similar. In the higher density 
regions the differences are minor, and mainly concern the coherence 
and anisotropy of the reconstructed filamentary (and sheetlike) 
features. Qualitatively speaking, the lower density regions do reveal more differences 
between the three methods. 

As expected, the volume-limited maps fail in reproducing the 
small-scale structure, while they trace the overall weblike outline 
seen in the full sample maps. 

\subsubsection{DTFE map}
The DTFE map is remarkably accurate in
outlining the tenuous weblike filamentary features, in particular in
the case of the full sample reconstruction. Amongst the three
reconstructions, the DTFE one looks more crispy than the NNFE and
Kriging maps.  It is slightly more capable in tracing the thin
filamentary and sheetlike features, while one might have a slight
worry with respect to the correct reproduction by NNFE and Kriging of
the shape of filaments and walls. Also, we find that the DTFE maps are
clearly marked by higher density contrasts, both within the overdense
regions as well as with respect to the underdense regions.

The downside of the detailed structural reconstruction of DTFE is the
more erratic nature of the DTFE contours, marked by sharp artifacts.
These artifacts are particularly prominent in the field reconstruction
of the volume-limited sample. They are manifestations of the linear
interpolation method: when two neighbouring grid cells are located in
different triangles, the field would appear to be discontinuous. The
latter is mostly an impression, as sampling at finer scales would show
it is just continuous. These artifacts occur in situations where the
point sample density is considerably sparser than the size of the
gridcells. This is also the reason why these artifacts are more
prominent in the DTFE reconstruction of the volume-limited sample than
in that of the full sample.

\subsubsection{NNFE and Kriging}
The reconstructed density maps of
the two higher order schemes, NNFE and Kriging, have a considerably
smoother appearance than the DTFE maps. The larger number of
neighbours involved in the NNFE and Kriging reconstructions translate
into the slightly more roundish contours of these structures. This
also reveals itself in the absence of artifacts such as seen in the
DTFE maps.

Part of the differences between the smoother higher order maps and the
DTFE maps is an expression of the number of points, and field values,
involved in the interpolation step (cf. eg.
equation~\ref{eq:log_krig_est}). DTFE uses 4 points, the vertices of a
Delaunay tetrahedron. NNFE involves on average 17 natural neighbours,
while the Natural Kriging scheme invokes 57 neighbours.

The smoother nature of the NNFE and Kriging maps is therefore an
expression of the somewhat lower information content of these filtered
maps. As a result, they also have less noisy low density regions than
those seen in the DTFE maps.

\subsubsection{Anisotropic Structure and Features}
One of the crucial benefits of DTFE is that it is able to identify 
anisotropic features, like walls and filaments, and successfully reproduce 
their shape and morphology \citep[see][]{Schaap07,Weyschaap09}. 

From the density maps we see that NNFE and Kriging find the same
filamentary structures. Overall, the impression is that DTFE and NNFE
produce maps in which the cosmic web is more coherent than in the
Kriging maps: the Kriging map mass concentrations have a slight
tendency to break up more easily into clumps. This is true for both
the full sample maps as well as the volume-limited sample maps. One
exception, in the volume-limited map, seems to be the filamentary
extension running from $(X,Y)=(170,160) \Mpch$ to $(X,Y)=(300,225)
\Mpch$. One reason for the somewhat more fragmentary nature of the
Kriging maps is its use of a radially symmetric covariance function,
while DTFE and NNFE are based on kernels that adapt to the local
shape.

\subsubsection{Underdensities \& Voids}
When turning to the
underdense regions, we find that in the full density map both DTFE and
Kriging delineate them at high contrast levels. The voids in the NNFE
map the voids have a lower contrast. This is partially the result of
the larger NNFE neighbourhood radius in low density areas. In this
respect, the Kriging correlations assure a better performance. DTFE
remains sensitive on behalf of its highly local character.

A comparison between the void population in the full sample
reconstruction and that in the volume-limited sample reconstruction
reveals a considerable contrast between the results for the different
reconstruction methods. None of the volume-limited reconstructions
contain the small voids visible in the full sample maps. This is a
reflection of the absence of such depressions in the diluted point
sample. Of the three reconstructions, the contrast between the two
DTFE maps is less distinct than that between the two NNFE and two
Kriging maps. DTFE at least manages to trace the large voids at
$(X,Y)=(180,140) \Mpch$, at $(X,Y)=(350,200) \Mpch$ and at
$(X,Y)=(460,120) \Mpch$. Kriging and NNFE hardly manage to find the
latter in the volume-limited maps, while the huge void complex near
$(X,Y)=(350,200) \Mpch$ is a largely uniform moderate underdensity in
the Kriging map. Only the DTFE map reveals its true nature, a region
marked by several deep voids embedded in a larger moderate undensity.

\begin{figure*}
  \centering
  \includegraphics[width=0.48\textwidth]{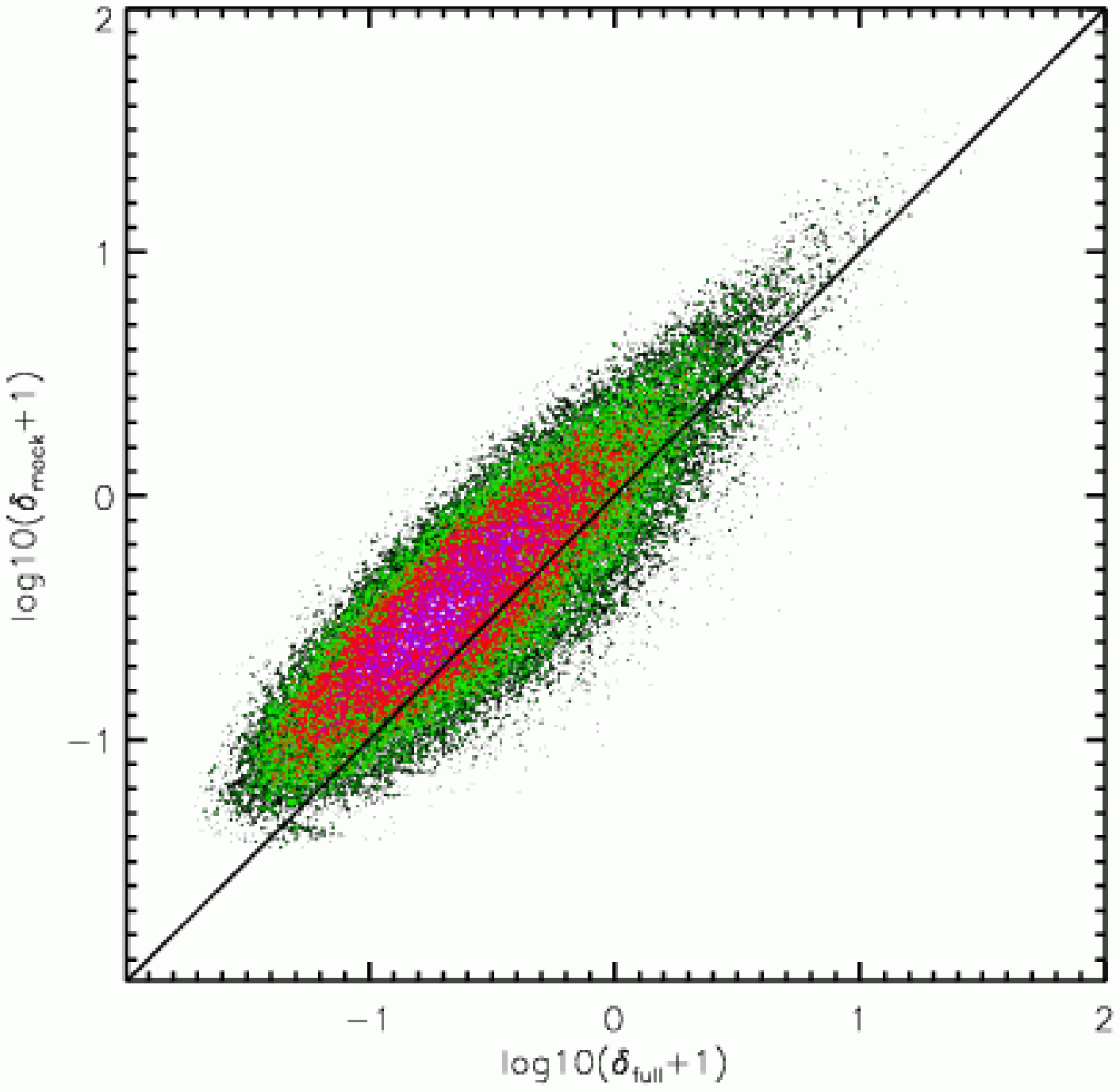}
  \includegraphics[width=0.48\textwidth]{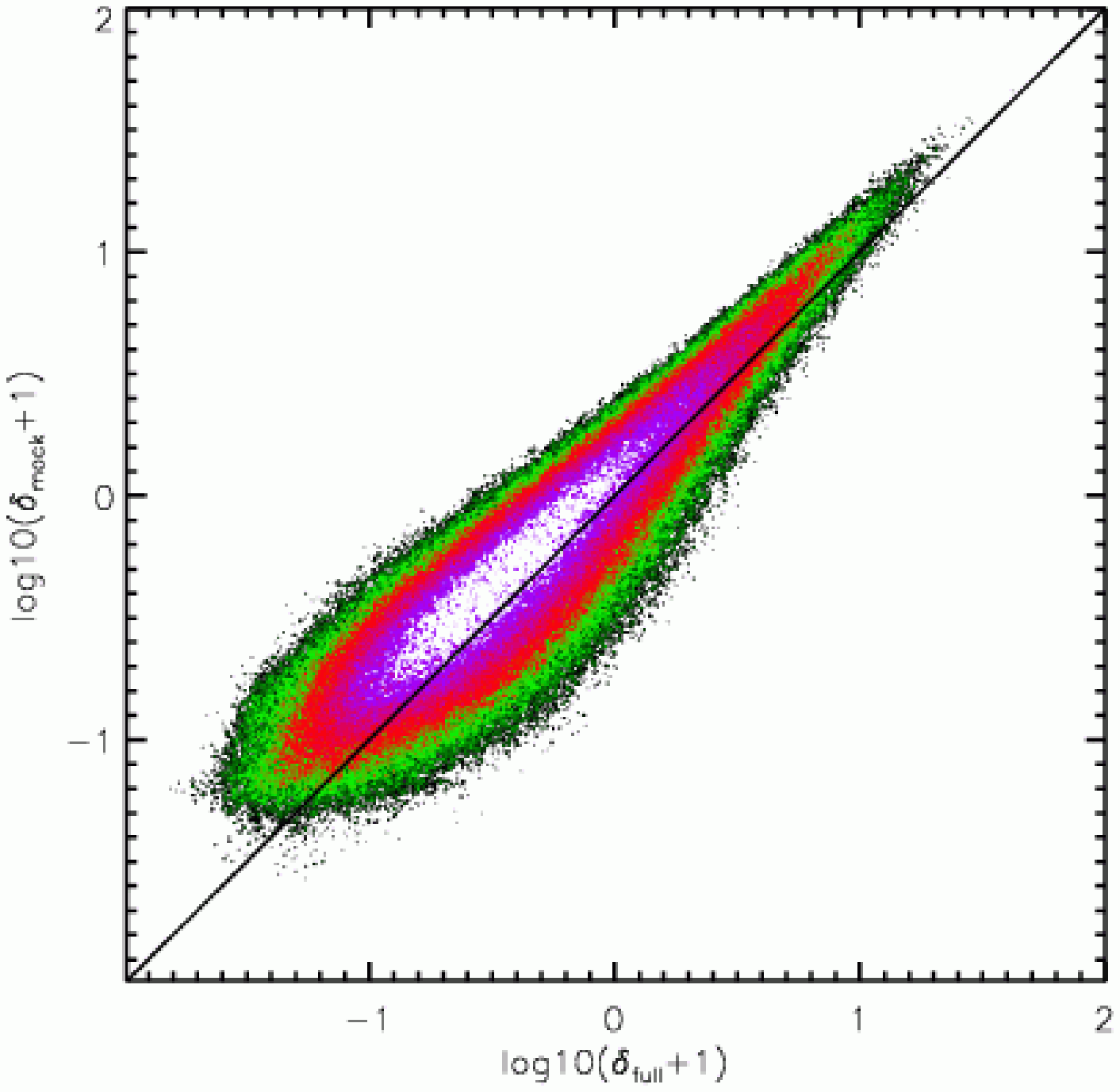}
  \includegraphics[width=0.48\textwidth]{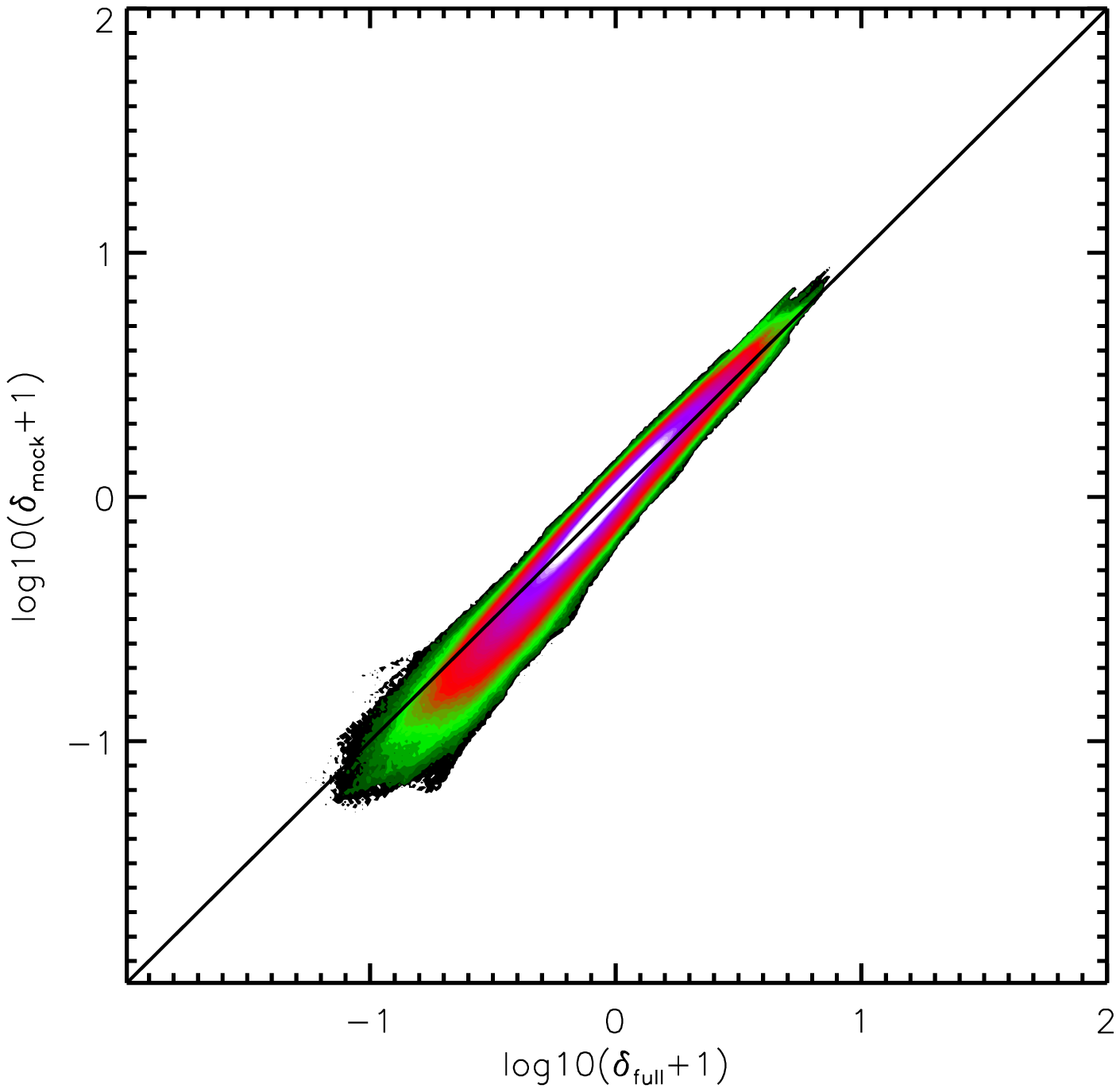}
  \includegraphics[width=0.48\textwidth]{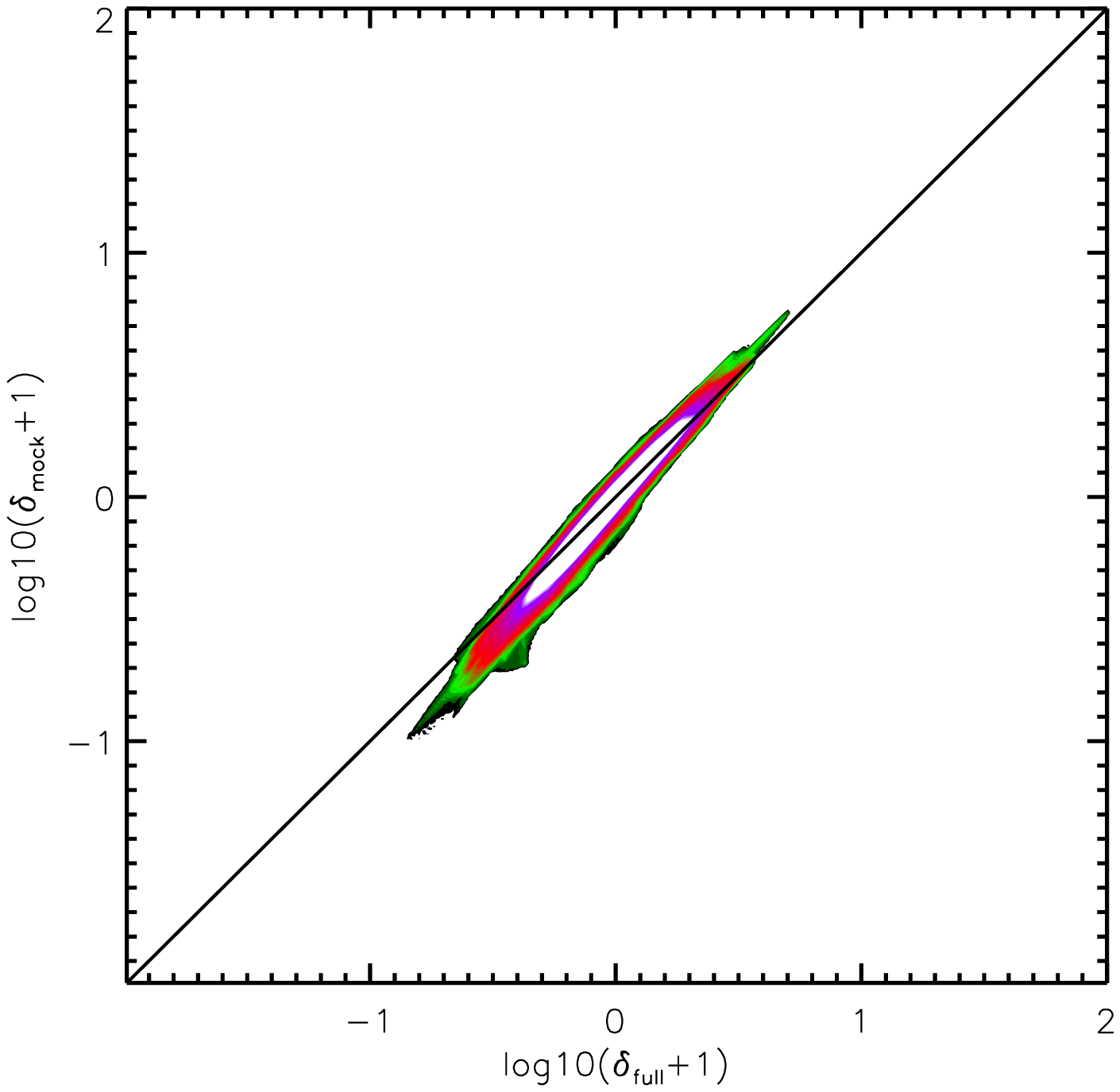}
  \caption{Correlation diagrams for DTFE density field reconstruction. Plotted 
  are the value of the density reconstruction from the full galaxy catalogue, 
  $\delta_{full}+1$ (abscissa), against the density reconstruction from the 
  magnitude-limited mock catalogues, $\delta_{mock}+1$ (ordinate). The colours indicate 
  the number of voxels that occupy the corresponding position in the correlation 
  diagram, with white indicating the density pairs with highest 
  occurrence and subsequent contour levels at logarithmic spacings of $\sim 2.5$ (see text). 
  {\it Top left}: unsmoothed density field reconstructions, 
  $R_f=0.0 \Mpch$. {\it Top right}: $R_f=1.0 \Mpch$. {\it Bottom left}: $R_f=6.0 \Mpch$. 
  {\it Bottom right}: $R_f=10.0 \Mpch$. Note that the apparent offset of the 
  small scale panels is a result of cosmic variance introduced by the locally 
  estimated density field normalization.}
  \label{fig:den_den}
\end{figure*}
\subsection{Density Profiles}
To appreciate the small-scale details in the density field
reconstructions, figure~\ref{fig:profile} displays a linear profile,
along a radial distance of $500 \Mpch$, through the density field
reconstructions. All profiles are taken along the Y-axis, at
$(X,Z)=(300,300)\Mpch$. Also cf. Fig.~\ref{fig:full_comparison} and
Fig.~\ref{fig:vollim_comparison}.

The panels, from top to bottom, concern the DTFE, NNFE and Kriging
reconstructions.  The black lines are linear profiles through the full
sample reconstructions, the superimposed red profiles concern the
volume-limited mock sample reconstructions.  We also show the linear
profiles through these density field reconstructions, Gaussian
filtered on a scale of $R_f=10\Mpch$. The gray lines are the linear
profiles through the filtered full sample density field, the orange
lines those through the filtered volume-limited sample density field.

The comparison between the linear profiles through the NNFE and Kriging 
reconstructions on the one hand, and the DTFE reconstruction on the other 
hand, leads to the following observation:
\begin{enumerate}
\item[$\bullet$] Underdense regions of NNFE and Kriging are less noisy. 
A nice example of this is the underdense region at $X=200\Mpch$. 
\item[$\bullet$] Maxima tend to be wider in the higher order schemes.
\item[$\bullet$] At larger radial distance, the topology of the unfiltered 
reconstructions is much smoother.
\end{enumerate}

\begin{figure*}
  \centering
  \includegraphics[width=0.32\textwidth]{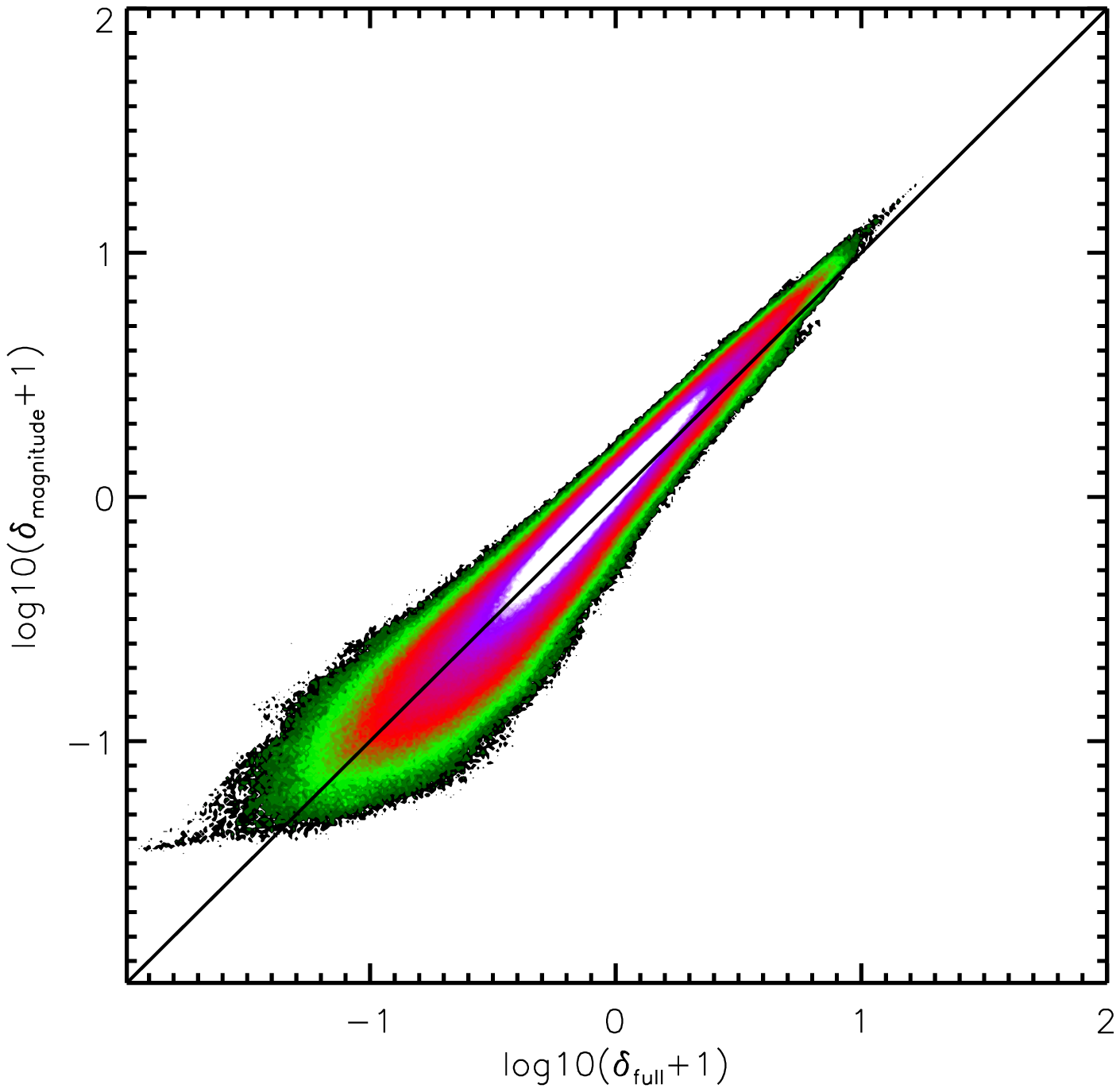}
  \includegraphics[width=0.32\textwidth]{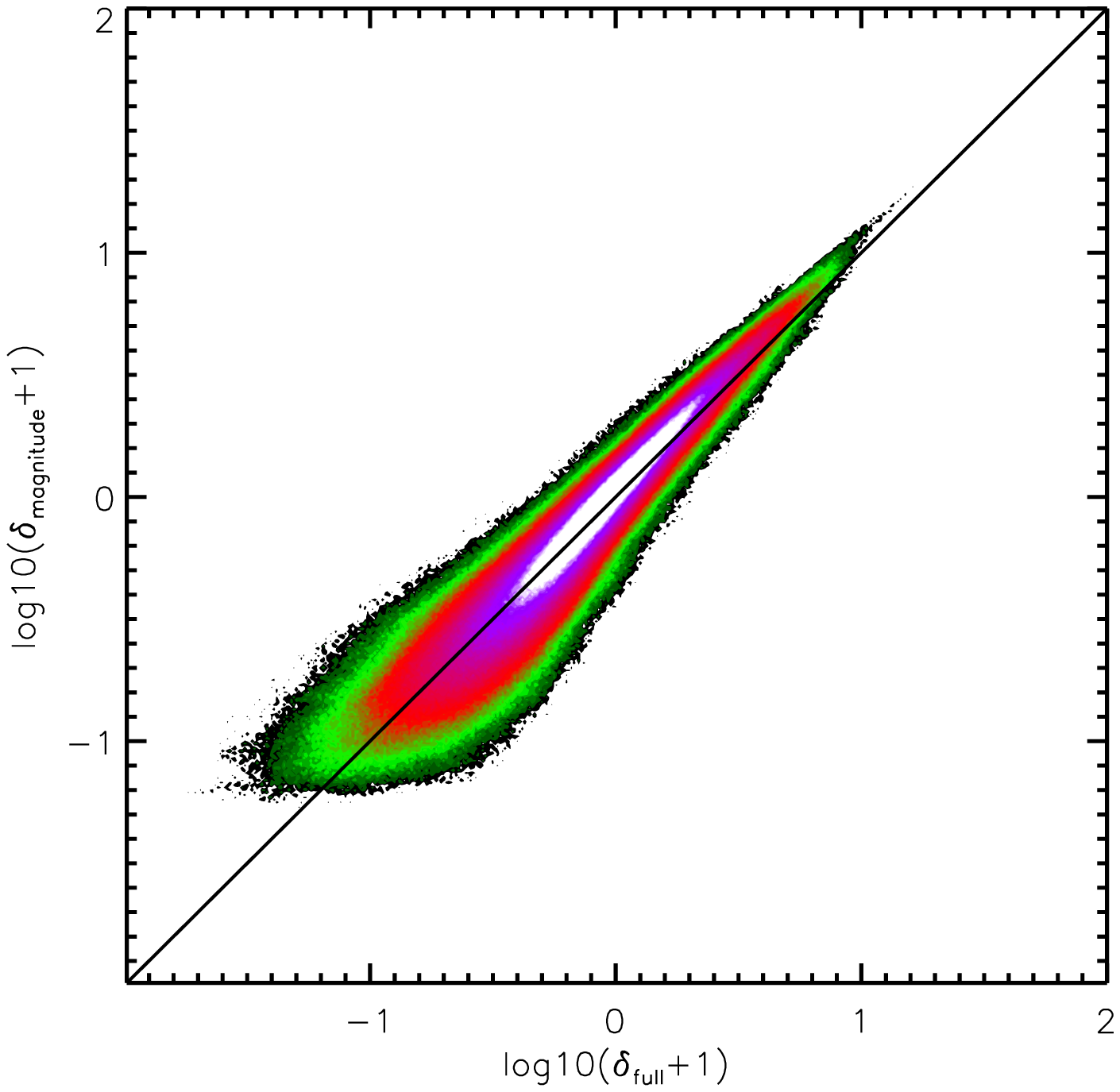}
  \includegraphics[width=0.32\textwidth]{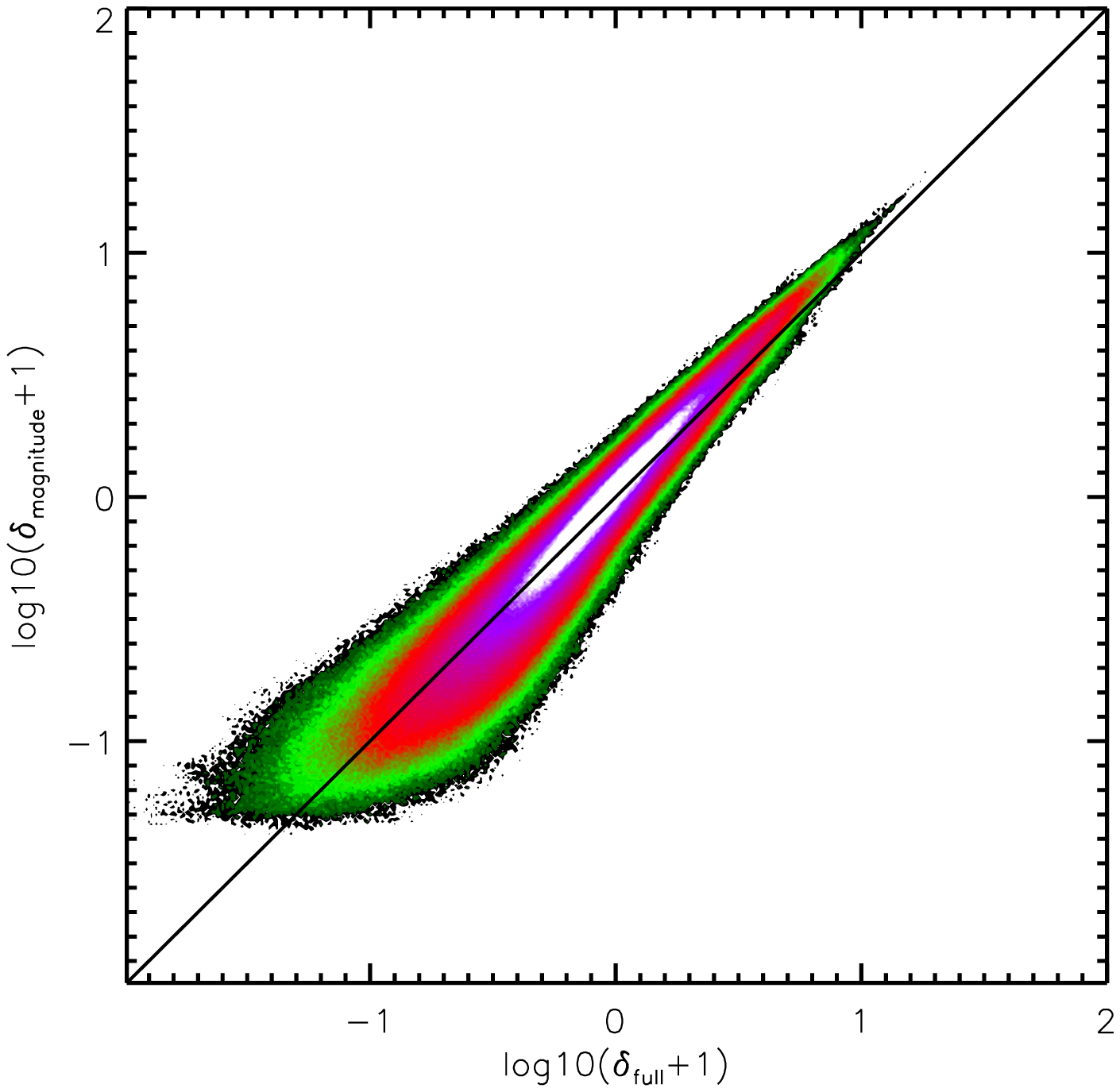}
  \caption{Correlation diagrams for DTFE, NNFE and Natural Lognormal Kriging density 
  field reconstructions. Plotted are the value of the density reconstruction from the full 
  galaxy catalogue, $\delta_{full}+1$ (abscissa), against the density reconstruction from the 
  magnitude-limited mock catalogues, $\delta_{mock}+1$ (ordinate). The density values are 
  those for the field filtered on a scale $R_f=3.0\Mpch$. The colours indicate 
  the number of voxels that occupy the corresponding position in the correlation 
  diagram, with white indicating the density pairs with highest occurrence and subsequent 
  contour levels at logarithmic spacings of $\sim 2.5$ (see text). 
  {\it Left}: DTFE reconstruction; {\it Centre}: NNFE reconstruction; {\it Right}: 
  Natural Lognormal Kriging.}
  \label{fig:den_den_3}
\end{figure*}

\medskip
In the cosmological context, the density is approximately lognormally distributed 
\citep[see][]{Coles91,Kofman94,Sheth95,Kayo01,Kitaura10}, which impels us to work 
the log of the inferred density field.  
Our adaption of the Kriging method to tesselations of lognormally distributed discrete 
point data is presented in detail in a separate paper \citep{Jones11}, where we also discuss 
the relationship of this procedure with the Constrained Random Field formalism developed by 
\cite{Bertschinger87} \citep[also see][]{Hoffman91,Rybicki92, Sheth95,Weyedb96}. We adopt the 
nomenclature used in the geostatistical literature.

\section{Quantitative Density Field analysis}
\label{sec:quan}
For the quantitative comparison and error evaluation of the three density field 
reconstruction methods, we are particularly interested in the ability to 
recover the underlying density field from a magnitude-limited or volume-limited 
survey. 

In this section we compare the DTFE, NNFE and Natural Lognormal Kriging 
reconstructions on the basis of magnitude-limited or volume-limited mock 
samples with that of the corresponding density fields determined from the 
full galaxy samples. The latter are galaxy redshift space samples that we 
would hypothetically obtain if we were able to observe all galaxies in the 
Millennium mock sample. The magnitude- and volume-limited samples are 
obtained from this Millennium mock sample by imposing the observational 
specifications of the SDSS survey (see sect.~\ref{sec:sdss}). An impression of 
the differences in structural resolution is provided by the visual 
comparison of the NNFE full, magnitude-limited and volume-limited sample 
reconstructions at the beginning of this study, in Fig.~\ref{fig:mag_vollim}. 

We will restrict the error analysis to the redshift space density maps, and 
not assess the errors introduced by peculiar velocity distortions. These are 
investigated in detail in the follow-up paper. 

Most of our analysis focuses on the quality of the density field reconstructions 
obtained from magnitude-limited samples, unless specifically stated otherwise. 

\bigskip
\bigskip
\subsection{Magnitude-limited survey reconstructions:\\ \ \ \ \ \ \ Correlation Diagrams}
\label{sec:corrdiagram}
The first comparison between the magnitude-limited sample density field reconstruction 
and the full sample density field concerns a purely local point-to-point comparison. This test 
involves an inspection of the correlation diagrams of the density field value 
$\delta_{full}({\bf r})$ of the full sample density field at location versus the corresponding 
density value $\delta_{mock}({\bf r})$. 

Since this is a strictly local comparison, and does not distinguish between 
systematic nonlocal offsets or random field fluctuations, we try to get an 
impression of environmental effects by simply studying a sequence of 
filtered density fields. Of each density field we study four Gaussian filtered 
versions, at filter radii of $R_f=0.0$, 1.0, 6.0 and 10.0$\Mpch$. These scales 
represent the transition from the non-linearity (1$\Mpch$) to quasi-linear and 
linear scales at 10.0$\Mpch$. 

If the survey-based reconstruction were perfect, the correlation diagram should 
be a perfect one-to-one line. The correlation diagrams show the level of 
scatter, and reveal whether the reconstruction errors are dependent on density 
and as well expose the presence of systematic offsets. 

\subsubsection{DTFE correlation diagrams}
Fig.~\ref{fig:den_den} presents the correlation diagrams for the DTFE reconstructed 
fields at four filter scales, $R_f=0.0, 1.0, 6.0, 10.0 \Mpch$. 

Instead of a pure scatter diagram, we depict the density of the pairs 
$(\delta_{full},\delta_{mock})$ by means of contour levels. The contour 
levels in Fig.~\ref{fig:den_den} depict the number density of pairs 
in logarithmic bins of size $\Delta \log(1+\delta_{full})\times\Delta \log(1+\delta_{full})=(0.02\times 0.02)$. 
The levels run from one pixel per bin (black) to the maximum number density 
(white), in logarithmic steps of $\approx 2.5$. In each frame the black diagonal line 
indicates the exact one-to-one relation between full sample density field and 
the density field following from the magnitude limited sample. 

For all filter radii, we find that over two orders of magnitude the correlation 
diagrams centre around a strict linear one-to-one relation. For the smaller 
filter radii ($R_f=0.0\Mpch$ and $R_f=1.0\Mpch$) the relation appears to be 
slightly offset. This is a reflection of the necessarily small volumes probed by the 
galaxy survey at distances close to the observer, introducing cosmic variance 
effects through the locally estimated density normalization (see 
section~\ref{sec:denval}).

Evidently, the scatter around the linear 1-1 relation decreases as the 
filter radius increases. For both $R_f=0.0\Mpch$ and $R_f=1.0\Mpch$, we 
find a more substantial scatter in the low density regions compared to 
the higher density areas. The scatter over the full density range turns into 
a more and more uniform level as we go to larger filter radii.  Indeed, for 
$R_f > 3.0\Mpch$ and $0.1 < \delta < 1$ the agreement is very good.

A related additional trend is that from a slight upturn at low density values for $R_f=1.0\Mpch$ towards 
a slight downturn at larger filter radii. In other words, at larger filter radii 
the reconstructions seem to have a systematic bias towards more underdense 
values. 

However, we need to be careful in drawing general conclusions with respect to the low 
and high density extremes. In particular, for the large filter radii, these tend to be 
dominated by only a few rare objects. Because the offsets are relatively minor, we 
do not consider it a serious problem. 

\subsubsection{DTFE, NNFE and Kriging correlation diagrams}
To compare the performance of the three methods, Fig.~\ref{fig:den_den_3} 
shows the correlation diagrams for the density field reconstructions at a 
filter scale of $R_f = 3.0\Mpch$. Each of the reconstructions shows an 
almost perfect one-to-one relation: all three methods yield unbiased 
reconstructions. 

We cannot detect any large differences between the methods. This suggests 
that the remaining deviations of the reconstructed density fields are due 
to the initial DTFE density estimate.   Again, for $R_f > 3.0\Mpch$ and $0.1 < \delta < 1$ the agreement is very good.

Alternative density estimators might lead to further improvements. 

\subsection{Magnitude-limited survey reconstructions: \\ \ \ \ \ \ \ \ Intrinsic Smoothing Scale}
A characteristic of the magnitude-limited survey is the change of
intrinsic spatial resolution as we proceed out to larger distances and
the galaxy sample includes only the most luminous objects. While one
may correct the density estimates for the accompanying dilution (see
sec.~\ref{sec:densdtfe}), the loss of small scale resolution and the
accompanying geometric resolution is impossible to correct. This
affects any study of filaments and walls on the basis of the
reconstructed density field maps.

The correction for this dilution effect is complicated by another
effect, the intrinsic density-dependent resolution of the sampled
galaxy surveys.  Higher density regions are sampled by more galaxies,
and are therefore more resolved than the low density void regions.

\begin{figure}
  \includegraphics[width=0.48\textwidth]{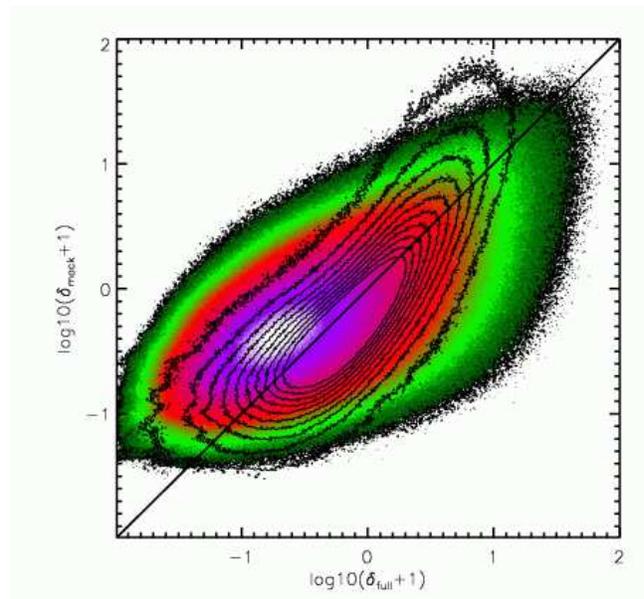}
  \caption{Intrinsic Smoothing Scale and Density. Black contours: the 
   correlation diagram between the $R_f=3.0\Mpch$ filtered DTFE density field 
   of the full galaxy sample (abscissa) and the unsmoothed DTFE density field reconstruction 
   on the basis of the magnitude-limited survey (ordinate). Superimposed on the 
   correlation diagram between the unfiltered full sample density field (abscissa) and 
   the unsmoothed magnitude-limited mock sample density field (ordinate). The colours 
   indicate the number of voxels that occupy the corresponding position in the correlation 
   diagram, with white indicating the density pairs with highest occurrence and subsequent 
   contour levels at logarithmic spacings of $\sim 2.5$.}
  \label{fig:den_den_banana}
\end{figure}

We may evaluate the intrinsic resolution of the density maps in terms
of the intrinsic smoothing scale $R_{int}$. Effectively, it is related
to the characteristic galaxy separation at a given redshift and
position. It is rather straightforward to incorporate the effect of
the increasing dilution as a function of redshift $z$, on the basis of
the radial selection function $\psi(z)$ (see
equation~\ref{eq:psiz}). However, the considerable density contrasts in
the inhomogeneous matter distribution render if far from trivial to
find a correct expression for $R_{int}(z)$. The mean galaxy separation
is biased to high density regions and seriously underestimates the
correct value $R_{int}(z)$: the intrinsic smoothing scale is smaller
than average in higher density areas, and larger in low density
regions.

\subsubsection{Intrinsic Smoothing Scale: Correlation Diagram}
To appreciate the effect and the dependence of intrinsic smoothing
length on density, we evaluate the correlation between the filtered
(DTFE) density field of the full galaxy sample and the DTFE density
field reconstruction on the basis of the magnitude-limited sample.

In particular, we are interested in the question in how far the {\it
  filtered} density field reconstructions are representative. We may
assume they are as long as the corresponding filter scale $R_f >
R_{int}(z)$. By evaluating the filtered density fields we may obtain
an idea of the intrinsic smoothing scale on the basis of the density
field error analysis: a sudden rise of error would indicate that
$R_{int}(z)>R_f$.

By means of the black contours in Fig.~\ref{fig:den_den_banana} we
show the correlation diagram of the full sample density field filtered
at $R_f=3.0\Mpch$ versus the unfiltered density field on the basis of
the magnitude-limited galaxy sample. It is superimposed on the
(colour) correlation diagram between the unfiltered full sample
density field and the unfiltered density field reconstruction from the
magnitude-limited mock sample. While the latter has substantial
scatter over the full density range, the correlation between the
filtered field and the magnitude-limited field appears to be much
tighter. It confirms the impression that reconstructed fields resemble
density fields that are filtered on a particular scale. The
reconstructed density maps in Fig.~\ref{fig:full_comparison} and
Fig.~\ref{fig:vollim_comparison} form a telling illustration of this
observation.

The most outstanding feature of the filtered field correlation diagram
is the curved nature of the contours, quite different from a regular
linear relation. In the higher density regions we find that the
densities in the magnitude-limited sample reconstruction are
systematically biased to higher values. In the low to moderate density
areas the trend is reverse: the densities of the raw survey density
field tend to be somewhat lower than in the filtered field. This is a 
direct illustration of the density dependent nature of the intrinsic
smoothing length $R_{int}(z)$. Because the effective smoothing length is 
small in high density regions the low density regions get relatively 
over-smoothed  and so the density is relatively over-estimated. More 
formally, it is an expression of the greater information content in the 
higher density regions of galaxy redshift surveys.

\subsubsection{Filtered density field reconstructions}
The filtering of a raw reconstructed density field will smooth out the
high density values.  The corresponding mass is redistributed to lower
density regions. We argue that this can only yield correct density
distributions if the nonlinear features in the mass distribution were
reproduced at the correct position and with the correct amplitude in
the raw sample density field reconstruction. Information from high
density features and nonlinear objects is crucial for obtaining the
correct large scale density field. 

\subsection{Radial Error Analysis}
\label{sec:denval}
To quantify the reconstruction errors, we compare the local density 
values $\widehat{f(\bmath{r})}_{mock}$ of the magnitude-limited survey 
reconstruction with that of the density $\widehat{f(\bmath{r})}$ of the 
full sample. We evaluate the absolute error $\epsilon_1({\bmath{r}})$, 
\begin{equation}
  \epsilon_1 (\bmath{r})\,=\,\left|\widehat{f(\bmath{r})}-\widehat{f(\bmath{r})}_{mock}\right|\,,
\label{eq:epsilon1}
\end{equation}
as well as the relative error $\epsilon_2({\bmath{r}})$,
\begin{equation}
  \epsilon_2 (\bmath{r})\,=\,\frac{\left|\widehat{f(\bmath{r})}-\widehat{f(\bmath{r})}_{mock}\right|}{\widehat{f(\bmath{r})}}\,.\\
  \label{eq:epsilon2}
\end{equation}

\subsubsection{Error Profiles}
\begin{figure}
  \centering
  \includegraphics[width=0.48\textwidth]{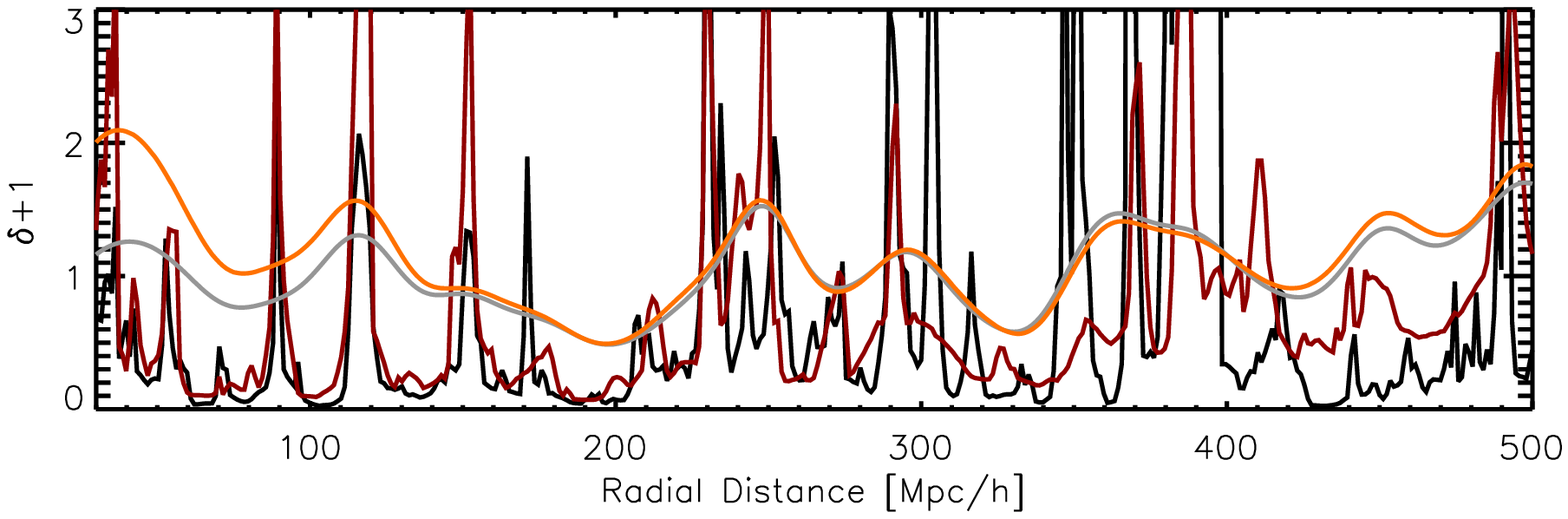}
  \includegraphics[width=0.48\textwidth]{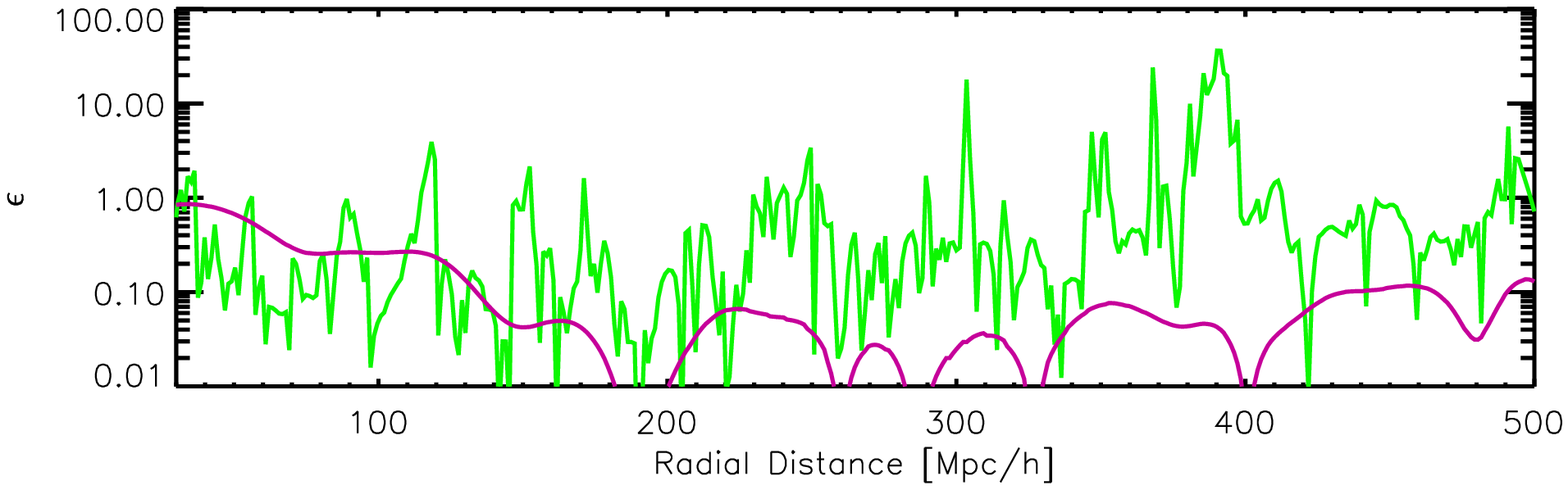}
  \includegraphics[width=0.48\textwidth]{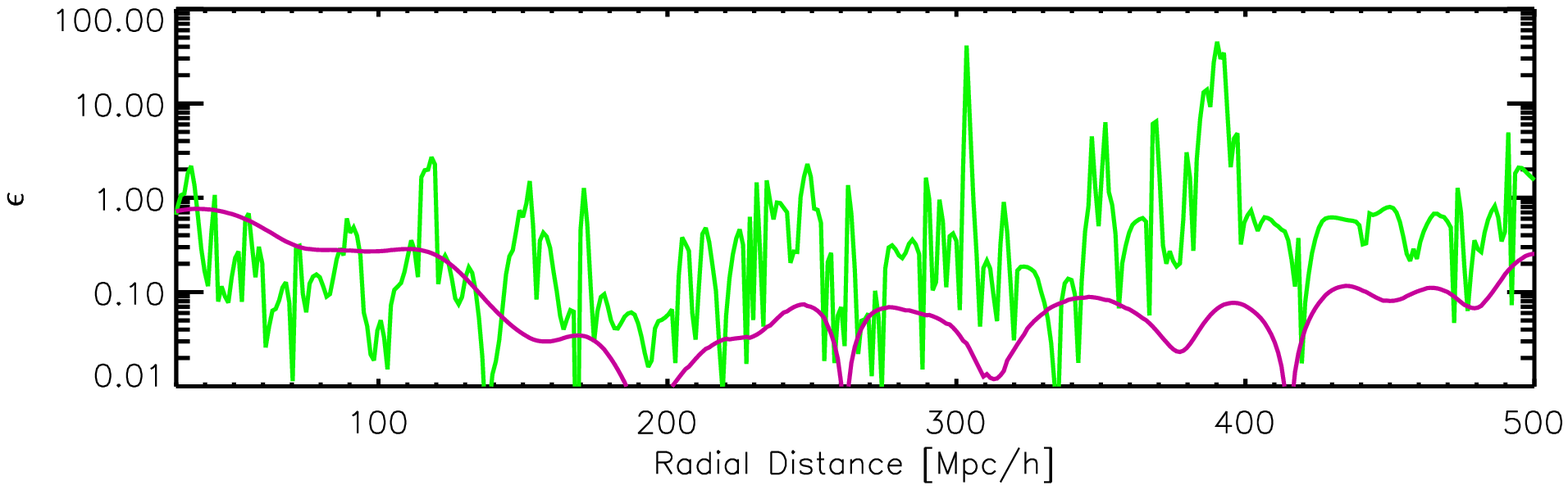}
  \includegraphics[width=0.48\textwidth]{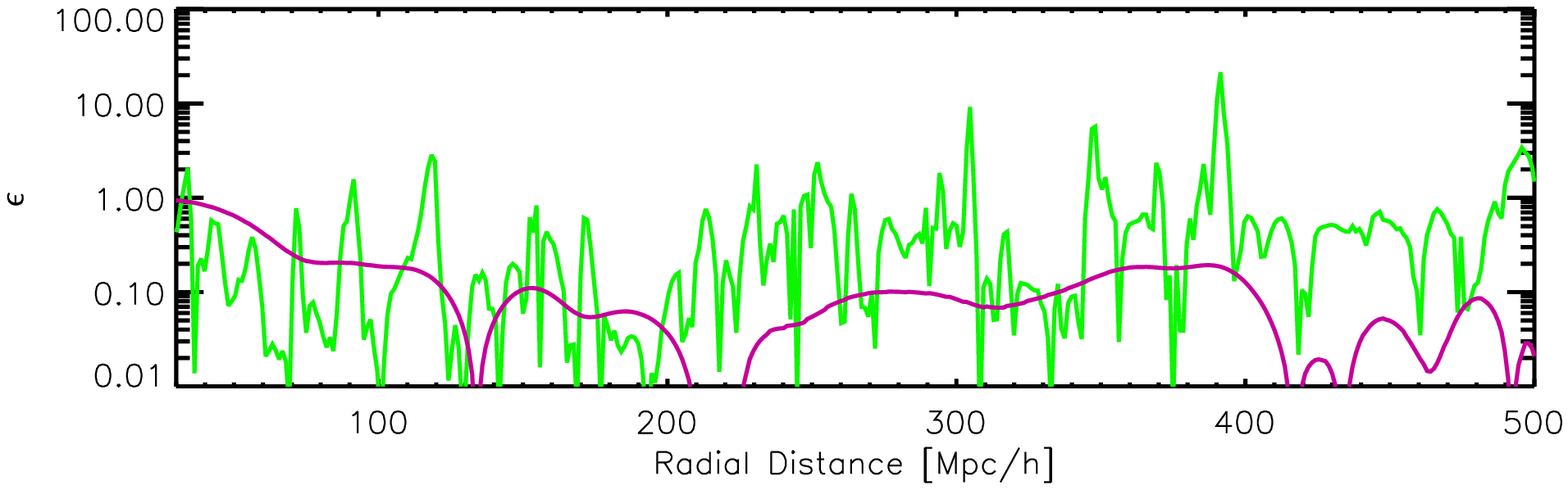}
  \caption{Linear Error profiles. The absolute error $\epsilon_1$ (equation~\ref{eq:epsilon1}) 
  as a function of distance, along the same axis as the density profiles in 
  Fig.~\ref{fig:profile}. {\it Upper panel}: the density profile through the 
  DTFE density field (see Fig.~\ref{fig:profile} for description). The subsequent 
  frames are the error profiles for the three reconstruction methods. The green line 
  represents the error of the unsmoothed density field, the purple line that of 
  the 10.0$\Mpch$ filtered field. {\it 2nd frame from top:} DTFE density field; 
  {\it 3rd frame from top:} NNFE density field; {\it bottom:} Natural Lognormal 
  Kriging density field.}
  \label{fig:eprofile}
\end{figure}

In figure~\ref{fig:eprofile} we plot the absolute errors
$\epsilon({\bmath{r}})$ along the same radial direction as in
Fig.~\ref{fig:profile}. The corresponding DTFE density profile is
plotted in the top panel, while the subsequent three panels depict the
$\epsilon_1$ profiles for the DTFE, NNFE and Natural Lognormal Kriging
density fields.

The green lines are the error profiles of the unfiltered density field (black line in 
the upper panel), the purple lines are the error profiles of the 10$\Mpch$ smoothed field 
(gray line in the upper panel). We find that the errors of all three reconstruction 
techniques are fairly similar, with a slightly increasing trend with distance for the 
unfiltered density field. At a distance of approximately 100 $\Mpch$ it is of the
order of unity. Beyond this distance the errors are characterised by
wide peaks that are mainly due to the magnitude-limited surveys undersampling of the 
density field at large distances.

The errors in the 10$\Mpch$ filtered field remain small over the whole reach 
of the linear profile, averaging an error level of around 10 percent. At 
intermediate distances, $200\Mpch<R<400\Mpch$, the DTFE errors are somewhat 
lower than those of the other methods. For all three methods, the largest errors are found at 
close distances. This is attributed to the survey mask, which is relatively thin in our immediate 
cosmic environment, ie. at distances $R<100\Mpch$. Note that the filtered field errors at large 
distances appear to converge to the error profile of the unfiltered field reconstruction. It is a 
reflection of the fact that the intrinsic smoothing length becomes comparable to the filter radius.

\subsubsection{Radially averaged error profiles}
By averaging the errors of the magnitude-limited survey density field 
reconstructions over radial shells we obtain an idea of 
error trends as function of distance. To minimize edge effects we 
exclude locations within 15 voxels from the edge of the survey volume. 

\begin{figure}
  \centering
  \includegraphics[width=0.48\textwidth]{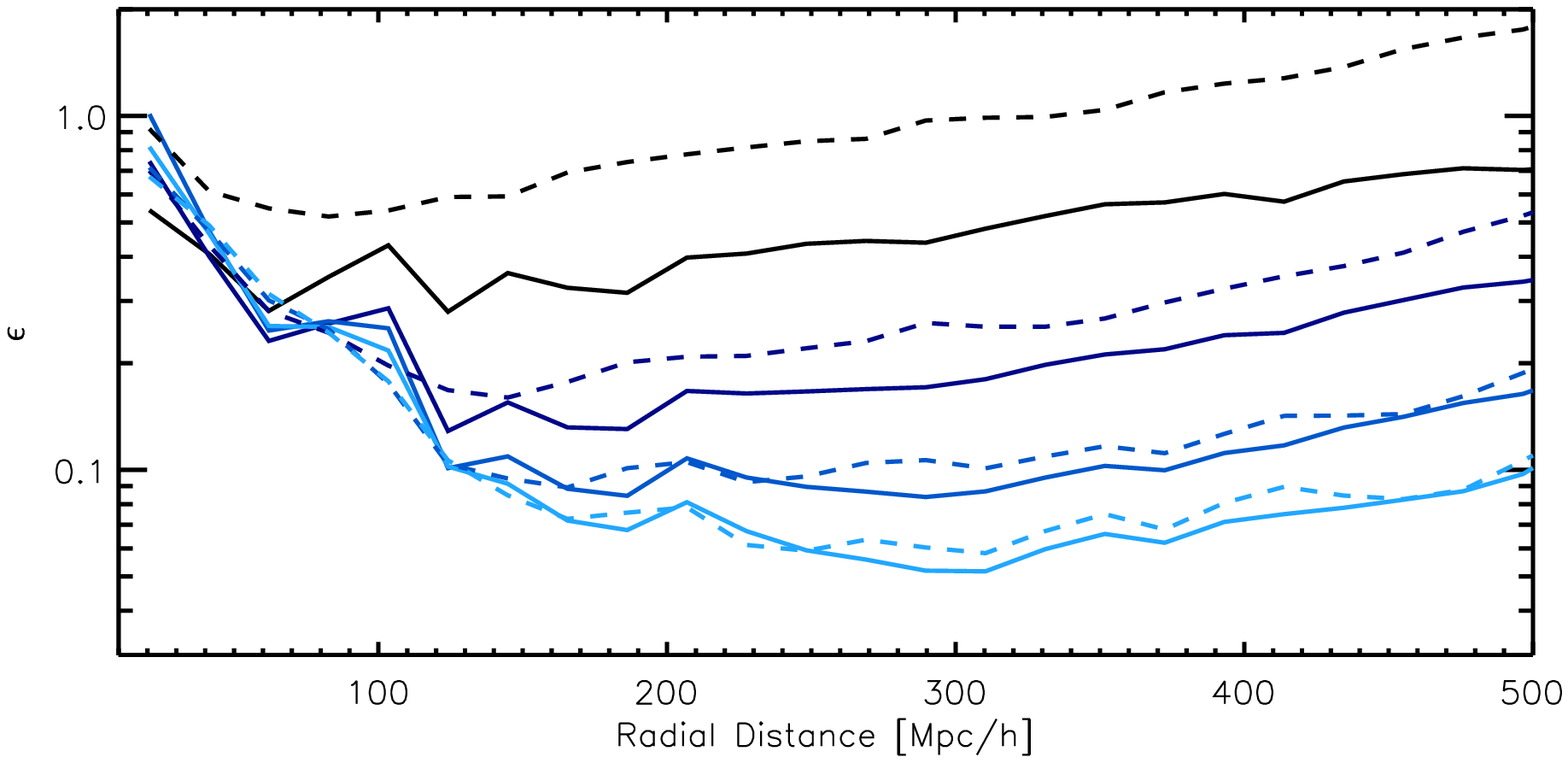}
  \includegraphics[width=0.48\textwidth]{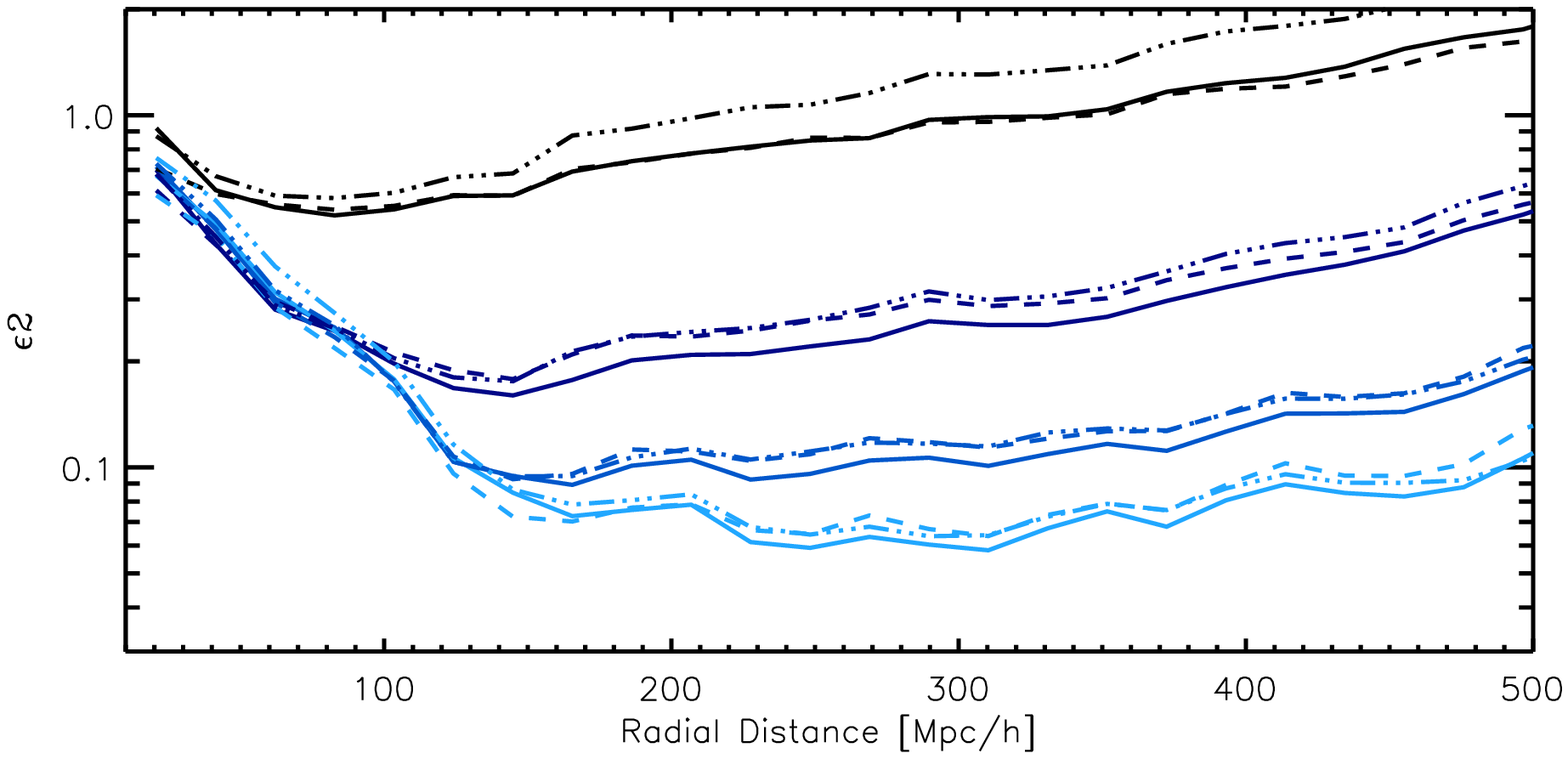}
\caption{Radially averaged error profiles. {\it Top:} the absolute error profile $\epsilon_1(r)$ (solid line) 
   and relative error $\epsilon_2(r)$ (dashed line) for the magnitude-limited 
   survey DTFE density field reconstruction. Going from dark to light blue, the profiles 
   represent the error profiles through a sequence of filtered density fields: 
   $R_f=1.0\Mpch$ (black), $R_f=3.0\Mpch$, $R_f=6.0\Mpch$ and $R_f=10.0\Mpch$ (light blue). 
   {\it Bottom}: The radially averaged relative error profile $\epsilon_2(r)$, for 
   the DTFE density field reconstruction (solid lines), the NNFE density field reconstruction 
   (dashed lines) and the Natural Lognormal Kriging reconstruction (dot-dashed lines). The 
   colours correspond to the same Gaussian filtered fields as specified for the top panel.}
 \label{fig:averr}
\end{figure}

The radially averaged error profiles for the DTFE density field reconstruction 
are shown in the top panel in Fig.~\ref{fig:averr}. It concerns the 
absolute error $\epsilon_1$ (solid line) and relative error $\epsilon_2$ 
(dashed line) for four different filtered DTFE density fields. These are Gaussian 
filtered fields at filter scales of $R_f=1.0$, 3.0, 6.0 and $10.0\Mpch$. 

\begin{figure*}
  \centering
  \includegraphics[width=0.48\textwidth]{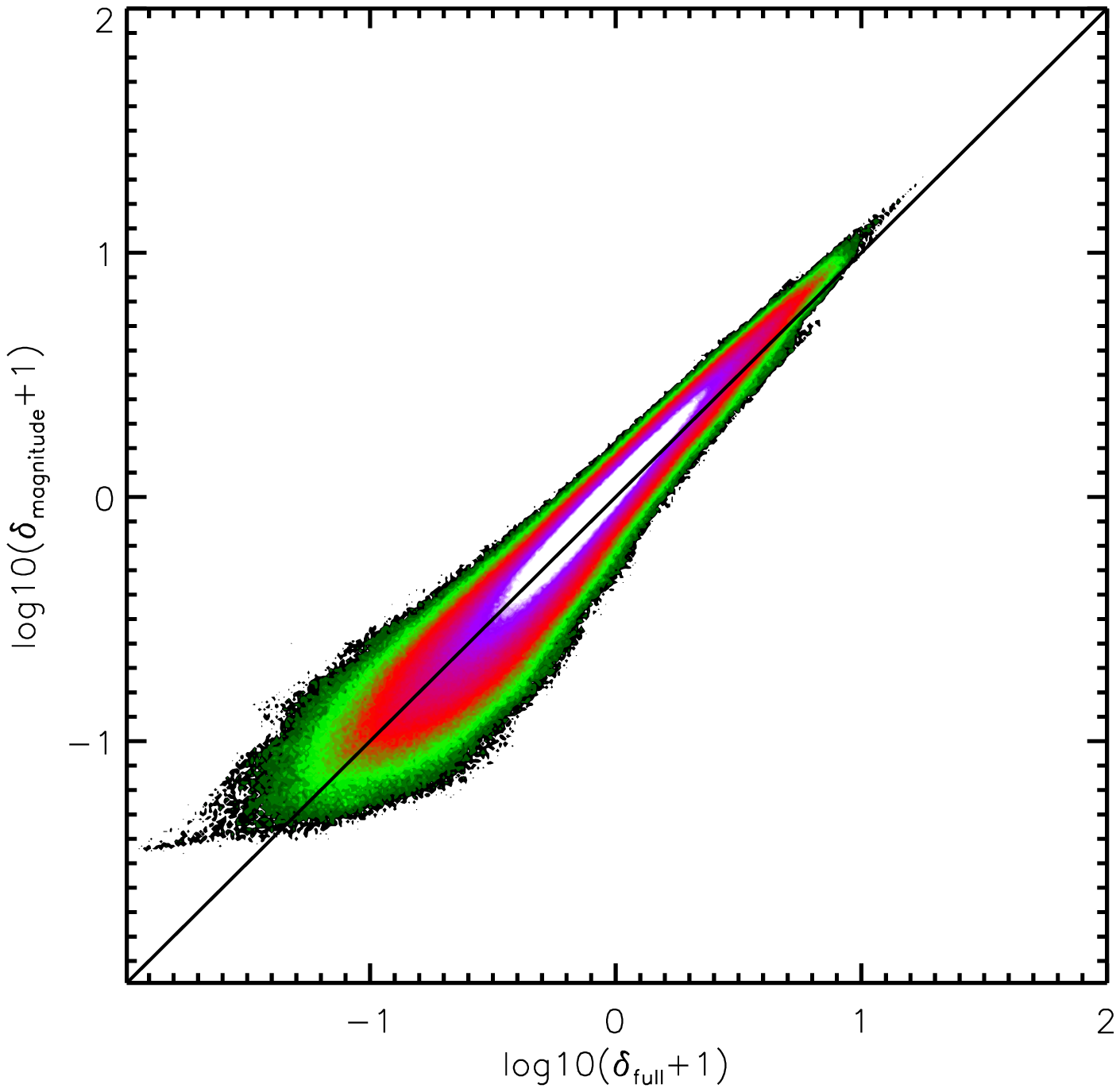}
  \includegraphics[width=0.48\textwidth]{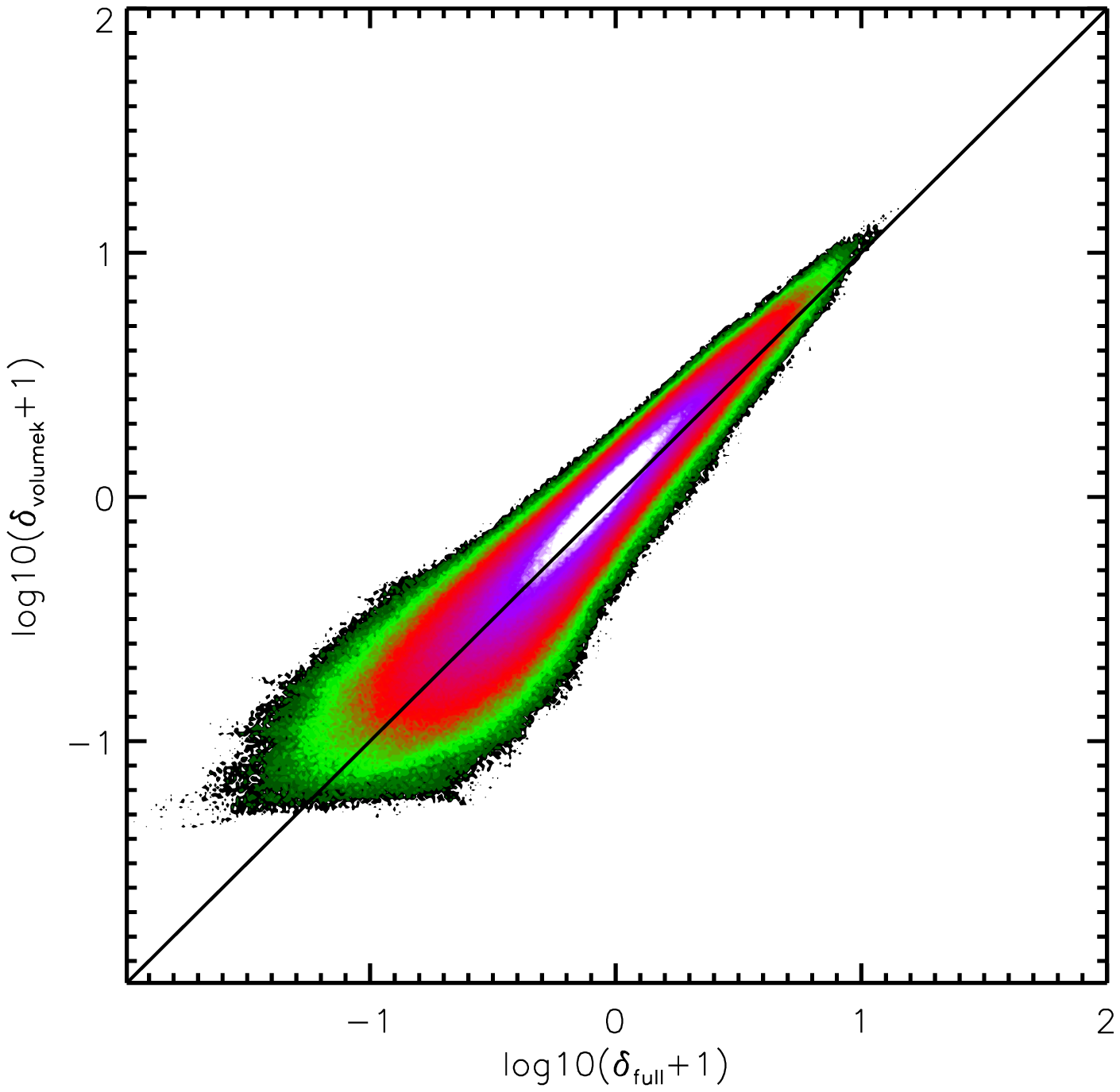}
  \caption{Density field Correlation Diagrams.{\it Left}: Correlation 
  diagram between full sample DTFE density field reconstruction (abscissa) and 
  the DTFE density field reconstruction on the basis of the magnitude-limited sample 
  (ordinate).  {\it Right}: Correlation diagram between full sample DTFE density field 
  reconstruction (abscissa) and the DTFE density field reconstruction on the basis of the 
  volume-limited sample (ordinate). All fields are Gaussian filtered on a scale 
  $R_f=3.0\Mpch$.}
  \label{fig:den_den_vol}
\end{figure*}

At nearby distances $R<100\Mpch$ the average error trend is that 
of decreasing absolute and relative errors. This is a reflection of the 
small and unrepresentative survey volumes involved. The fact that this 
effect appears to be most prominent for the largest filter scale, $R_f=10.0\Mpch$, 
is indicative. At this scale the reconstruction involves only 
a few independent wavevectors that constitute the resulting density 
field. Errors are enhanced by the way in which the weight factors 
$w(z)$ (equation~\ref{eq:surselect}) are determined. We do this by fitting 
the selection function on the basis of the data (see sect.~\ref{sec:weights}). 
At close distances this fit is heavily influenced by the presence or absence 
of large superstructures, which easily evokes systematic offsets in the 
local density estimate. It would probably be better to deal with such 
systematics on the basis of volume limited samples at close distances 
$R < 150\Mpch$.

\subsubsection{DTFE, NNFE and Kriging error profiles}
The bottom panel of Fig.~\ref{fig:averr} compares the relative error trend 
for the three different reconstruction formalisms. The radial error profiles 
are determined are shown for the four filter scales $R_f=1.0, .0, 
6.0$ and $10.0\Mpch$. The DTFE error profiles are marked by the solid 
lines, the NNFE ones by the dashed lines and the Lognormal Kriging ones 
by the dot-dashed lines. 

For all three methods the errors at the smallest filter scale, $R_f=1.0\Mpch$, 
are substantial, hardly ever better than 50\%. The errors decrease 
rapidly with filter scale, such that at scales $R_f>6\Mpch$ the density
field can be reconstructed with reasonable accuracy. Beyond a distance 
of 100$\Mpch$ the relative errors throughout the whole survey volume do 
not exceed the 10$\%$ level. 

Overall, we find strikingly small differences between the three 
methods. None performs distinctly better than any of the others. 
On a more detailed level, we may observe two differences. Firstly, 
we see that the Kriging errors at small scales are relatively high. 
Secondly, DTFE appears to perform somewhat better at nearly all scales, 
except the smallest scale $R_f=1.0\Mpch$. In terms of density 
errors, the relatively simple and direct DTFE method seems to 
work best. 

\subsection{Volume limited survey density field reconstructions}
\label{sec:magvol}
A volume limited survey circumvents the resolution complications encountered 
in magnitude limited surveys. Instead of having to correct for the steadily 
decreasing resolution at larger distances, volume limited surveys have the 
advantage of a uniform sample resolution. This renders the geometric and 
topological analysis of structures considerably more straightforward. 
Moreover, in addition to removing the problem of varying spatial resolution, 
the major advantae of using volume-limited samples is that one is using the 
same type and luminosity of galaxies at different distance. 
The major disadvantage of volume-limited galaxy samples is their relatively 
low resolution, resulting from the necessity to uniformly sample galaxies 
throughout the sample volume. In essence, it involves a trade-off between 
spatial resolution and a sufficiently large and representative sample volume.

To test the performance of density field reconstructions on the basis of 
a volume-limited galaxy redshift survey, we take a volume limited sample 
from the DeLucia mock sample based on the Millennium simulation \cite{DeLucia07}, 
restricted to a volume in between redshifts $z=0.02-0.1$ and containing 
galaxies brighter than $M_r=-20.45$. 

\begin{figure}
  \centering
  \includegraphics[width=0.48\textwidth]{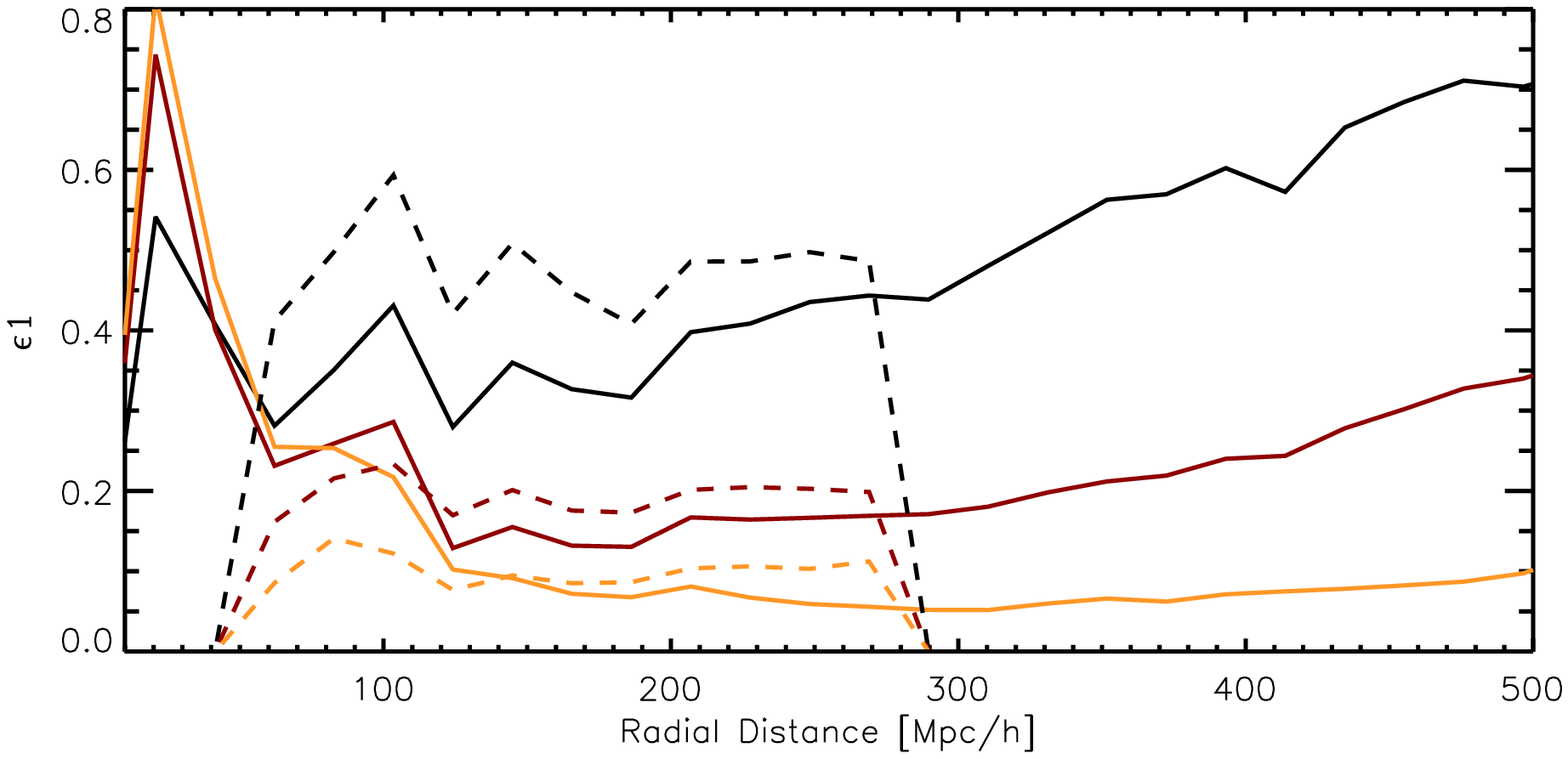}
  \includegraphics[width=0.48\textwidth]{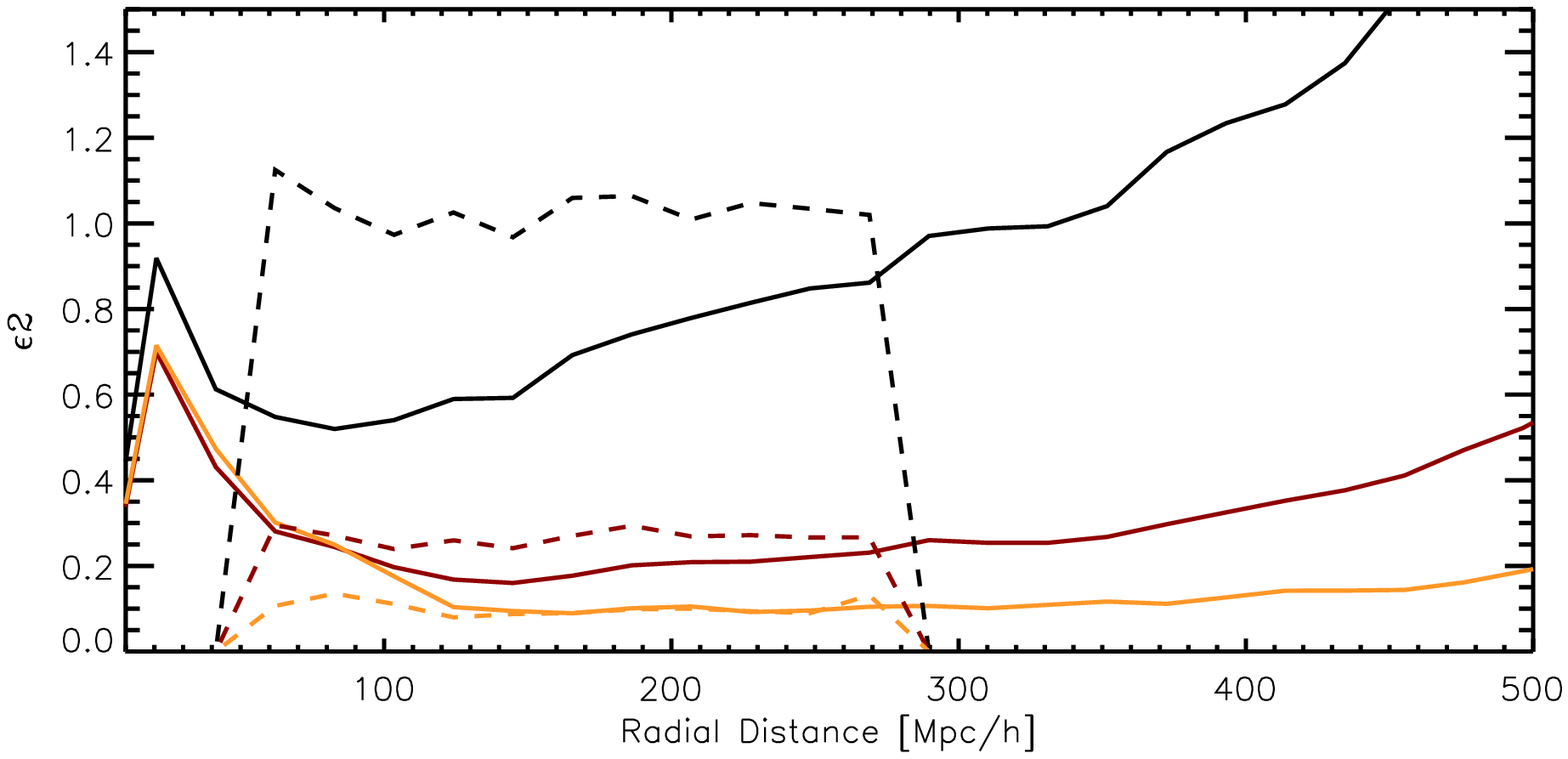}
  \caption{Radially averaged error profiles. {\it Top}: radially averaged profile of 
  absolute error $\epsilon_1(r)$. {\it Bottom}: radially average profile of relative error 
  $\epsilon_2(r)$. Solid lines, both panels: the error of the DTFE density field reconstruction 
  on the basis of the magnitude-limited sample. Dashed lines, both panels: the error of the 
  DTFE density field reconstruction on the basis of the volume-limited sample (within the 
  limits of the volume-limited sample volume). Each panel shows the error profiles for three 
  different filter scales: $R_f=1.0$ (orange), 3.0 (red) and 10.0$\Mpch$ (black).}
  \label{fig:averr_mock}
\end{figure}
\begin{figure}
  \includegraphics[width=0.48\textwidth]{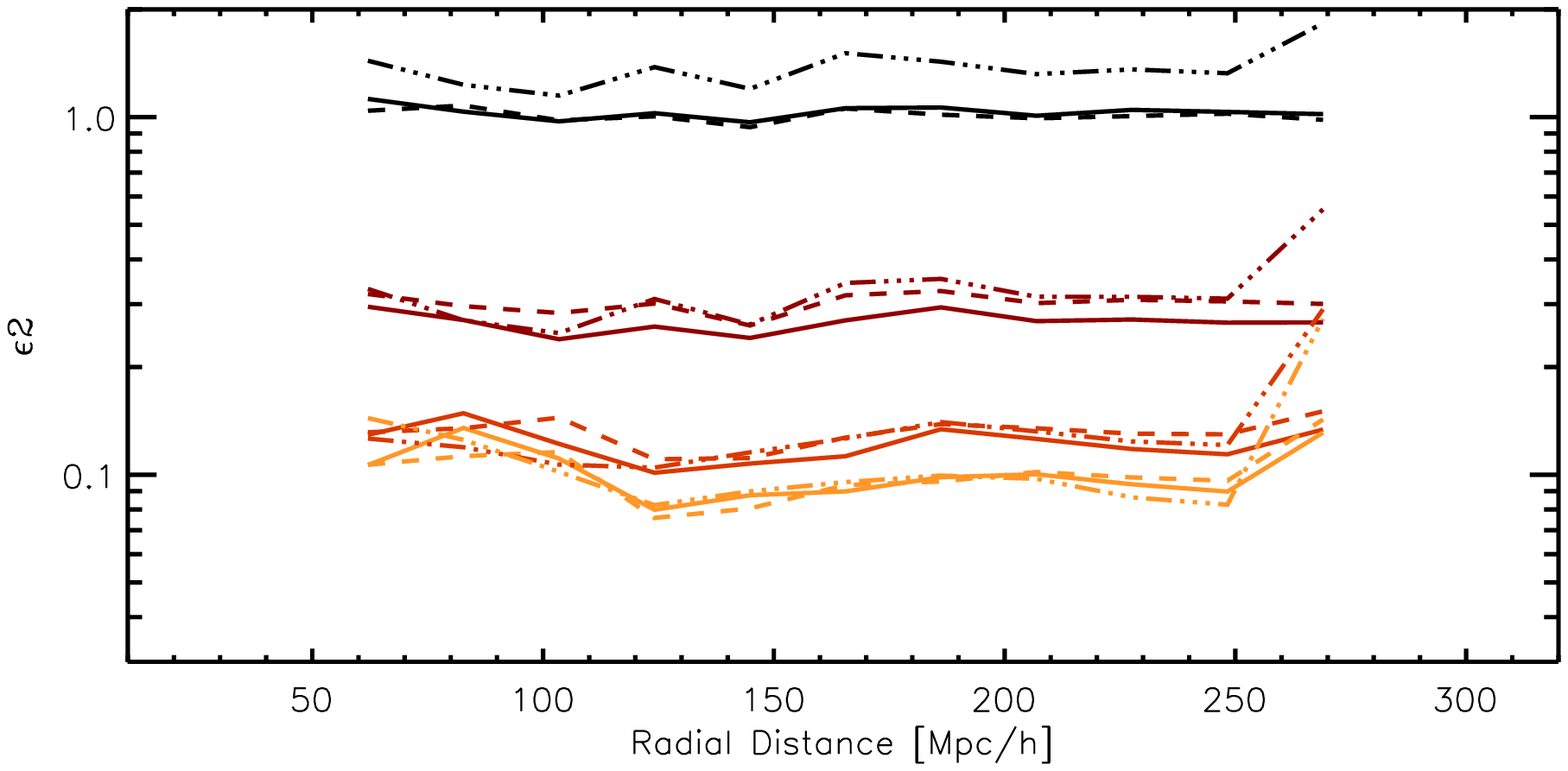}
  \caption{Radially averaged error profiles for DTFE, NNFE and Kriging 
  density field reconstructions on the basis of volume-limited galaxy sample. The relative 
  error $\epsilon_2(r)$ is plotted for three filter scales: $R_f=1.0$ (orange), 3.0 (red)and 
  10.0$\Mpch$ (black). 
  Solid lines: DTFE error profiles. Dashed lines: NNFE error profiles. Dot-dashed lines: 
  Natural Lognormal Kriging profiles.}
  \label{fig:averr_vollim_comp}
\end{figure}

The resulting galaxy sample can be seen in the bottom panel of 
Fig.~\ref{fig:mag_vollim} (along with the corresponding NNFE density field 
contours). With respect to the magnitude-limited sample 
(central panel), let alone the full sample, it clearly lacks spatial 
resolution. This may also be appreciated from the three corresponding 
density field reconstructions in Fig.~\ref{fig:vollim_comparison}.  
The loss of small scale details is obvious: close inspection and 
comparison with the full sample reveals the loss of fine filamentary 
features separating small voids. They either merge into larger 
surrounding overdensities or get lost in enlarged void regions.

\subsubsection{Density Field Correlation Diagrams}
Figure~\ref{fig:den_den_vol} compares the correlation diagrams for the
DTFE density field reconstructions on the basis of magnitude (left) and volume 
limited samples (right). Overall, they occur rather similar, although there 
are some important details in which they differ. One particular one 
concerns the more extended maximum of the correlation diagram in the case of the 
magnitude limited sample density field. In other words, the errors of the 
magnitude limited sample density field are usually smaller than those 
for the volume limited sample density field.

\subsubsection{Radially averaged error profiles}
Figure~\ref{fig:averr_mock} compares the profiles of the radially averaged 
absolute errors $\epsilon_1(r)$ (equation~\ref{eq:epsilon1}, and of the radially averaged 
relative errors $\epsilon_2(r)$ (equation~\ref{eq:epsilon2}) for the DTFE 
density field reconstructions on the basis of the magnitude-limited (solid lines) and 
on the basis of the volume-limited surveys (dashed lines). The error profiles are 
assessed for three Gaussian filter radii, $R_f=1.0$, $3.0$ and $10.0\Mpch$. 

The first observation from Fig.~\ref{fig:averr_mock} is that the errors of the 
volume-limited sample are uniformly distributed throughout the survey volume, 
as might be expected for a statistically uniform sample. We also find that the 
errors of the magnitude-limited sample density field reconstructions are systematically 
lower than those for the volume limited sample. This remains so up to the edge 
of the volume limited sample, at $R \approx 300\Mpch$, where the sampling density of the 
volume limited and magnitude limited sample are equal. 

\subsubsection{DTFE, NNFE and Kriging error profiles}
When comparing the error profiles for the three different density field 
reconstructions we hardly find any significant differences. This is 
confirmed by Fig.~\ref{fig:averr_vollim_comp}, which present the 
radially averaged relative error profiles $\epsilon_2(r)$ for 
the DTFE, NNFE and Kriging reconstructions at three different 
filter radii. 

The uniformity of the errors appears to suggest that 
a major source for the observed errors has to be found in the 
density estimate itself, rather than in the interpolation technique. 
If there are any differences, we may argue that these concern a 
slightly better performance of DTFE.

\section{Topological Analysis}
\label{sec:topo}
To probe the global pattern of the mass distribution we turn to 
the topological structure of the SDSS survey. This is largely 
dependent on the higher order correlations in the density field. 
The error analysis described in the previous subsections would not 
necessarily be able to detect key differences in the large 
scale topology. In this subsection we will seek to evaluate the 
quality of the topological renderings of the density field, where 
we focus on the structure defined by the voids in the cosmic 
desnsit field. 

There are various approaches to studying the topological structure of 
the large scale mass distribution. One option for characterizing the global 
topology of the cosmic matter distribution is in terms of four Minkowski 
functionals \citep{Mecke94,Schmalzing99}. These are solidly based on the 
theory of spatial statistics and also have the great advantage of being known 
analytically in the case of Gaussian random fields. In particular, the 
\textit{genus} of the density field has received substantial attention as a 
strongly discriminating factor between intrinsically different spatial patterns 
\citep{Gott86,Gott89,Park92,Hoyvog02,Gott08,Zhang09}. An attempt to extend the scope of the Minkowski 
functionals towards locally defined topological measures of the density field 
has been developed in the SURFGEN project defined by Sahni and Shandarin and 
their coworkers \citep{Sahni98,Shandarin04}. The main problem for these 
formalisms remains the user-defined, and thus potentially biased, nature of 
the continuous density field inferred from the sample of discrete objects. 

Here we specifically address the topology of the SDSS galaxy distribution 
on the basis of the void population. To this end, we segment the 
galaxy distribution into void patches by means of the Watershed Void 
Finder \citep{Platen07}. We test the watershed segmentation of the 
DTFE density field obtained from the magnitude-limited mock sample 
by comparing it to the watershed segmentation of the full galaxy 
sample density field. The topological errors are quantified according to the
watershed segmentation of both fields.

\begin{figure*}
  \centering
  \includegraphics[width=0.8\textwidth]{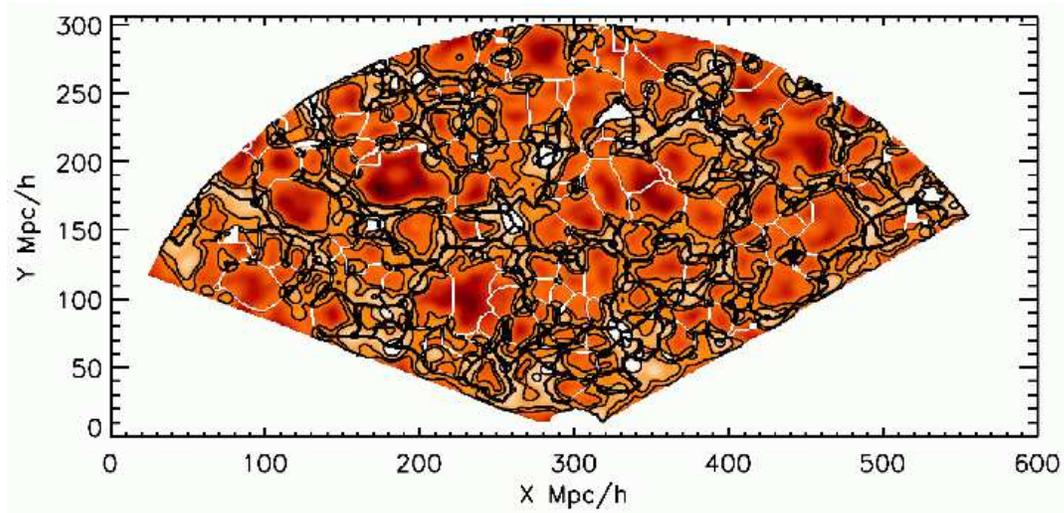}
  \caption{Watershed Segmentation. The WVF watershed boundaries (white - 
   and black edged) inferred from the Millennium magnitude-limited 
   SDSS mock sample. They are superimposed on the density field of the full 
   SDSS mock galaxy sample (coloured), and the contours for the overdense 
   regions in the magnitude-limited survey density field, filtered on a 
   $R_f=2.0 \Mpch$ scale.}
  \label{fig:topo}
\end{figure*}
\begin{figure*}
  \centering
  \includegraphics[width=0.8\textwidth]{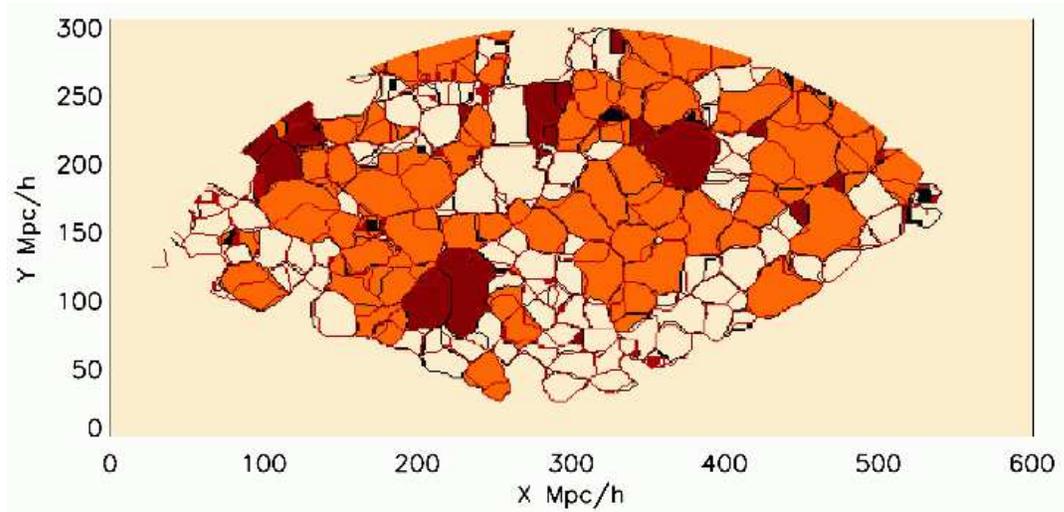}
  \mbox{\hskip 0.65truecm\includegraphics[width=0.7\textwidth]{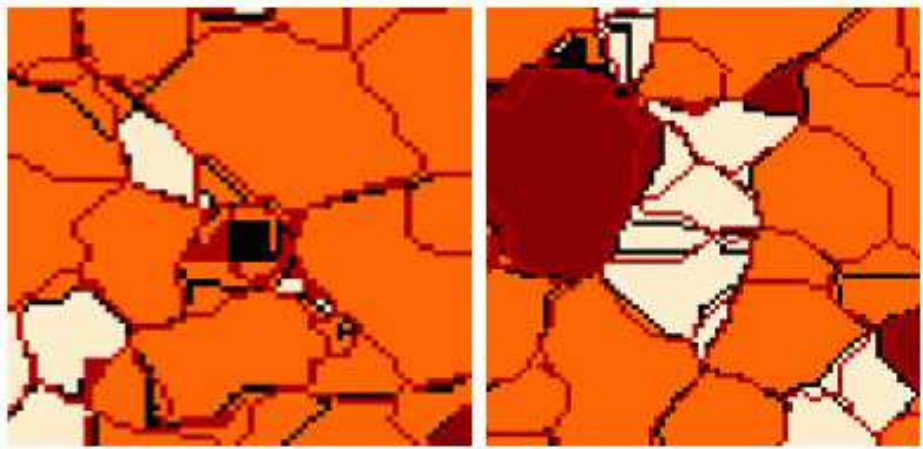}}
  \caption{WVF Segmentation Comparison. The WVF segmentation boundaries of the 
   full SDSS mock galaxy sample are indicated by red, the ones for the magnitude-limited 
   sample by black lines. The colour of the segments indicate the nature of the 
   topological errors. Orange: {\it false mergers}; Red: {\it false splits}. 
   Below: The zoom-ins show two regions with a serious mismatch between the WVF segmentations  
   of full and magnitude-limited sample.}
  \label{fig:topoerr}
\end{figure*}

\subsection{Watershed Void Finder (WVF)}\ \\
The Watershed Void Finder (WVF) is an implementation of the {\it Watershed 
Transform} for segmentation of images of the galaxy and matter distribution 
into distinct regions and objects and the subsequent identification of voids 
\citep{Platen07}. 

The basic idea behind the watershed transform finds its origin in geophysics. It 
delineates the boundaries of the separate domains, the {\it basins}, into which 
yields of e.g. rainfall will collect. The analogy with the cosmological context is 
straightforward: {\it voids} are to be identified with the {\it basins}, while 
the {\it filaments} and {\it walls} of the cosmic web are the ridges separating 
the voids from each other. 

With respect to the other void finders the watershed algorithm has several advantages. 
Because it is identifies a void segment on the basis of the crests in a density field 
surrounding a density minimum it is able to trace the void boundary even though it 
has a distorted and twisted shape. Also, because the contours around well chosen minima 
are by definition closed the transform is not sensitive to local protrusions between 
two adjacent voids. The main advantage of the WVF is that for an ideally smoothed 
density field it is able to find voids in an entirely parameter free fashion. 

The WVF consists of eight steps, which are extensively outlined in \cite{Platen07}. 
For the success of the WVF it is of utmost importance that the density field retains 
its morphological character. To this end, the two essential first steps relate 
directly to DTFE, which guarantees the correct representation of the hierarchical 
nature, the weblike morphology dominated by filaments and walls, and the presence of voids 
\citep{Weyschaap09}. Because in and around low-density void regions the raw density field 
is characterized by a considerable level of noise, a second essential step 
suppresses the noise by an adaptive smoothing algorithm which in a consecutive sequence 
of repetitive steps determines the median of densities within the {\it contiguous Voronoi cell} 
surrounding a point. The determination of the median density of the natural 
neighbours turns out to define a stable and asymptotically converging smooth 
density field fit for a proper watershed segmentation. The subsequent central 
step of the WVF formalism consists of the application of the discrete watershed 
transform on this adaptively filtered density field.

A related tessellation-based method for void identification, ZOBOV \cite{Neyrinck08}, 
does yield similar results as WVF \citep[see][]{Colberg08}. It demonstrates the successful 
application of tessellation-based techniques to identify structures within the cosmic 
matter distribution. In addition to the WVF and ZOBOV there is an array of void 
identification procedures \citep[see e.g.][]{KauFair91,ElAdPir97,Hoyvog02,ArbMul02,PliBas02,Pati06,
ColShe05,ShaFel06,Colberg08}. The ``voidfinder'' algorithm of \cite{ElAdPir97} has been at the basis 
of most voidfinding methods. However, this succesful approach is not able to analyze
complex spatial configurations in which voids may have arbitrary shapes and contain a range 
and variety of substructures, which lies at the heart of our analysis.

\subsection{WVF void population maps}\ \\
Figure~\ref{fig:topo} shows the watershed segmentation generated by
the mock catalogue. It is marked by the watershed boundaries superimposed 
on the contour maps of the density field of the full galaxy sample. These 
boundaries are visible as the white or thick black cell edges (dependent on 
the width of the boundary). The softer contour levels indicate the overdense regions 
in the $R_f=2.0\Mpch$ filtered full sample density field. 

For an impression of the ability to infer the correct watershed segmentation 
- and void population - from the magnitude-limited survey, we compare it 
with the segmentation for the full galaxy sample. In Fig.~\ref{fig:topoerr} 
the watershed boundaries of the latter are marked by red edges, while the 
ones for the magnitude-limited sample are marked by the black lines. 

At distances up to $R \sim 200\Mpch$ there is overall a reasonable agreement between 
the two void segmentations. Beyond that distance, we find that the larger 
segments of the magnitude-limited sample - a consequence of the decreasing 
structural resolution of the survey - encompass several smaller void segments 
from the full sample. Beyond that distance we also find some regions with 
strong differences between the two segmentations. The segmentation within the 
large voids in the two zoom-ins illustrate this. In the magnitude-limited survey 
segmentation, the weaker bridges between the smaller voids visible in the 
full sample have vanished. 

It is a clear illustration of the fact that at large distances the only 
topological information retained in the magnitude-limited density field is the 
skeleton defined by the strongest and most overdense filaments and walls, 
locations traced by the brightest galaxies. The more tenuous filigree of 
smaller filaments within the low density regions is lost.  

\subsection{Topological Error Definition:\\ \ \ \ \ \ \ \ False-splits and False-mergers.}
We evaluate the performance of the magnitude-limited survey watershed segmentation 
by comparing its void patches with those of the full galaxy sample. 

As described extensively in \cite{Platen07}, the errors can be classified, 
to first order, by {\it false splits} and {\it false mergers}. 
A {\it false split} is the situation in which a void segment from the 
reference field splits into two or more voidpatches. The reverse 
situation is that of a {\it false merger}, where two void segments 
in the reference field merge into one voidpatch in the segmentation 
of the magnitude-limited survey. 

\begin{figure}
   \includegraphics[width=0.4\textwidth]{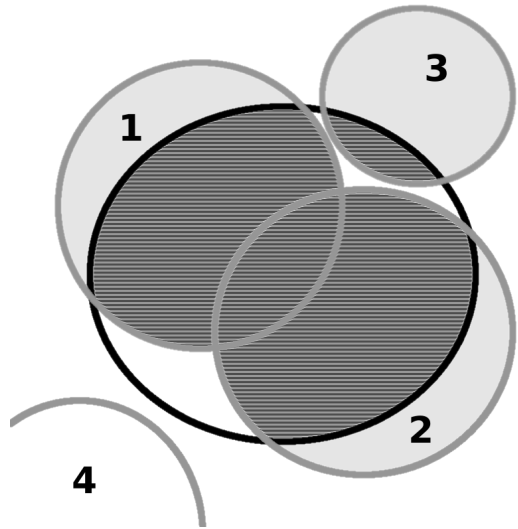}
    \caption{Definition of the false split measure. The four gray circles
    represent the patches from the reconstructed segmentation, the
    black circles would refer to a patch in the original
    segmentation. The gray shaded areas ${\cal A}_j$ belong to the areas of the
    three reconstructed patches that have an intersection with
    the black circle. The dashed areas represent intersections
    between the original and the reconstructed segmentation ${\cal A}_{j\cap O}$.}
  \label{fig:topo_errors}
 \end{figure}

\medskip
\noindent {\it False Split measure}\\
We define a measure for the false split errors. Starting from the 
void segmentation in the full sample reconstruction, we find for 
each void segment the overlapping void patches in the magnitude 
limited segmentation. The relative fraction of overlap of these 
survey void patches with the original void segment is used as a 
measure of significance of the overlap. If there are two or more 
significant overlaps, we classify the configuration as a 
false split.

Figure~\ref{fig:topo_errors} illustrates the above. 
The large black circle represents a voidpatch in the original segmentation, 
while the four gray circles are void segments in the survey segmentation. 
The gray shaded areas belong to the areas of the three reconstructed patches 
$j$ that intersect the black circle, while the resulting dashed areas 
represents the intersection of the original with each of these void patches. 
If the surface area of circle $j$ is equal to ${\cal A}_j$, and that of the 
overlap with the black original segment ${\cal A}_{j \cap O}$, then the 
false split measure
\begin{equation}
f_{j}^{FS}\,=\,{\displaystyle {\cal A}_{j \cap 0} \over \displaystyle {\cal A}_j}\,,
\end{equation}
\noindent decides on whether this is a significant overlap or not. We 
deem this to be so if $f_{j}^{FS} \geq 0.6$. In Fig.~\ref{fig:topo_errors},
void1 and void2 would correspond to significant overlaps. Void3 would 
be excluded as such and would only correspond to a slight shift of the boundary. 
Because of the two significantly overlapping voids, this configuration 
would be classified as a false split. 

\medskip
\noindent {\it False Merger measure}\\
An almost equivalent measure can be defined for a false merger. Since by 
symmetry, a false merger can be considered a false split in the full (original) 
sample segmentation, we may simply reverse the definition of the false split 
measure. If $l$ is a voidpatch in the original field, with area/volume ${\cal A}_l$, 
and its relative overlap with a large voidpatch in the magnitude-limited survey is 
\begin{equation}
f_{l}^{FM}\,=\,{\displaystyle {\cal A}_{l \cap O} \over \displaystyle {\cal A}_l}\,,
\end{equation}
we deem it a significant overlap when $f_{l}^{FM}>0.6$. When there are at least two original 
void segments that overlap significantly with one in the magnitude-limited survey, 
we may consider it a false merger. 

\medskip
\noindent {\it Correctly identified voids}\\
Having defined the false split and false merger errors, we may specify the 
meaning of a {\it correctly} identified patch. A {\it correct} void patch in the 
magnitude-limited survey is a void segment which overlaps for at least 60$\%$ with 
a void segment in the original field, as well as the other way around. These 
two conditions prevent a void from being either a false split or a false merger.

\subsection{Spatial Distribution Topological Errors}
Figure~\ref{fig:topoerr} shows the spatial distribution of the topological 
errors. The image contains the watershed segmentation for the full galaxy 
sample, indicated by the black solid lines. These are superimposed on the 
watershed segmentation for the magnitude-limited survey, indicated by the red 
solid lines. 

The false mergers are indicated by orange patches. The dark red 
patches represent the false splits, while the correctly reproduced 
patches are represented as white cells. It is directly apparent that 
false mergers are far more abundant than false splits.

A visual comparison between the two segmentations (also cf. Fig.~\ref{fig:topo})  
reveals the disappearance of void boundaries seen in the original full sample 
matter distribution. They disappear within the large void regions found 
in the magnitude limited survey. In other words, these minima are absorbed 
by one large encompassing void. It is a result of the tenuous and usually 
underdense nature of the walls in these regions, so that only a few galaxies 
have to fall out of the survey to evoke a merging of voids. 

\begin{figure}
  \centering
  \includegraphics[width=0.48\textwidth]{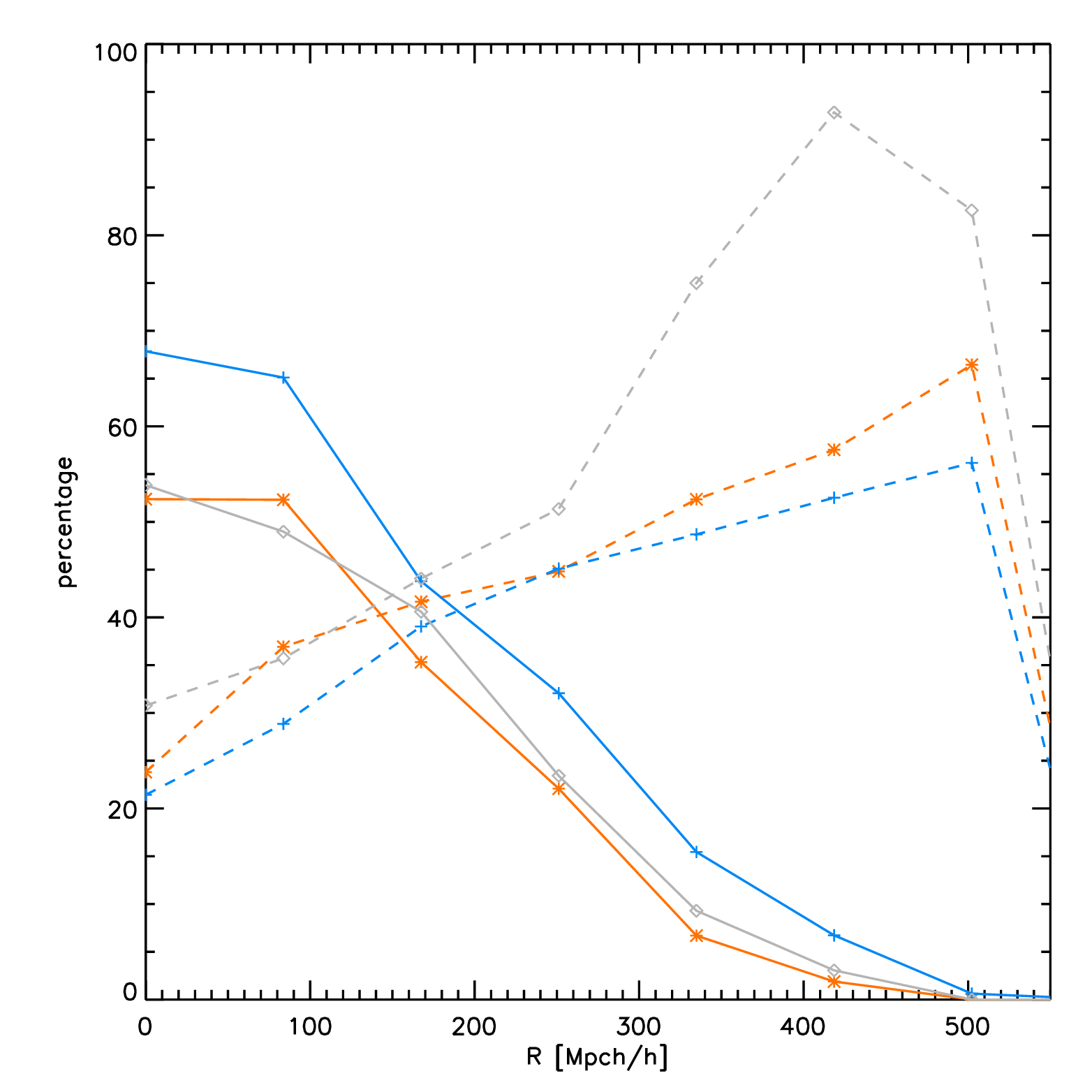}
  \includegraphics[width=0.48\textwidth]{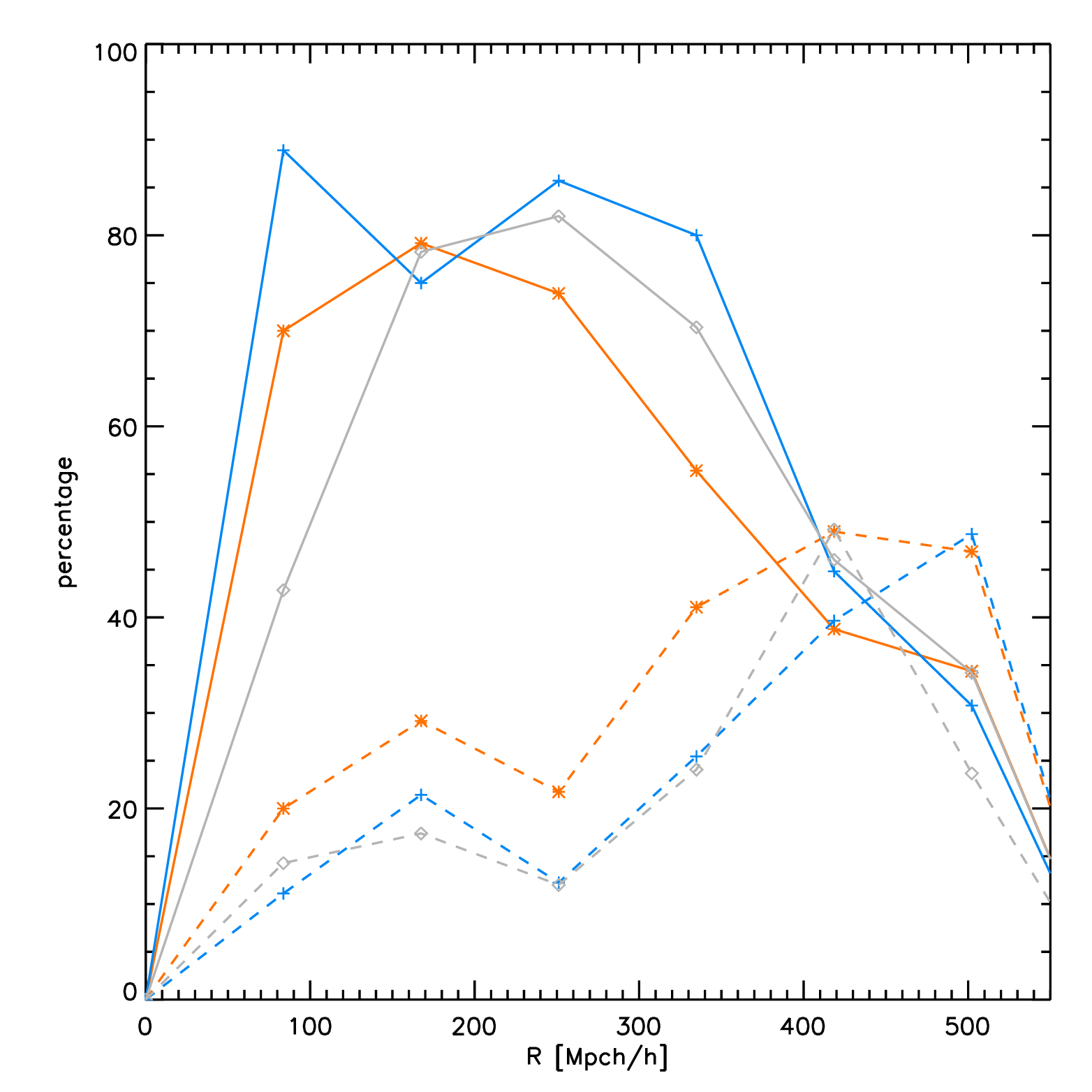}
  \caption{Correct and erroneous void identifications. Solid lines: 
   (radially averaged) percentage of correct voidpatch identifications as 
   a function of distance $R$. Dashed lines: (radially averaged) percentage 
   of erroneous voidpatch identifications. Blue: DTFE; Orange: NNFE; 
   Gray: Kriging. {\it Top}: density field reconstructions smoothed at 
   $R_f=3.0\Mpch$. {\it Bottom}: density field reconstructions at 
   $R_f=10.0\Mpch$.}
  \label{fig:averr3}
\end{figure}
\begin{figure}
  \includegraphics[width=0.48\textwidth]{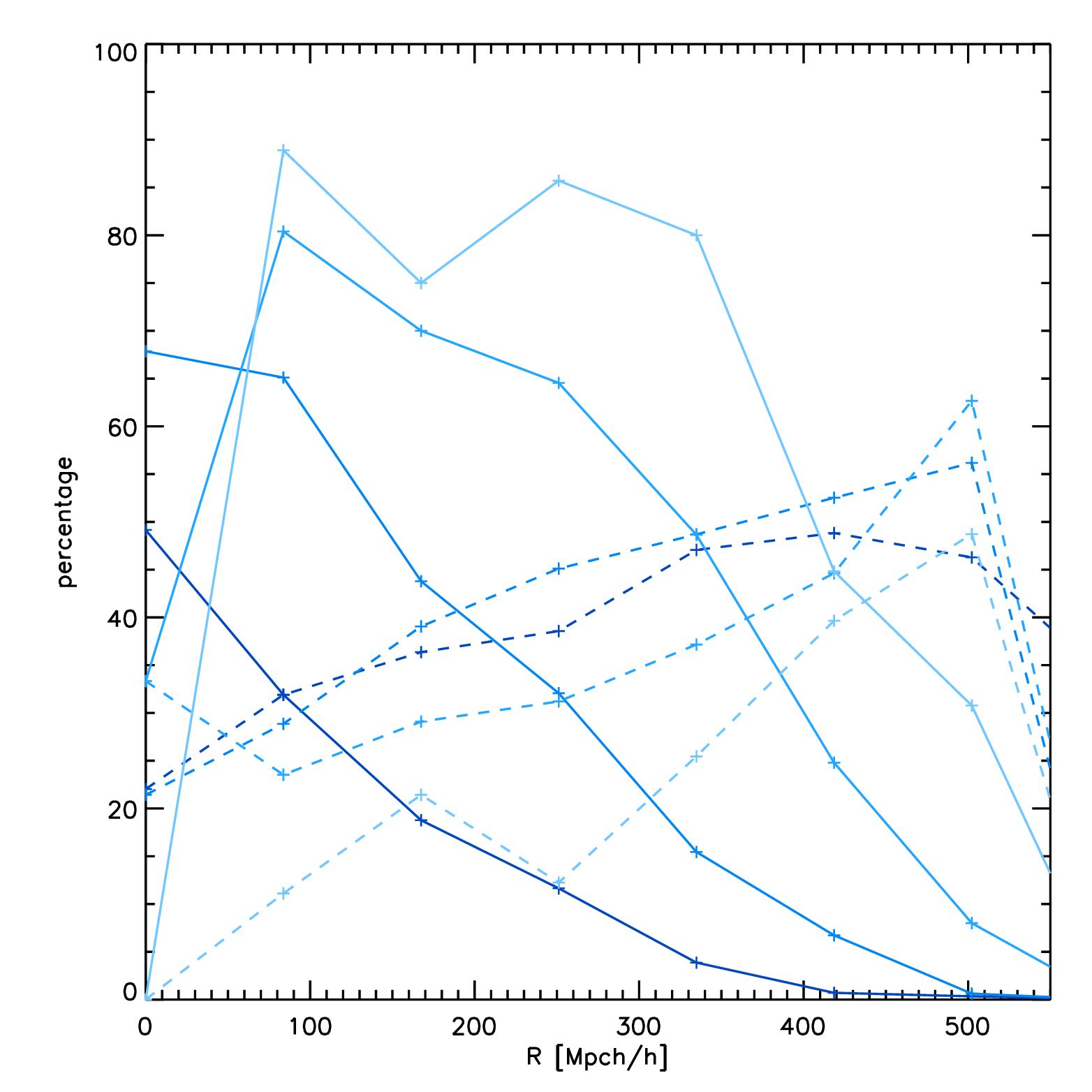}
    \caption{Correct and erroneous void identifications by DTFE. Solid lines: 
   (radially averaged) percentage of correct voidpatch identifications as 
   a function of distance $R$. Dashed lines: (radially averaged) percentage 
   of erroneous voidpatch identifications. Various line textures indicate 
   a range of filter scales: $R_f=1.0\Mpch$ (dark blue), 
   3.0$\Mpch$ , 6$\Mpch$ and 10$\Mpch$ (lightest blue).}
  \label{fig:averr2}
\end{figure}

Nonetheless, the coherence of the large persistent watershed lines remains 
strong throughout the volume. The defining skeleton of the cosmic mass distribution 
remains largely intact in a magnitude-limited survey. 

\subsection{Topological Error Characteristics}
The quality of the void representation of the survey can be inferred from 
the percentage of correctly identified voidpatches, as well as from the 
percentage of topological errors, ie. the total number of false splits 
and false mergers. Here we assess the correct identifications and topological 
errors as a function of distance $R$. We accomplish this by counting the number 
correctly identified void patches in radial shells, along with the 
number of error patches in the same shells. 

Note that the percentage of correct and of incorrect void identification do not 
always add up to 100 percent, since topological errors may be far more complex 
than just a false split and/or false merger. This happens with multiple additions 
or disappearances of void-walls. 

Fig.~\ref{fig:averr3} plots the identification percentages as function of radial 
distance $R$, 
for all three reconstruction techniques (DTFE: blue; NNFE: orange; Kriging: gray). 
The top panel concerns the $3\Mpch$ filtered field, the bottom panel the $10\Mpch$ 
filtered field. In the case of the $3\Mpch$ field we see a gradual and continuous 
decrease of correct void identifications as we move outward, while there is a 
corresponding increase of erroneous identifications. The $10\Mpch$ filtered 
shows an increase of correct identifications at a very close range. This is a 
consequence of the rapidly rising ability to outline the corresponding larger 
underdensities as we expand towards a larger volume starting from the small 
nearby cosmic volume. 

\begin{figure*}
  \centering
  \includegraphics[width=0.9\textwidth,clip]{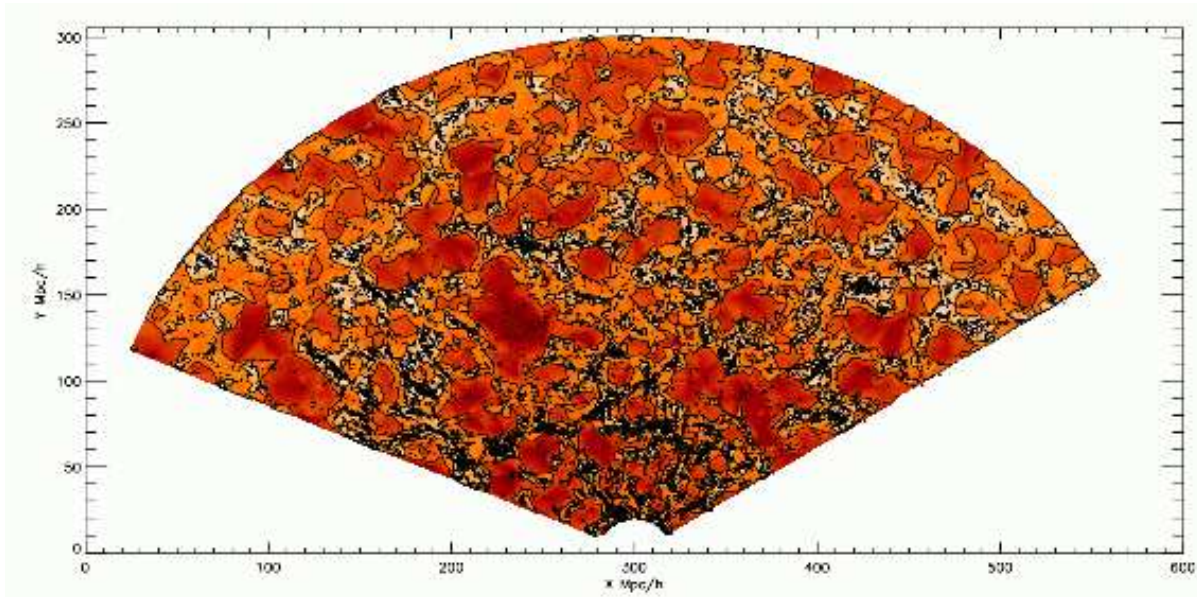} 
  \caption{Galaxy Distribution SDSS DR6. The magnitude-limited galaxy sample of the 
   SDSS DR6 survey is superimposed on the DTFE density field contours. The density field 
   section is infinitely thin, the galaxies lie within a slice of $10\hmpc$ width. The red  
   coloured contour levels represent the underdense regions, at $\rho/\rho_u=[0.001, 0.002, 0.005, 0.01,
    0.02, 0.03, 0.05, 0.07, 0.1, 0.2, 0.3, 0.6, 0.7, 0.8, 1]$. The heavy black 
    solid line is the mean cosmic density level, $\delta=0$. Within its contour 
    are the overdense regions, marked by black contour lines at density $\rho/\rho_u=[1., 3., 10.]$.}
  \label{fig:sdss_dtfe}
\end{figure*}
\begin{figure*}
  \centering
  \includegraphics[width=0.8\textwidth,clip]{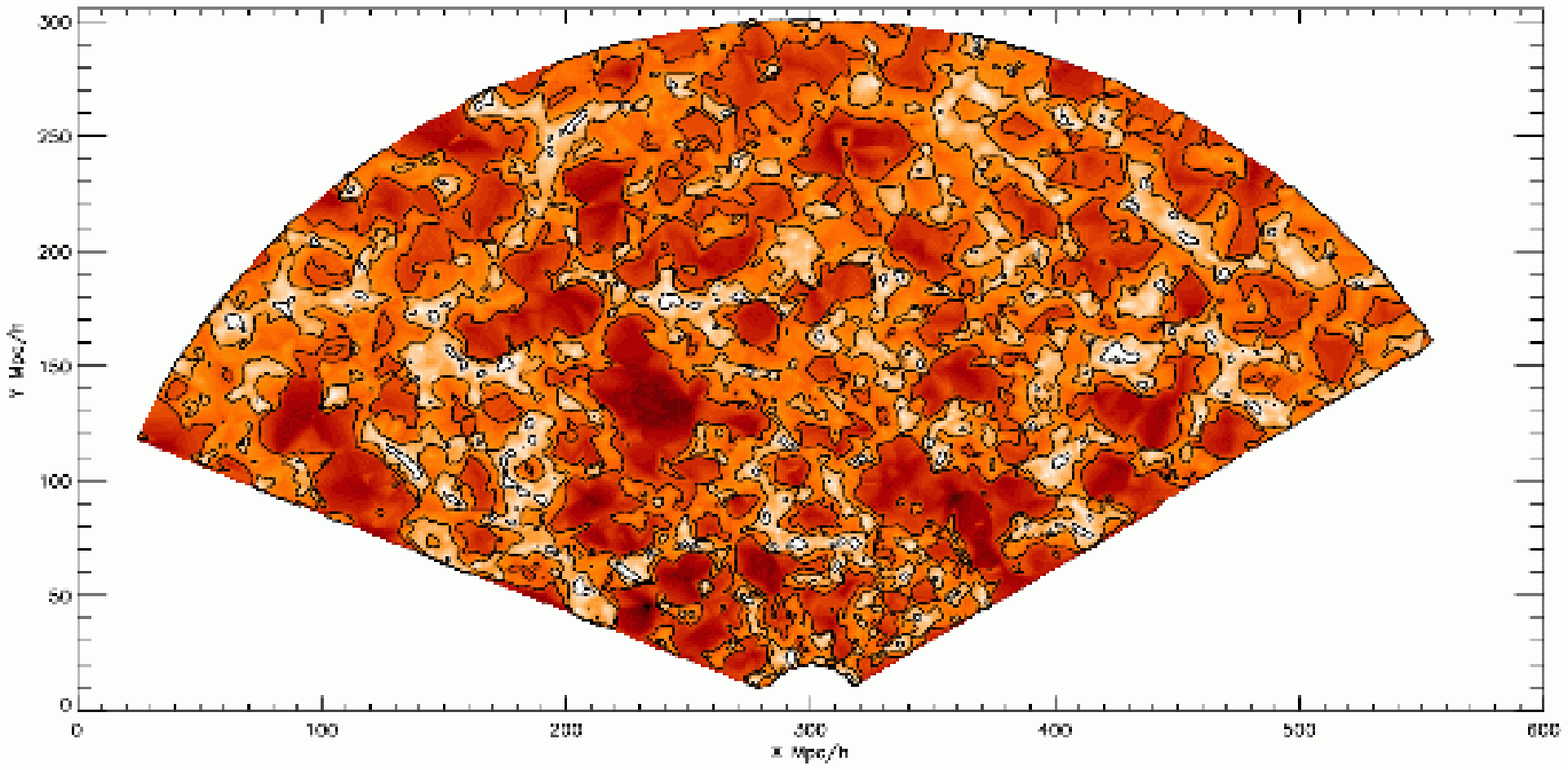}
  \includegraphics[width=0.8\textwidth]{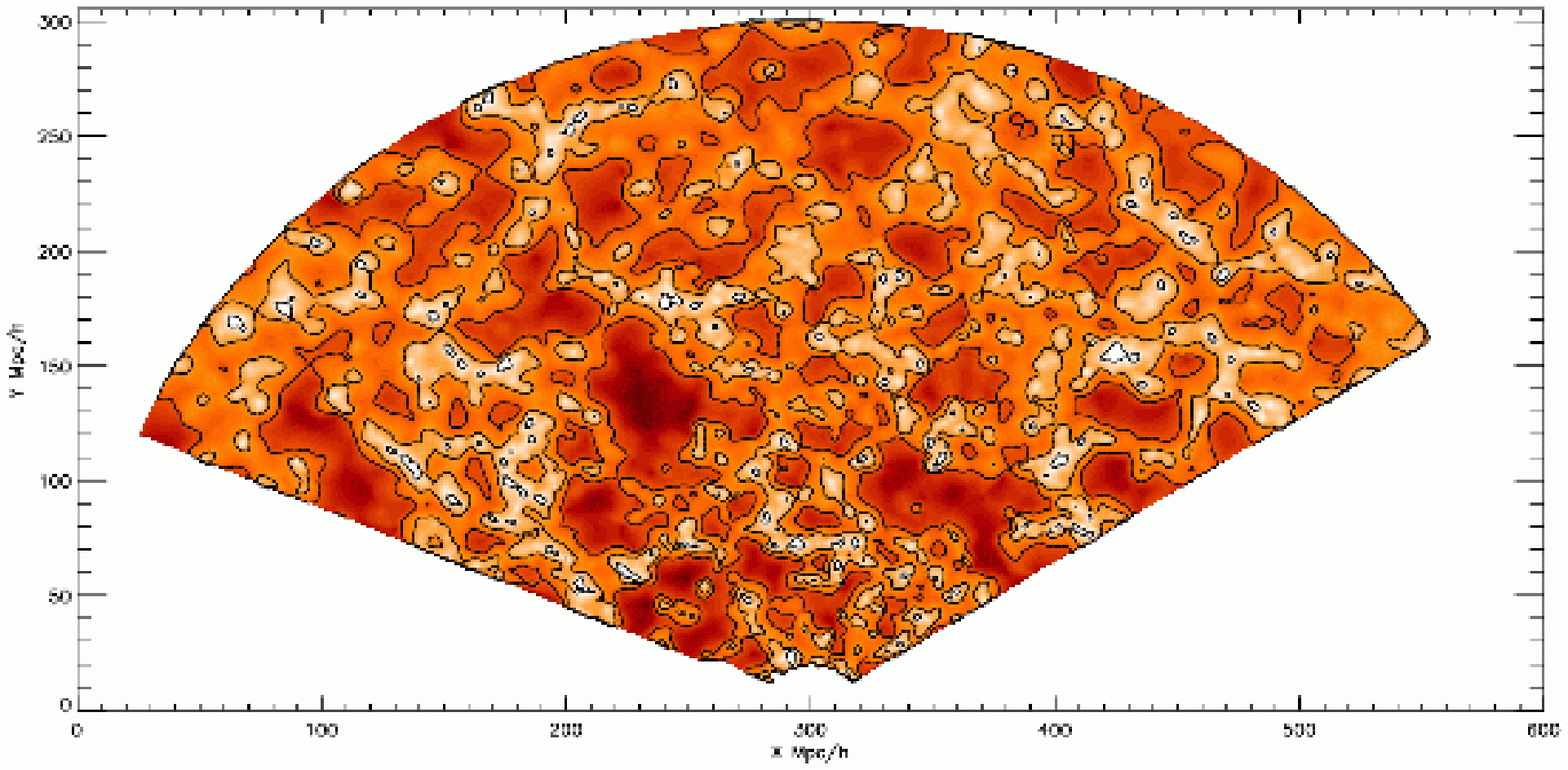}
  \includegraphics[width=0.8\textwidth]{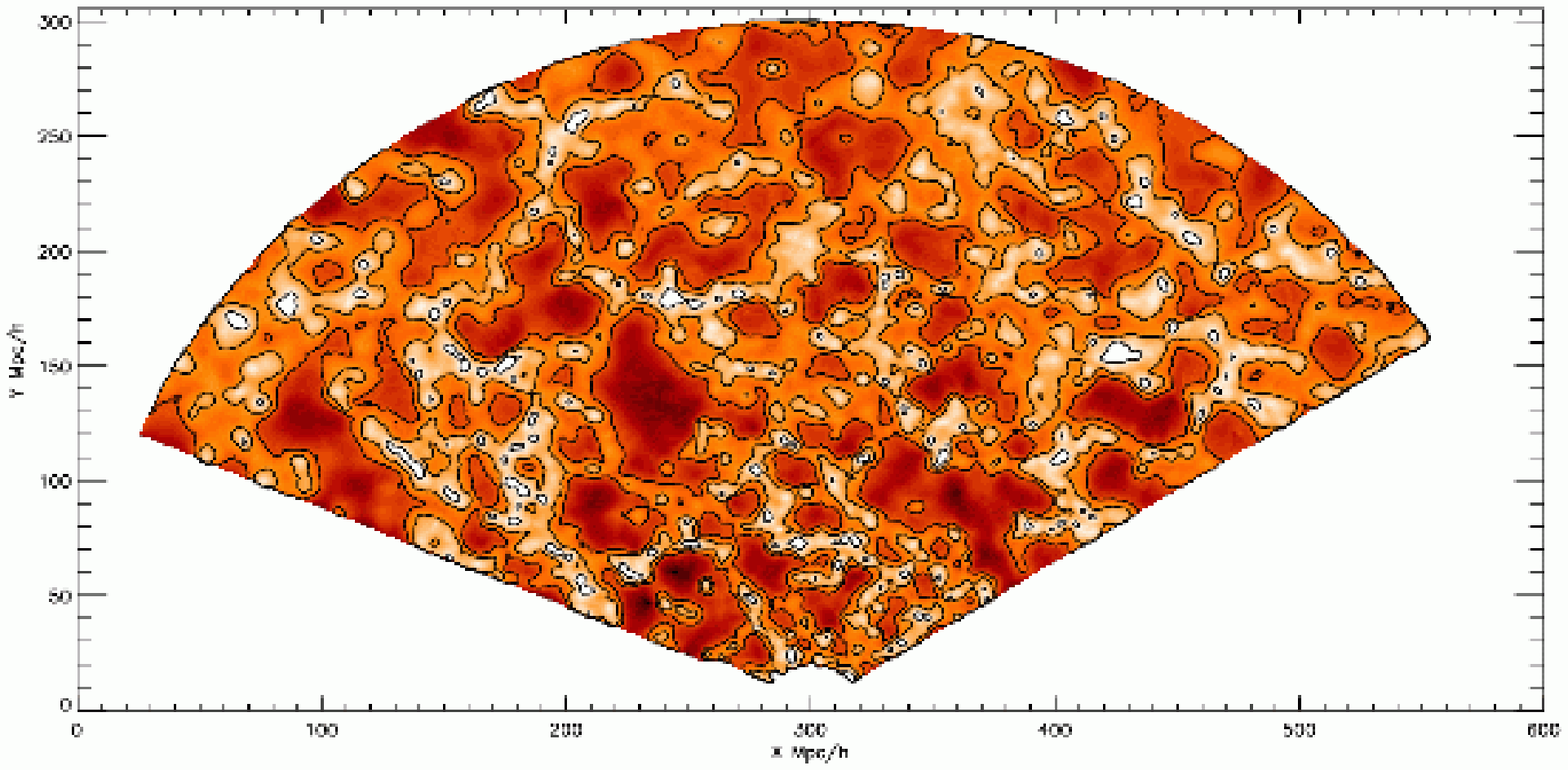}
  \caption{SDSS DR6 density field reconstructions. {\it Top}: DTFE; {\it Centre}: NNFE; 
  {\it Bottom}: Kriging. Shown are thin - zero width - slice sections through the reconstructed 
    density field. The red coloured contour levels represent the underdense regions, at 
    $\rho/\rho_u=[0.001, 0.002, 0.005, 0.01, 0.02, 0.03, 0.05, 0.07, 0.1, 0.2, 0.3, 0.6, 
    0.7, 0.8, 1]$. The heavy black solid line is the mean cosmic density level, $\delta=0$. Within 
    its contour are the overdense regions, marked by black contour lines at density 
    $\rho/\rho_u=[1., 3., 10.]$.} 
  \label{fig:sdss_high}
\end{figure*}

\medskip
\noindent {\it Filter Scale \& Void finding}\\
The DTFE reconstruction has been studied in somewhat more detail in 
Fig.~\ref{fig:averr2}. It plots the percentage of correctly and incorrectly 
identified void segments at a range of filter scales, running from
$R_f=1.0\Mpch$, $R_f=3.0\Mpch$, $R_f=6.0\Mpch$ to $R_f=10.0\Mpch$. 

For the larger filter scales, we find that the topology is 
reliably reconstructed within a range of $R \approx 100-300\Mpch$, while 
for the smaller filter scales this seems to at a much closer range from
$R\approx 0-200 \Mpch$. It is also interesting to see that on small 
scales, over the whole radial range, the number of correct identifications 
remains rather low and hardly exceeds $60\%$. On filter scales of 6$\Mpch$
and 10$\Mpch$ the performance is certainly better and consistently acceptable  
over a 200$\Mpch$ range. Only beyond $R\sim 300-400\Mpch$ the number of 
correct identifications drops below the 50$\%$ mark. 

\subsection{Topology Range}
In general, we find that the number of correctly identified voids decreases 
at larger distances, corresponding with a related increase of topological 
errors with distance. Dependent on the filter scale, we may define a distance 
$R_{max}$ out to which DTFE/WVF manages to identify a reasonable amount of voids from a
magnitude-limited survey. From Fig.~\ref{fig:averr2} we find:
\begin{itemize}
\item $R_f=$1$\Mpch$: $\qquad$\ \ \,$R_{max} \approx 100$$\Mpch$ 
\item $R_f=$3$\Mpch$: $\qquad$\ \ \,$R_{max} \approx 200$$\Mpch$
\item $R_f=$6$\Mpch$: $\qquad$\ \ \,$R_{max} \approx 300$$\Mpch$
\item $R_f=$10$\Mpch$: $\qquad$ $R_{max} \approx 400$$\Mpch$
\end{itemize}
Within this range, the fraction of erroneous void identifications remains well 
below unity: $R_{max}$ is roughly estimated from Fig.~\ref{fig:averr2} from 
the scale R at which the solid curves (correct identification) and dashed 
curves (erroneous identification) cross. 

Also, overall we find a slightly better performance by the DTFE reconstruction. 
Even though the higher order NNFE and Kriging techniques produce smoother 
void regions, this does not result in a higher number of correctly 
identified voidpatches.

\section{SDSS-DR6 density reconstruction}
\label{sec:sdssden}
Following the extensive analysis discussed in previous sections, we arrive 
at the application of the assessed technologies to the real world 
galaxy distribution in the 6th data release of SDSS. These resulting density 
maps form the starting point of the extensive statistical study - starting 
with a study of the one-point density distribution function and galaxy 
bias - presented in the subsequent papers discussing the density field 
and its cosmography. 

The DTFE, NNFE and Kriging density field reconstructions based on the 
(magnitude-limited) galaxy distribution in the 6th data release of SDSS are shown 
in Fig.~\ref{fig:sdss_high}. The density field has been reconstructed 
on a $512^3$ grid, representing a spatial resolution of $1.17$$\Mpch$. 

For reference, in Fig.~\ref{fig:sdss_dtfe} we show the spatial distribution 
of the SDSS DR7 galaxies on which the reconstruction is based. They are 
superimposed as black dots on the DTFE density field. 

Besides some minor differences, all three density maps show the same prominent 
features. The higher order NNFE and Kriging maps look smoother than the 
DTFE map. It is easy to recognize the almost one-to-one correspondence 
between the DTFE and NNFE map, with the DTFE map having the appearance 
of more noisy version. The Kriging map not only contains the same 
structures and features, it also has a more coherent appearance in which 
we can more easily recognize the global weblike morphology of the density 
field. Both the DTFE and NNFE maps have a more fragmented appearance. 

\section{Summary and Discussion}
\label{sec:summary}
This study is the first in a series in which we analyze the structure and 
topology of the Cosmic Web as traced by the Sloan Digital Sky Survey. 

In this study we investigate the ability of three reconstruction
techniques to analyze and investigate weblike features and geometries
in a discrete distribution of objects. The three methods are the
Delaunay Tessellation Field Estimator (DTFE), Natural Neighbour Field
Estimator (NNFE) and a local ``natural'' implementation of Kriging,
the {\it Natural Lognormal Kriging} technique. DTFE and NNFE are based on 
the local geometry defined by the Voronoi and Delaunay tessellations of 
the galaxy distribution. The Kriging formalism 
is adapted and optimized for the the approximate lognormal density 
distribution encountered in the mildly nonlinear cosmic web and 
is based on the logarithm of the measured density values. Also, we have
chosen to restrict the evaluations to a localized neighbourhood, the
{\it tetrahedral natural neighbourhood} based on the Delaunay
triangulation of the point sample. 

The three reconstruction methods are analysed and compared using mock
magnitude-limited and volume-limited SDSS redshift surveys, obtained
on the basis of the Millennium simulation. We investigate error
trends, biases and the topological structure of the resulting
fields. The reconstructed density fields, mainly from the
magnitude-limited mock survey samples but also from volume-limited
ones, are compared with the density field of the total simulation
galaxy sample. The differences between the various field 
reconstructions are investigated on the basis of an error analysis, 
mostly involving point-to-point comparisons. Environmental
effects are addressed by evaluating the density fields on a range of
Gaussian filter scales. With respect to the topology of the survey
fields, we concentrate on the void population identified by the
Watershed Void Finder \citep{Platen07}.  The void population in the
full density field and in the magnitude-limited survey density fields
are compared. The number of {\it false mergers} - in which original
voids emerge as a part of a larger void in the survey field - and of
{\it false splits} - in which an original void splits up in one or
more voids in the survey field - forms the basis of the topological
quality evaluation. 

By investigating the quality of the resulting density field estimates, over 
a range of scales and in different environments, as well as the more global
topological structure of the weblike network, we wish to identify and
understand the qualities of these techniques for the different
purposes and post-processing steps which are the subject of the
following papers in this series. The following observations were 
made:

\begin{itemize}
\item[$\bullet$] In most tests, DTFE, NNFE and Kriging have largely similar density 
and topology error behaviour. 
\item[$\bullet$] Cosmetically, higher order NNFE and 
Kriging methods produce more visually appealing reconstructions. 
\item[$\bullet$] Quantitatively, DTFE performs (marginally) better. Part 
of this at first sight surprising finding is a consequence of the 
higher sensitivity of the higher order NNFE and Kriging interpolation 
to intrinsic errors in the galaxy sample. An additional factor is the 
smaller natural neighbourhood of DTFE and NNFE with respect to the 
3 to 4 times larger neighbourhood of Natural Lognormal Kriging, which 
restricts density errors to smaller volumes. 
\item[$\bullet$]  With respect to the topological properties of the reconstructed density 
fields, it has become clear that a successful recovery of the void 
population on small scales is rather difficult. On these scales, the 
removal of only a couple of void galaxies leads to the spurious
merging of observed voids. 
\item[$\bullet$] The void recovery rate improves significantly 
at filter scales $> 3\Mpch$. The immediate repercussion is that a 
proper analysis of small scale voids, and void galaxies, has to be necessarily 
restricted to the local Universe out to at most 100$\Mpch$. As the environmental 
influences on the galaxy formation process seem to be mostly determined 
on these scales \citep{Park07b}, our study within this project, subject 
of a forthcoming paper in this series, will restrict itself mainly to this volume. 
\end{itemize}

\bigskip
A variety of technical improvements of our DTFE, NNFE and Kriging implementations 
may lead to a better performance. One immediate option is to invoke an image grid 
which has a more natural character for the galaxy survey context of our study 
than the cubic grid we have used in the evaluations described in this paper. 
Because of the flexibility of their definition, the Kriging formalism will 
be form a promising context for further adaptations and improvements. Of immediate 
importance is the implementation of non-local techniques to optimize the matrix 
inversion calculations while retaining the influence of large scale correlations and 
predictor-corrector methods to deal with oscillatory instabilities. While we 
have not yet investigated its performance, results of studies in other fields 
emphasize the importance of evaluating the performance of Radial Basis Function 
techniques as a possible alternative. 

\bigskip
The DTFE, NNFE and Kriging density field reconstructions form the basis of a series 
of studies in which we analyze the cosmography, void population and spinal structure 
of the local Cosmic Web in the Sloan Digital Sky Survey. Given the ease and efficiency 
of calculation, and its good quantitative behaviour, for most of these studies we 
use the DTFE formalism. However, the Natural Lognormal Kriging results looks very 
promising and appears to produces a well-behave coherent weblike density map of the 
SDSS survey. With the large advantage of controlling the error behaviour and 
properties of the reconstructed map, along with the large potential for extensions and 
optimizations of the method, the Natural Lognormal Kriging maps will play a 
dominant role in our study of the Cosmic Web. 

\bigskip
The density field reconstructions within the SDSS DR6, and subsequently SDSS DR7, 
volumes will be subjected to a statistical study in the second paper of this 
series. We will focus in particular on the 1-point probability function of the 
SDSS density field. With the help of corresponding magnitude-limited mock 
catalogues we will infer in how far the lognormal probability function, as well as 
related higher order versions, form a proper description of its statistical 
character. Also, the mock catalogues will allow us to assess the bias of 
the galaxy population in high density areas and, in particular, the low 
density void region. A cosmographic description of the reconstructed Local Universe, 
along with the identification of filamentary supercluster complexes, voids and 
supervoid complexes is the subject of the third paper in this series. The study of 
the void population in the SDSS density field, concerning a detailed assessment 
of the void size and shape properties, forms an important rationale behind the 
development of the density field reconstructions described in this study. This has 
become an interesting area of research, in particular as recent studies have emphasized 
the potential of extracting cosmological information from cosmic voids, in particular 
that concerning the dark energy equation of state \citep{LeePark09,Lavaux09,Biswas10}. 
In recent years, there has also been a strong interest in large scale environmental 
influences on galaxy properties and on the galaxy formation process. The tools 
described in the present study, will allow an assessment on the basis of 
a properly defined density field on quasi-linear scales. 

\section*{Acknowledgements}
EP likes to thank Alex Szalay for his hospitality and financial
support at JHU. Also we are thankful for the help from Gerard Lemson
and Sebastien Heinis with the Millennium Database queries. We would
like to thank Istvan Szapudi and Mark Neyrinck for their helpful
discussion. We are grateful to the referee, Michael Vogeley, for 
the many incisive and helpful remarks and recommendations. 

The Millennium Simulation databases used in this paper and the web
application providing online access to them were constructed as part
of the activities of the German Astrophysical Virtual Observatory
(GAVO).

Funding for the SDSS and SDSS-II has been provided by the Alfred P. Sloan Foundation, 
the Participating Institutions, the National Science Foundation, 
the U.S. Department of Energy, the National Aeronautics and Space Administration, 
the Japanese Monbukagakusho, the Max Planck Society, and the Higher Education Funding 
Council for England. The SDSS Web Site is http://www.sdss.org/.

The SDSS is managed by the Astrophysical Research Consortium for the Participating Institutions. 
The Participating Institutions are the American Museum of Natural History, 
Astrophysical Institute Potsdam, University of Basel, University of Cambridge, 
Case Western Reserve University, University of Chicago, Drexel University, Fermilab, 
the Institute for Advanced Study, the Japan Participation Group, Johns Hopkins University, 
the Joint Institute for Nuclear Astrophysics, the Kavli Institute for Particle Astrophysics 
and Cosmology, the Korean Scientist Group, the Chinese Academy of Sciences (LAMOST), 
Los Alamos National Laboratory, the Max-Planck-Institute for Astronomy (MPIA), 
the Max-Planck-Institute for Astrophysics (MPA), New Mexico State University, 
Ohio State University, University of Pittsburgh, University of Portsmouth, 
Princeton University, the United States Naval Observatory, and the University of Washington.

\appendix
\section{\\SDSS Coordinate System}
\label{app:XYZ}
For the analysis of our datasample, we transform the equatorial 
coordinates $(\alpha,\delta,z)$ of the DR6 NGP and DR7 galaxy sample 
to a grid based coordinate system (X,Y,Z). The observer is located 
at $(X,Y,Z)=(300.,0.,300.)$ $\Mpch$, while the centre of the 
northern strip is rotated to lie parallel to the Y-axis, starting 
at $(X,Z)=(300,300) \Mpch$. The corresponding transformation for 
an object with a comoving distance $R(z)=cz/100$ $\Mpch$ is 
defined by:
\begin{eqnarray}
  X &=& R(z)\cos(\delta)\cos(\alpha-90)\nonumber\\ 
  Y &=& R(z)\cos(\delta)\sin(\alpha-90)\\ 
  Z &=& R(z)\sin(\delta),\nonumber
\end{eqnarray}

\bigskip
\section{\\DTFE: Delaunay Tessellation Field Estimator}
\label{app:dtfe}
An extensive outline of the full DTFE procedure can be found in 
\cite{Weyschaap09}. For the specific application to the SDSS 
density field reconstruction, we follow the following steps of 
the DTFE procedure:
\begin{enumerate}
\item[$\bullet$] {\bf Point sample}\\ The (mock) galaxy samples are supposed to 
represent an unbiased sample of the underlying density field. It is 
therefore considered to be a general Poisson process of the 
underlying density field. 
\medskip
\item[$\bullet$] {\bf Boundary Conditions}\\ We assume {\it vacuum 
boundary conditions}: outside the galaxy sample volume we take the 
minimal assumption of having no points. 
\medskip
\item[$\bullet$] {\bf Delaunay Tessellation}\\ Construct the Delaunay
tessellation from the point sample using the \cite{cgal} library.
\medskip
\item[$\bullet$] {\bf Field values point sample}\\ 
The density values at the sampled points are determined from the corresponding Voronoi tessellations.
The estimate of the density at each sample point is the normalized inverse of the volume of its {\it contiguous} 
Voronoi cell ${\cal W}_i$ of each point $i$. The {\it contiguous Voronoi cell} of a point $i$ is the union of 
all Delaunay tetrahedra of which point $i$ forms one of the four vertices (see Fig.~\ref{fig:nnbr} for an 
illustration). We recognize two applicable situations:\\
\begin{enumerate}
\item[+] {\it Uniform sampling process}:\\ 
the point sample is an unbiased sample of the underlying density field. Typical 
example is that of $N$-body simulation particles. For $D$-dimensional space the density estimate is, 
\begin{equation}
{\widehat f}({\bf x}_i)\,=\,(1+D)\,\frac{m_i}{V({\cal W}_i)} \,.
\label{eq:densvor}
\end{equation}
\noindent with $m_i$ the ``mass'' of sample point $i$. This situation concerns the ``full'' mock galaxy 
samples and the volume-limited galaxy samples. \\
\item[+] {\it Systematic non-uniform sampling process}:\\ sampling density according to specified 
selection process quantified by an a priori known selection function $\psi({\bf x})$, varying as 
function of sky position $(\alpha,\delta)$ as well as depth/redshift. For $D$-dimensional space the 
density estimate is, 
\begin{equation}
{\widehat f}({\bf x}_i)\,=\,(1+D)\,\frac{m_i}{\psi({\bf x}_i)\,V({\cal W}_i)} \,.
\label{eq:densvornu}
\end{equation}
This situation is relevant for the magnitude- or flux-limited SDSS and mock galaxy samples. 
\end{enumerate}
\medskip
\item[$\bullet$] {\bf Field Gradient}\\ Calculation of the field
gradient estimate $\widehat{\nabla f}|_m$ in each $D$-dimensional
Delaunay simplex $m$ ($D=3$: tetrahedron; $D=2$: triangle) by solving
the set of linear equations for the field values at the positions of
the $(D+1)$ tetrahedron vertices,\\
\begin{eqnarray}
\widehat{\nabla f}|_m \ \ \Longleftarrow\ \ 
\begin{cases}
f_0 \ \ \ \ f_1 \ \ \ \ f_2 \ \ \ \ f_3 \\
\ \\
{\bf r}_0 \ \ \ \ {\bf r}_1 \ \ \ \ {\bf r}_2 \ \ \ \ {\bf r}_3 \\
\end{cases}\,
\label{eq:dtfegrad}
\end{eqnarray}
\medskip
\item[$\bullet$] {\bf Interpolation}.\\ The final basic step of the DTFE
procedure is the field interpolation. The processing and
post-processing steps involve numerous interpolation calculations, for
each of the involved locations $\widehat{\bf r}$.
\medskip
Given a location $\widehat{\bf r}$, the Delaunay tetrahedron $m$ in which it
is embedded is determined. On the basis of the field gradient
$\widehat{\nabla f}|_m$ the field value is computed by (linear)
interpolation,
\begin{equation}
{\widehat f}({\bf r})\,=\,{\widehat f}({\bf r}_{i})\,+\,{\widehat {\nabla f}} \bigl|_m \,\cdot\,({\bf r}-{\bf r}_{i}) \,.
\label{eq:fieldval}
\end{equation}

\medskip 
\item[$\bullet$] {\bf Processing}.\\ We make a distinction between straightforward 
processing steps concerning the production of images and simple smoothing filtering 
operations on the one hand, and more complex post-processing on the other hand. 
Basic to the processing steps is the determination of
field values following the interpolation procedure(s) outlined
above. Straightforward ``first line'' field operations are {\it
``Image reconstruction''} and, subsequently, {\it
``Smoothing/Filtering''}.  \\
\begin{enumerate}
\item[+] {\it Image reconstruction}:\\ 
For a set of {\it image points}, usually grid points, determine the {\it image
value}: formally the average field value within the corresponding
gridcell. For that purpose in this study we use the {\it Monte Carlo approach}: 
approximate the integral by taking the average over a number of (interpolated) field 
values probed at randomly distributed locations within the gridcell around an 
{\it image point}. The final estimate is obtained by averaging over the interpolated 
field values within a gridcell.

\smallskip 
For image reconstruction we need to assure ourselves that we obtain a 
sensible density estimate within a voxel element. In a spatially irregular 
sample, the Voronoi cell defines the natural Nyquist interval. To avoid 
aliasing, the number of interpolation points therefore needs to oversample 
the voxels of the image grid. One may also take the alternative and exact 
option of piecewise integrating the density of the Delaunay tetrahedra that 
are (partially) overlapping with the image voxel. For DTFE this can be 
accomplished exact and relatively fast, although the computational and 
geometric aspects are far from trivial. 
\medskip
\item[+] {\it Smoothing} and {\it Filtering}:\\
Linear filtering of the field ${\widehat f}$: convolution of the field 
${\widehat f}$ with a filter function $W_f({\bf r},{\bf y})$, usually user-specified, 
   \begin{equation}
     f_s({\bf r})\,=\,\int\,{\widehat f}({\bf r'})\, W_f({\bf r'},{\bf y})\,d{\bf r'}     
   \end{equation}
\end{enumerate}
\medskip
\item[$\bullet$] {\bf Post-processing}.\\ The real potential of DTFE
fields may be found in sophisticated applications, tuned towards
uncovering characteristics of the reconstructed fields.  An important
aspect of this involves the analysis of structures in the density
field. This can be finding voids, identifying cosmic structures or 
advanced statistical analysis of the density field. 
\end{enumerate}

\label{lastpage}

\end{document}